%% file: main.tex
\documentclass[journal,onecolumn]{IEEEtran}
\input{defns}

\usepackage[margin=1in]{geometry}
\usepackage{float}
\usepackage{hyperref}
\usepackage{booktabs}
\usepackage{tabularx}
\usepackage{array}
\usepackage{centernot}
\usepackage{xcolor}

\definecolor{DarkRed}{rgb}{0.8, 0.0, 0.0} 
\definecolor{DarkGreen}{rgb}{0.0, 0.7, 0.0}
\hypersetup{
    colorlinks=true,
    linkcolor=DarkRed,
    citecolor=DarkGreen
}

\newcommand{\rank}{\operatorname{rank}}

\newif\ifproofreading
\proofreadingfalse

\ifproofreading
    \newenvironment{unread}
        {\color{red}}
        {}
    \newenvironment{finalized}
        {\color{green!50!black}}
        {}
\else

\fi

\begin{document}
\title{On Dependence Measures Based on $f$-Divergence and $\Phi$-Entropy
\thanks{
This work was supported by the Hong Kong Research Grants Council (RGC) under grant number 14310025, and by the 1+1+1 CUHK–CUHK(SZ)–GDST Joint Collaboration Fund.}
}

\author{Chenyu Wang, Amin Gohari\\
\emph{\small Department of Information Engineering}\\
\emph{\small The Chinese University of Hong Kong}\\
{\small Email: \{wc025, agohari\}@ie.cuhk.edu.hk}
}

\maketitle

\begin{abstract}
We study dependence measures induced by $\Phi$-divergence (also known as $ f$-divergence in the literature) for finite random variables. We define conditional $\Phi$-mutual information through an exact chain rule and develop its structural properties and variational representations. But in general it lacks conditional symmetry and deterministic self-reduction. To tackle this, we introduce max-$\Phi$-mutual information $\wIPhi$, a variational envelope obtained by optimizing over $\IPhi$ quantities containing auxiliary random variables, which restores both properties and satisfies a subadditive chain rule. Both quantities reduce to Shannon's mutual information when $\Phi(t)=t\log t$. For smooth convex generators with $\Phi(0)<\infty$ and concave $t\mapsto t^2\Phi''(t)$, we establish an optimizer characterization and obtain a closed-form expression for the unconditional quantity in terms of $\Phi$-self-information. For the same class, we show that its private-randomness-assisted version is either an affine function of Shannon's mutual information, with coefficients depending only on $\Phi$, or identically infinite. This dichotomy is determined by the asymptotic growth of $\Phi$ and clarifies the sensitivity of max-$\Phi$-mutual information to local randomness. 

We further derive inner bounds on $\Phi$-ribbons from prescribed independence structures and prove that these bounds are optimal when $\Phi(t)=t\log t$. We also establish a multipartite $\Phi$-generalization of the Zhang--Yeung inequality and derive corresponding ribbon regions. Finally, we give sufficient conditions for reducing the computation of $\Phi$-strong data processing constants for Z-channel sources to a boundary optimization.
\footnote{This work was presented in part at the IEEE Symposium on Information Theory 2025 \cite{wangphient} and 2026 \cite{wangribbons}.}

\end{abstract}

\section{Introduction}

Shannon's mutual information plays several roles simultaneously. It is
the Kullback--Leibler (KL) divergence between a joint distribution and the
product of its marginals, it admits an exact chain rule, it
characterizes conditional independence, and it satisfies 
data processing. These structural
features make mutual information useful as a tool for manipulating information inequalities. 

A natural question is how much of this structure survives when the
KL divergence is replaced by a general
$\Phi$-divergence (commonly called $f$-divergence). The $\Phi$-divergence was originally proposed by Ali and Silvey \cite{ali1966general} as well as Csiszár \cite{csiszar2011information,csiszar1967information} and defined as 
\begin{align*}
    \DPhi(q_X\|p_X)
    &\defeq \sum_{x} p(x) \Phi\left(\frac{q(x)}{p(x)}\right)-\Phi(1),
\end{align*}
where $\Phi:(0,\infty)\to \R$ is convex.  
Here we do not impose the usual normalization $\Phi(1)=0$. When $\Phi(t)=t\log t$, the $\Phi$-divergence recovers the KL divergence $D(q_X\|p_X)$. Throughout, all random variables take values in finite sets. 

Following \cite{ziv1973functionals} and \cite[Eq. 3.10.1]{cohen1998comparisons}\footnote{For a broader generalization, refer to \cite{zakai1975generalization}.}, we define the \emph{$\Phi$-mutual information} by 
\begin{align}
    I_{\Phi}(A;B)\defeq D_{\Phi}(p_{AB}\|p_A\otimes p_B).\label{If1}
\end{align}

The above definition of $\IPhi(X;Y)$ has been extensively analyzed in various works, such as \cite{merhav2011data,tridenski2015ziv,hsu2018generalizing,cohen1993relative}. 
This quantity is symmetric $I_{\Phi}(X;Y)=I_{\Phi}(Y;X)$, nonnegative $I_{\Phi}(X;Y)\geq 0$, satisfies the data processing inequality $I_{\Phi}(X;YZ)\geq I_{\Phi}(X;Z)$, and is naturally related to the $\Phi$-entropy. When $\Phi(t)=t\log t$, the $\Phi$-mutual information recovers Shannon's mutual information $I(X;Y)$. 

In this paper, we define the conditional $\Phi$-mutual information as 
\begin{align}
    I_\Phi(X;Y| Z)
\defeq
I_\Phi(X;YZ)-I_\Phi(X;Z),\label{eqndefconfphi}
\end{align}
when $I_\Phi(X;YZ),I_\Phi(X;Z)<\infty$. 
To the best of our knowledge, this definition and its associated calculus have not been systematically studied.

This definition preserves the exact chain rule on the second coordinate, i.e., $$I_\Phi(X;YZ)=I_\Phi(X;Y| Z)+I_\Phi(X;Z),$$ and is therefore
suitable for telescoping and information-inequality arguments. Additionally, definition \eqref{eqndefconfphi} is  natural when we view $I_\Phi$ as a $\Phi$-entropy (as discussed later in Section \ref{subsec:Phi-MI-Phi-ent}; see equation \eqref{eqnDefCond}). In the special case of $Z$ being independent of $(X,Y)$, $I_\Phi(X;Y| Z)$ reduces to $I_\Phi(X;Y)$ which is symmetric. However, 
$I_\Phi(X;Y| Z)$ is
not generally \emph{conditionally symmetric}, i.e.,   
\begin{align*}
    \IPhi(X;Y|Z)\neq \IPhi(Y;X|Z),
\end{align*}
and does not generally admit the averaging
representation 
$$I_\Phi(X;Y|Z)\neq \sum_z p(z)I_\Phi(X;Y|Z=z).$$ 
Moreover, it fails \emph{conditional
deterministic self-reduction} on the second coordinate, i.e., when $Y=g(X,Z)$, $\IPhi(X;Y|Z)\neq \IPhi(Y;Y|Z)$. 

We introduce a variational envelope, denoted as $\wIPhi$, that restores several of the missing properties. Define the \emph{max-$\Phi$-mutual information} by 
\[
\wIPhi(X;Y)
\defeq
\sup_{U}
\left\{
I_\Phi(U;X)+I_\Phi(U;Y)-I_\Phi(U;XY)
\right\},
\]
where the supremum is over all conditional distributions
$p_{U|XY}$.  Its conditional counterpart is defined analogously.
Compared to $\IPhi$, the quantity $\wIPhi$ is conditionally symmetric
\begin{align*}
    \wIPhi(X;Y|Z)=\wIPhi(Y;X|Z),
\end{align*}
and
satisfies conditional deterministic self-reduction, although its chain rule is only subadditive. Observe that when $\Phi(t)=t\log t$, $\wIPhi(X;Y|Z)=I_\Phi(X;Y|Z)=I(X;Y|Z)$ recovering Shannon's conditional mutual information. 

Our contributions fall into three themes.
\begin{enumerate}[\rm (i)]

\item
\emph{Conditional calculus of $\IPhi$.} 
We develop a calculus for $I_\Phi$, establishing self-information bounds, invariance properties, monotonicity, and subadditivity under suitable assumptions on $\Phi$. Conditional $I_\Phi$ preserves an exact chain rule, but does not satisfy  conditional symmetry and deterministic self-reduction. For $\Phi\in\mathscr F$, we also derive a direct variational representation of conditional $I_\Phi$ that can be useful in machine learning. We also characterize the exceptional role of Shannon's mutual information: under the regularity assumptions of Theorem~\ref{thm:CDFI-forces-MI}, requiring either conditional symmetry or conditional deterministic self-reduction forces $I_\Phi$ to be a positive multiple of Shannon's mutual information. 

\item
\emph{Max-information $\wIPhi$ and private-randomness-assisted $\widehat{I}_{\Phi}$.}
We establish structural properties of the max-$\Phi$-mutual information $\wIPhi$ and characterize 
an optimizer when $\Phi\in\mathscr G$ and $\Phi(0)<\infty$,
where $\mathscr G$ is the class of convex functions such that $t\mapsto t^2\Phi''(t)$ is concave. 
In particular, for $\Phi\in \mathscr{G}$, we obtain the closed-form identity 
\[
    \wIPhi(X;Y)
    =
    I_\Phi(X;X)+I_\Phi(Y;Y)-I_\Phi(XY;XY).
\]
Under the same assumptions, we characterize the effect of
adjoining local private randomness. Writing
$\widehat I_\Phi(X;Y)$ for the supremum of $\wIPhi$ over such
extensions, we prove a dichotomy: either
\[
    \widehat I_\Phi(X;Y)
    =
    c_\Phi(\ln 2) I(X;Y)+K_\Phi,
\]
where $c_\Phi,K_\Phi\geq0$ depend only on $\Phi$, or
$\widehat I_\Phi$ is identically infinite.
The asymptotic growth of $\Phi$ determines which regime occurs.

\item
\emph{Ribbon inequalities and strong data processing.}

For $\Phi\in\mathscr F$, we derive a universal inner bound on
the $\Phi$-ribbon given by the convex hull of the incidence
vectors of mutually independent subsets.
For the HC ribbon, this bound is optimal among those determined
solely by the prescribed independence relations.
We also establish a multipartite $\Phi$-generalization of the
Zhang--Yeung inequality involving $I_\Phi$ and $\wIPhi$,
which yields additional ribbon regions from conditional
independence relations. In a concurrent work, this result can be generalized to the matrix case \cite{wangmatrix}. 

For Z-channel sources, we give sufficient conditions on $\Phi$
under which the optimization defining the $\Phi$-SDPI constant
can be restricted to functions satisfying $f(1)=0$ when
$\Phi(0)<\infty$. When $\Phi(0)=+\infty$, we evaluate the
corresponding boundary limit.

\end{enumerate}

\subsection{Organization}

The remainder of the paper is organized as follows.
Section~\ref{sec:Phi-MI} introduces the 
$\Phi$-mutual information $\IPhi$, its conditional form, and its basic
calculus. Section~\ref{sec:max-Phi-MI} studies the variational quantity
$\wIPhi$, including its structural properties, closed-form
expression, and private-randomness-assisted $\widehat I_\Phi$. Section~\ref{sec:characterization-ribbon} derives ribbon
regions from independence structures. Section~\ref{sec:non-Shannon} derives ribbon regions from
non-Shannon-type information inequalities. In Section~\ref{sec:SDPI}, we establish sufficient conditions on $\Phi$ under which, when $\Phi(0)<\infty$, the optimization defining the $\Phi$-SDPI constant of a Z-channel source can be restricted to functions satisfying $f(1)=0$.

\subsection{Notations}

We write $\R_+\defeq[0,\infty)$ and $\R_{++}\defeq(0,\infty)$.
For a nonnegative integer $n$, let $[n]\defeq\{1,\ldots,n\}$. For a finite set $S$, $|S|$ denotes its cardinality, and $\cP(S)$ (also written $2^S$) denotes its power set. We use $\Conv(S)$ for the convex hull of a subset $S$ of a real vector space and $\F_q$ for the finite field with $q$ elements, where
$q$ is a prime power. The notation $\rank(L)$ denotes the rank of a matrix or linear map $L$. For $t\in\R$, let $t_+\defeq\max\{t,0\}$. 

All random variables are assumed to take values in finite sets. Capital letters denote random variables, lowercase letters their realizations, and calligraphic letters their alphabets. For $S\subseteq[n]$, set
$X_S\defeq(X_i:i\in S)$, and $\cX_S\defeq\prod_{i\in S}\cX_i$; $x_S$ denotes a realization of $X_S$. In particular, $X_{[n]}=X^n=(X_1,\ldots,X_n)$. 
We use the same indexing convention for other families,
such as $A_{[n]}$, $B_{[n]}$, $\Pi_{[n]}$, and $\Omega_{[n]}$. 
Vectors of parameters are written in bold, e.g., $\vec\lambda=(\lambda_1,\ldots,\lambda_n)$. We write $\vec e_i$ 
for the $i$-th standard basis vector and
$\vec e_S\defeq\sum_{i\in S}\vec e_i$. 
The notation $X\sim p_X$ means that $X$ has law $p_X$, and we denote 
\[
\supp(X)=\supp(p_X)\defeq\{x\in\cX:p_X(x)>0\}. 
\]
We write $p\ll q$ if $\supp(p)\subseteq\supp(q)$, and
$p_X\otimes p_Y$ for the product law with mass function $p(x)p(y)$.
The relation $X\independent Y$ denotes independence,
$X\independent Y|Z$ denotes conditional independence given $Z$, 
and $U\to X\to Y$ denotes a Markov chain, equivalently
$U\independent Y|X$. We use $\Bern(\theta)$  for the distribution on $\{0,1\}$ assigning 
probability $\theta$ to $1$, and $\Unif(S)$ for the uniform distribution on a nonempty finite set $S$. 

Let $X$ be a random variable taking values in the finite set $\cX$.
Then a function $f:\cX\to\R$ can be thought of as a random variable. 
The expectation and variance of $f$ are denoted by 
$\E[f]$ and $\Var(f)$, respectively. 
We sometimes denote the expectation by $\E_X[f]$ to emphasize that it is computed 
with respect to $X$.

Let $f=f_{XY}=f(X,Y)$ be a function of two random variables
$(X,Y)$ with the joint distribution $p_{XY}$.
Then the conditional expectation of $f$ over $Y$ given a fixed value for $X$, denoted as $\E[f|X]$, is a function of $X$. To be precise, 
\[
\E[f| X](x)
=
\E[f| X=x]
=
\sum_y p(y| x)f(x,y).
\]
For simplicity, we sometimes denote $f_X=\E[f|X]$. 

We use $\log$ for logarithm of base 2 and $\ln$ for natural logarithm. 

Unless a different domain is specified, $\Phi:(0,\infty)\to\R$
is convex and is extended to zero by
\[
\Phi(0)\defeq\lim_{t\downarrow0}\Phi(t)
\in\R\cup\{+\infty\}.
\]
Convexity ensures that this limit exists. This convention gives a
lower semicontinuous convex extension to $[0,\infty)$, which is
continuous at zero when $\Phi(0)<\infty$. We write $\dom(\Phi)$
for the domain of $\Phi$.

\section{The $\Phi$-mutual information $\IPhi$} 
\label{sec:Phi-MI}

\subsection{Definition and $\Phi$-entropy representation} 
\label{subsec:Phi-MI-Phi-ent}

By definition, we have 
\begin{align}
    I_\Phi(X;Y)
    &=
    D_\Phi(p_{XY}\|p_X\otimes p_Y) \label{eq:IPhi-def}\\
    &=\sum_x p(x) D_\Phi(p_{Y|x}\|p_Y). \nonumber
\end{align}
In the literature, there are other definitions of $\Phi$-mutual information, shown in Table \ref{tab:Phi-MI-definitions}. 
We adopt this definition since \eqref{eq:IPhi-def} admits a natural representation in terms
of $\Phi$-entropy. This representation is essential for developing the calculus of $\IPhi$. 

\begin{table}[H]
\centering
\caption{Divergence-based notions of mutual information}
\label{tab:Phi-MI-definitions}
\renewcommand{\arraystretch}{1.35}
\setlength{\tabcolsep}{5pt}
\begin{tabularx}{\textwidth}{
>{\raggedright\arraybackslash}p{0.15\textwidth}
>{\centering\arraybackslash}X
>{\raggedright\arraybackslash}p{0.25\textwidth}}
\toprule
\textbf{Notation} & \textbf{Definition} & \textbf{Authors and references} \\
\midrule

$I_\Phi(A;B)$
&
$\displaystyle
D_\Phi\!\left(
p_{AB}\middle\|p_A\otimes p_B
\right)$
&
Ziv and Zakai \cite{ziv1973functionals};
Cohen, Kemperman, and Zbaganu
\cite[Eq.~3.10.1]{cohen1998comparisons}.
See also \cite{zakai1975generalization}.
\\

$I_\Phi^{\mathsf{PV}}(A;B)$
&
$\displaystyle
\inf_{q_B}
D_\Phi\!\left(
p_{AB}\middle\|p_A\otimes q_B
\right)$
&
Polyanskiy and Verdú
\cite[Eq.~79]{polyanskiy2010arimoto}.
\\

$I_\Phi^{\mathsf{MBGYA}}(A;B)$
&
$\displaystyle
\inf_{q_B}
\left\{
\begin{aligned}
& D_\Phi\!\left(
p_{AB}\middle\|p_A\otimes q_B
\right) \\
& \quad - D_\Phi\!\left(p_B\middle\|q_B\right)
\end{aligned}
\right\}$
&
Mojahedian, Beigi, Gohari, Yassaee, and Aref
\cite[Appendix~B]{mojahedian2019correlation}.
\\

$I_\alpha^{\mathsf{LP}}(A;B)$
&
$\displaystyle
\inf_{q_A,q_B}
D_\alpha\!\left(
p_{AB}\middle\|q_A\otimes q_B
\right)$ where $D_\alpha$ denotes the R\'enyi divergence of order $\alpha$
&
Lapidoth and Pfister
\cite{lapidoth2018testing}.
\\

\bottomrule
\end{tabularx}
\end{table}

Given a convex function $\Phi:\R_+\to\R$, a function $f:\cX\to \R_+$, and a random variable $X$ on $\cX$, the $\Phi$-entropy of $f$ is defined to be (see \cite{Boucheron2013concentration,chafai2004entropies,chafai2006binomial})
\begin{align*}
    \HPhi(f) &\defeq \E[\Phi(f)] - \Phi(\E[f]).
\end{align*}
Given some arbitrary $p_{XY}$, define $f_x(y)\defeq \frac{p(x,y)}{p(x)p(y)}$. Note that $f_x(y)\geq 0$ for any $x,y$ and $\E_Y[f_x]=1$ for any $x$. Then we have 
\begin{align}
    \IPhi(X;Y)
    &=\sum_{x,y} p(x) p(y) \Phi \left(\frac{p(x,y)}{p(x)p(y)}\right) -\Phi(1) \nonumber\\
    &=\sum_{x} p(x) \HPhi(f_x) \label{rel:Phi-MI-Ent},
\end{align}
where $\HPhi(f _x)$ is calculated using $p_Y$ for every $x$.  

Given a function $f$ of two random variables $(X,Y)$, we define the \emph{conditional $\Phi$-entropy} as 
\begin{align}
    H_\Phi(f|Y)& \defeq\E[\Phi(f)]-\mathbb{E}_Y[\Phi(\E[f|Y])]\nonumber\\
    &= \sum_{y}p(y)\big(\E[\Phi(f)|Y=y] - \Phi(\E [f|Y=y])\big)\nonumber\\
    &=\HPhi(f)-\HPhi(\E[f|Y]) \label{Phi-ent-CR}.
\end{align}
Note that \eqref{Phi-ent-CR} is the chain rule for the $\Phi$-entropy:
\begin{align*}
    \HPhi(f)
    &=H_\Phi(f|Y)
    +\HPhi(\E[f|Y]).
\end{align*}
This definition of the conditional $\Phi$-entropy has the chain rule. Since $\IPhi$ can be represented in $\Phi$-entropy, we can analogously define the conditional $\Phi$-mutual information in order to preserve the chain rule.

Consider the functions $(f_x(y,z))_{x\in \cX}$ where $f_x(y,z)\defeq \frac{p(x,y,z)}{p(x)p(y,z)}$. Then $\E_Y[f_x|Z=z]=\sum_{y} p(y|z)\frac{p(x,y,z)}{p(x)p(y,z)}=\frac{p(x,z)}{p(x)p(z)}$. By \eqref{rel:Phi-MI-Ent}, we have 
\begin{align*}
    \IPhi(X;YZ)=\sum_x p(x) \HPhi(f_x), 
\end{align*} 
and 
\begin{align*}
    \IPhi(X;Z)=\sum_x p(x) \HPhi(\E[f_x|Z]). 
\end{align*}

This chain rule and the \(\Phi\)-entropy representation of \(I_\Phi\) suggest the following definition of conditional \(\Phi\)-mutual information. 
Define the \emph{conditional $\Phi$-mutual information} as
\begin{align*}
    \IPhi(X;Y|Z) \defeq \IPhi(X;YZ)-\IPhi(X;Z). 
\end{align*}
Thus it can be represented as 
\begin{align}
    \IPhi(X;Y|Z) 
    &=\sum_x p(x) \Big(\HPhi(f_x)
    -\HPhi(\E[f_x|Z])\Big)\nonumber\\
    &=\sum_x p(x) \HPhi(f_x|Z). \label{eqnDefCond}
\end{align}

This definition preserves the chain rule
\[
    I_\Phi(X;YZ)=I_\Phi(X;Z)+I_\Phi(X;Y|Z).
\]
Furthermore, 
\begin{align*}
    I_\Phi(U;XY|Z)=I_\Phi(U;X|YZ)+I_\Phi(U;Y|Z).
\end{align*}

\subsection{Basic properties and calculus of $\IPhi$}
In addition to the chain rule property on the second coordinate \begin{align*}
    I_\Phi(U;XY|Z)=I_\Phi(U;X|YZ)+I_\Phi(U;Y|Z),
\end{align*}
we have the following properties:
\begin{proposition} \label{prop:properties-Phi-MI}
Let \(X,Y,Z,W,U\) be finite random variables and $\Phi:\R_+\to \R$ be convex. The following properties hold.

\begin{enumerate}[\rm (i)]
    \item \label{Phi-MI-self-reduction} \textbf{Deterministic self-reduction.} (\cite[Theorem 5]{zvarova1974measures})
    If \(Y=g(X)\), then 
    \begin{align}
        I_\Phi(X;Y)=I_\Phi(Y;Y).
        \label{Phi-MI-func}
    \end{align}
    Moreover, if $\Phi$ is strictly convex, 
    \begin{align*}
        I_\Phi(X;Y)=I_\Phi(Y;Y)
        \quad
        \Longleftrightarrow\quad 
        Y=g(X).
    \end{align*}

    \item \textbf{Markov faithfulness.}
    If \(X\to Z\to Y\) is a Markov chain, then
    \begin{align}
        I_\Phi(X;Y|Z)=0.
        \label{Phi-MI-Markov}
    \end{align}
    If, in addition, \(\Phi\) is strictly convex, then the converse also holds:
    \begin{align}
        I_\Phi(X;Y|Z)=0
        \quad\Longleftrightarrow\quad
        X\to Z\to Y .
        \label{Phi-MI-Markov2}
    \end{align}

    \item \textbf{Idempotence.}
    We have
    \begin{align}
        I_\Phi(X;XY|Z)=I_\Phi(X;X|Z).
        \label{Phi-MI-redundant-RV}
    \end{align}

    \item \textbf{Permutation invariance.} 
    Suppose $X,Y,Z$ take values in $\cX,\cY,\cZ$ respectively. 
    Let $f:\cX\to \cA$, $g:\cY\to \cB$ and $h:\cZ\to \cC$ be one-to-one maps. Then 
    \begin{align} \label{permutation-invariance}
        \IPhi(X;Y|Z)=\IPhi(f(X);g(Y)|h(Z)). 
    \end{align}
    We have
    \begin{align}
        I_\Phi(X;YY|Z)=I_\Phi(X;Y|Z).
        \label{Phi-MI-redundant-RV-2}
    \end{align}

    \item \textbf{Self-information bound.} (\cite[Theorem 3]{zvarova1974measures})
    We have
    \begin{align}
        I_\Phi(X;Y)
        \le
        \min\{I_\Phi(X;X),I_\Phi(Y;Y)\}.
        \label{Phi-MI-UB-self}
    \end{align}
    
    \item \textbf{Conditional absorption on the second argument.} \label{absorption}
    \begin{align} \label{Phi-MI-arg-cond}
        \IPhi(X;YZ|Z)=\IPhi(X;Y|Z). 
    \end{align}

    \item \textbf{Data processing on the second argument.} \label{IPhi-data-processing}
    We have
    \begin{align}
        I_\Phi(W;X|Z)\le I_\Phi(W;XY|Z).
        \label{Phi-MI-more-RV}
    \end{align}
    
\end{enumerate}

\end{proposition}

The proof is in Appendix \ref{app:proof-prop:properties-Phi-MI}. 
For $\Phi(t)=\left(\sqrt{t}-1\right)^2$, one can verify that 
$$s_\Phi=\lim_{t\to \infty}\frac{\Phi(t)-\Phi(0)}{t}<\infty.$$
Thus, $\IPhi(X;Y)$ is universally upper bounded in this case. 

In Proposition \ref{prop:properties-Phi-MI}, there are two properties, i.e., \ref{absorption} and \ref{IPhi-data-processing}, that only hold for the second argument. We provide examples showing these properties may not hold for the first argument in the following proposition. 

\begin{proposition} \label{prop:IPhi-counterexample}
    \begin{enumerate}[\rm (i)]
        \item Suppose $\Phi(t)=(t-1)^2$. There exist random variables $X,Y,Z$ such that 
        $$\IPhi(XZ;Y|Z)> \IPhi(X;Y|Z).$$
        \item Suppose $\Phi(t)=(\sqrt{t}-1)^2$. There exist random variables $W,X,Y,Z$ such that 
        $$\IPhi(X;Y|Z)>\IPhi(WX;Y|Z).$$      
    \end{enumerate}
\end{proposition}

The proof is in Appendix \ref{app:proof-prop:IPhi-counterexample}.

Conditional deterministic self-reduction generally fails. Under the assumptions below, it characterizes positive multiples of Shannon’s mutual information.

\begin{theorem} \label{thm:CDFI-forces-MI}
    Suppose $\Phi:\R_+\to \R$ is strictly convex, twice continuously differentiable on $(0,\infty)$, and $\Phi(0)<\infty$.
    \begin{enumerate}[\rm (i)]
        \item Assume that $\IPhi$ satisfies conditional deterministic self-reduction, i.e., for arbitrary $X,Y,Z,f$ such that $Y=f(X,Z)$, we have $\IPhi(X;Y|Z)= \IPhi(Y;Y|Z)$. 
        Then $\IPhi$ is conditionally symmetric, i.e., for any $X,Y,Z$, $\IPhi(X;Y|Z)=\IPhi(Y;X|Z)$. 
        \item If $\IPhi$ is conditionally symmetric, $\IPhi$ is a positive multiple of Shannon's mutual information. 
    \end{enumerate}

\end{theorem}

The proof is in Appendix \ref{app:proof-thm:CDFI-forces-MI}.

Suppose we have access to $P_X$ and $P_Y$, but not $P_{XY}$. We have the following upper bounds on $I_\Phi(X;Y)$:
$$I_\Phi(X;Y)\leq \min\{I_\Phi(X;X),I_\Phi(Y;Y)\}.$$
In Appendix~\ref{app:max-IPhi-coupling}, we show that one can get strictly better upper bounds on $I_\Phi(X;Y)$ that depend only on the marginal distributions of $X$ and $Y$.

\subsection{Bounds on self-information}
\label{subsec:self-information-bounds}

For Shannon's mutual information, self-information equals entropy:
$\max_{p_X}I(X;X)=\max_{p_X}H(X)=\log|\cX|$, where the maximum is achieved by the uniform distribution on $\cX$. We study analogous optimization problems for $I_\Phi$, first on a fixed
alphabet and then over all finite alphabets.

If $\Phi(0)=+\infty$, then $I_\Phi(X;X)=+\infty$ for every
nonconstant $X$. Throughout the remainder of this subsection, we
therefore assume that $\Phi(0)<\infty$. By definition,
\begin{align}
    I_\Phi(X;X)
    &=
    \sum_{x:p_X(x)>0}
    p_X(x)^2\Phi\left(\frac{1}{p_X(x)}\right)
    +
    \left(1-\sum_{x\in\cX}p_X(x)^2\right)\Phi(0)
    -\Phi(1).
    \label{eq:self-info}
\end{align}
In particular, if $A$ is uniform on a set of size $m$, then
\begin{equation}
    I_\Phi(A;A)
    =
    \frac{1}{m}\Phi(m)
    +\left(1-\frac{1}{m}\right)\Phi(0)
    -\Phi(1).
    \label{eq:self-uniform}
\end{equation}

\subsubsection{Optimization on a fixed alphabet}

For a fixed finite alphabet $\cX$, consider
\begin{equation}
    \sup_{p_X} I_\Phi(X;X),
    \label{Phi-MI-self-info}
\end{equation}
where the supremum is over all probability distributions on $\cX$.
To express this problem in separable form, define
\begin{equation}
    h_\Phi(p)
    \defeq
    \begin{cases}
        p^2\left[
            \Phi\left(\dfrac{1}{p}\right)-\Phi(0)
        \right],
        & p\in(0,1],\\[1ex]
        0,
        & p=0.
    \end{cases}
    \label{eq:self-information-h-def}
\end{equation}
Equation~\eqref{eq:self-info} becomes
\begin{equation}
    I_\Phi(X;X)
    =
    \Phi(0)-\Phi(1)
    +\sum_{x\in\cX}h_\Phi(p_X(x)).
    \label{eq:self-information-h-representation}
\end{equation}
Thus \eqref{Phi-MI-self-info} is equivalent to
\begin{equation}
    \sup_{\substack{
        p(x)\geq0,\ x\in\cX\\
        \sum_{x\in\cX}p(x)=1
    }}
    \left\{
        \sum_{x\in\cX}h_\Phi(p(x))
        +\Phi(0)-\Phi(1)
    \right\}.
    \label{opt:self-information}
\end{equation}

It is also useful to define
\begin{equation}
    G_\Phi(t)
    \defeq
    \frac{\Phi(t)-\Phi(0)}{t},
    \qquad t>0.
    \label{eq:self-information-G-def}
\end{equation}
Since
\begin{equation}
    h_\Phi(p)
    =
    pG_\Phi\left(\frac{1}{p}\right),
    \qquad p\in(0,1],
    \label{eq:self-information-h-G}
\end{equation}
we obtain the alternative representation
\begin{equation}
    I_\Phi(X;X)
    =
    \sum_{x:p_X(x)>0}
    p_X(x)G_\Phi\left(\frac{1}{p_X(x)}\right)
    +\Phi(0)-\Phi(1).
    \label{eq:self-info-GPhi}
\end{equation}
The following result of Zv\'arov\'a gives a sufficient condition
for the uniform distribution to be optimal.

\begin{theorem}[{\cite[Theorem~4]{zvarova1974measures}}]
\label{thm:unif-max}
Suppose that $\Phi:[0,\infty)\to\R$ is convex and
$\Phi(0)<\infty$. If $G_\Phi$ is concave on $(0,\infty)$, then,
for every finite alphabet $\cX$,
\begin{align}
    I_\Phi(X;X)
    &\leq
    \frac{1}{|\cX|}\Phi(|\cX|)
    +
    \left(1-\frac{1}{|\cX|}\right)\Phi(0)
    -\Phi(1).
    \label{eq:self-information-uniform-bound}
\end{align}
Equality holds when $X$ is uniformly distributed on $\cX$.
\end{theorem}

To study \eqref{opt:self-information} without assuming concavity
of $G_\Phi$, we first examine the objective near the boundary of
the probability simplex. For $p>0$,
\begin{equation}
    h_\Phi(p)
    =
    \frac{\Phi(1/p)-\Phi(0)}{(1/p)^2}.
    \label{eq:self-information-boundary-growth}
\end{equation}
Thus its behavior as $p\downarrow0$ depends on the growth of $\Phi$
relative to $t^2$.

If $\Phi(t)=\omega(t^2)$ as $t\to\infty$, then
$h_\Phi(p)\to+\infty$ as $p\downarrow0$. Consequently, for every
alphabet with $|\cX|\geq2$,
\[
    \sup_{p_X}I_\Phi(X;X)=+\infty.
\]
If instead $\Phi(t)=Ct^2+o(t^2)$ for some $C>0$, then
$h_\Phi(p)\to C$ as $p\downarrow0$, whereas $h_\Phi(0)=0$.
For example, if $Z_p\sim\Bern(p)$, then
\[
    \lim_{p\downarrow0}I_\Phi(Z_p;Z_p)
    =
    C
    \neq
    0
    =
    I_\Phi(Z_0;Z_0).
\]
Hence self-information can be discontinuous as probabilities vanish.

We therefore focus on the subquadratic regime
\begin{equation}
    \lim_{t\to\infty}\frac{\Phi(t)}{t^2}=0.
    \label{eq:self-information-subquadratic}
\end{equation}
In this regime, $h_\Phi(p)\to0=h_\Phi(0)$ as $p\downarrow0$.
The objective in \eqref{opt:self-information} is therefore continuous
on the probability simplex, which ensures that a maximizer exists.
The next theorem also shows that every maximizer has full support.

\begin{theorem}
\label{thm:self-information-optimality}
Suppose that $\Phi:[0,\infty)\to\R$ is convex, non-affine,
differentiable on $(0,\infty)$, and satisfies $\Phi(0)<\infty$ and $\lim_{t\to\infty} \Phi(t)/t^2=0$. 
For every finite alphabet $\cX$ with $|\cX|\geq2$,
the optimization problem \eqref{opt:self-information} admits a
maximizer, and every maximizer $p_X^\ast$ lies in the interior of
the probability simplex.

Moreover, there exists $\lambda\in\R$ such that for every $x\in \cX$, 
\begin{equation}
    h_\Phi'(p_X^\ast(x))
    =
    \lambda.
    \label{eq:self-information-interior-KKT}
\end{equation}
Equivalently, for every $x\in\cX$,
\begin{align}
    &2p_X^\ast(x)
    \left[
        \Phi\left(\frac{1}{p_X^\ast(x)}\right)-\Phi(0)
    \right]
    -
    \Phi'\left(\frac{1}{p_X^\ast(x)}\right)
    =
    \lambda.
    \label{eq:self-information-interior-KKT-explicit}
\end{align}
\end{theorem}

The proof is given in
Appendix~\ref{app:proof-thm:self-information-optimality}.

\begin{corollary}
\label{cor:uniform-maximizer}
Under the assumptions of
Theorem~\ref{thm:self-information-optimality}, the following hold:
\begin{enumerate}[\rm (i)]
    \item
    If $h_\Phi'$ is nonincreasing on $(0,1]$, then the uniform
    distribution maximizes \eqref{opt:self-information}.

    \item
    If $h_\Phi'$ is injective on $(0,1]$, then the uniform
    distribution is the unique maximizer.
\end{enumerate}
\end{corollary}

\begin{proof}
If $h_\Phi'$ is nonincreasing, then $h_\Phi$ is concave on $[0,1]$.
Writing $n=|\cX|$, Jensen's inequality gives
\[
    \sum_{x\in\cX}h_\Phi(p_X(x))
    \leq
    n h_\Phi\left(\frac{1}{n}\right).
\]
The uniform distribution therefore maximizes
\eqref{opt:self-information}.

If $h_\Phi'$ is injective, then
\eqref{eq:self-information-interior-KKT} forces every maximizer
to assign the same probability to each $x\in\cX$.
Existence of a maximizer then implies that the uniform distribution
is the unique maximizer.
\end{proof}

The connection with Theorem~\ref{thm:unif-max} follows from
\[
    h_\Phi''(p)
    =
    \frac{1}{p^3}G_\Phi''(1/p),
    \qquad p\in(0,1),
\]
whenever $\Phi$ is twice differentiable. Thus concavity of $G_\Phi$
implies concavity of $h_\Phi$, and the first assertion of
Corollary~\ref{cor:uniform-maximizer} recovers uniform optimality
under the assumptions of
Theorem~\ref{thm:self-information-optimality}.

\subsubsection{Optimization over all finite alphabets}

We next allow the alphabet size to vary. The resulting supremum
will be useful in Section~\ref{subsec:private-randomness}.

By convexity of $\Phi$, for every $0<s<t$,
\[
    \frac{\Phi(s)-\Phi(0)}{s}
    \leq
    \frac{\Phi(t)-\Phi(0)}{t}.
\]
Hence $G_\Phi$ is nondecreasing, and the limit
\begin{equation}
    s_\Phi
    \defeq
    \lim_{t\to\infty}G_\Phi(t)
    =
    \lim_{t\to\infty}\frac{\Phi(t)-\Phi(0)}{t}
    \in\R\cup\{+\infty\}
    \label{eq:self-information-asymptotic-slope}
\end{equation}
exists. This asymptotic slope determines the exact supremum.

\begin{theorem}
\label{thm:self-information-universal-bound}
Suppose that $\Phi:[0,\infty)\to\R$ is convex and
$\Phi(0)<\infty$. Then
\begin{equation}
    \sup_{\substack{
        X:\,|\operatorname{supp}(X)|<\infty
    }}
    I_\Phi(X;X)
    =
    s_\Phi+\Phi(0)-\Phi(1),
    \label{eq:self-information-universal-supremum}
\end{equation}
where the equality is understood in the extended real sense.
Moreover, if $X_n$ is uniformly distributed on $[n]$, then
\begin{equation}
    \lim_{n\to\infty}I_\Phi(X_n;X_n)
    =
    s_\Phi+\Phi(0)-\Phi(1).
    \label{eq:self-information-uniform-asymptotic}
\end{equation}
\end{theorem}

\begin{proof}
Since $G_\Phi(t)\leq s_\Phi$ for every $t>0$,
\eqref{eq:self-info-GPhi} yields
\begin{equation}
    I_\Phi(X;X)
    \leq
    s_\Phi+\Phi(0)-\Phi(1)
    \label{eq:self-information-universal-upper}
\end{equation}
for every finitely supported random variable $X$.

On the other hand, if $X_n$ is uniform on $[n]$, then
\eqref{eq:self-uniform} gives
\[
    I_\Phi(X_n;X_n)
    =
    G_\Phi(n)+\Phi(0)-\Phi(1).
\]
Letting $n\to\infty$ proves
\eqref{eq:self-information-uniform-asymptotic} and shows that the
upper bound in \eqref{eq:self-information-universal-upper} is
approached by uniform distributions.
\end{proof}

\begin{remark}
If $s_\Phi<\infty$, then $G_\Phi$ is bounded on $[1,\infty)$, so
\[
    \Phi(t)=\Phi(0)+tG_\Phi(t)=O(t)
    \qquad\text{as }t\to\infty.
\]
Thus a finite supremum over all finite alphabets implies the
subquadratic growth condition
\eqref{eq:self-information-subquadratic}.
\end{remark}

\subsection{Special classes of functions}

\begin{definition}[
\cite{Boucheron2013concentration}] \label{def:class-F}
    We define $\mathscr{F}$ to be the class of all convex functions $\Phi:\R_{+}\to \R$  that are twice differentiable on $(0,\infty)$, non-affine, and such that $1/\Phi''$ is concave. 
\end{definition}

\begin{definition} \label{def:class-G}
    We define $\mathscr{G}$ to be the class of smooth convex functions $\Phi:(0,\infty)\to \R$ that are not affine and such that $t\mapsto t^2\Phi''(t)$ is concave.
\end{definition}

\begin{table}[h!]
\centering
\caption{Examples of $\Phi$-divergences and their membership in
$\mathscr F$ and $\mathscr G$.}
\label{tab:divergences}
\renewcommand{\arraystretch}{1.35}
\setlength{\tabcolsep}{6pt}
\small

\begin{tabularx}{\textwidth}{
    @{}
    >{\raggedright\arraybackslash}p{0.26\textwidth}
    >{\centering\arraybackslash}X
    >{\centering\arraybackslash}p{0.12\textwidth}
    >{\centering\arraybackslash}p{0.12\textwidth}
    @{}
}
\toprule
\textbf{Divergence}
& $\boldsymbol{\Phi(x)}$
& $\boldsymbol{\mathscr F}$
& $\boldsymbol{\mathscr G}$ \\
\midrule

KL
& $x\log x$
& Yes
& Yes \\
\midrule

Reverse KL
& $-\log x$
& No
& Yes \\

Squared Hellinger
& $(\sqrt{x}-1)^2$
& No
& Yes \\

Jensen--Shannon
& $\displaystyle
  \frac{1}{2}
  \left(x\log x-(1+x)\log\!\left(\frac{1+x}{2}\right)\right) $
& No
& Yes \\

Jeffreys
& $(x-1)\log x$
& No
& Yes \\

Power divergence, $0<\alpha<1$
& $\displaystyle
  \frac{x^\alpha-\alpha x+\alpha-1}{\alpha(\alpha-1)}$
& No
& Yes \\
\midrule

Power divergence, $1<\alpha<2$
& $\displaystyle
  \frac{x^\alpha-\alpha x+\alpha-1}{\alpha(\alpha-1)}$
& Yes
& No \\

Pearson $\chi^2$
& $(x-1)^2$
& Yes
& No \\
\midrule

Neyman $\chi^2$
& $\displaystyle\frac{(x-1)^2}{x}$
& No
& No \\

Triangular discrimination
& $\displaystyle\frac{(x-1)^2}{x+1}$
& No
& No \\

\bottomrule
\end{tabularx}
\end{table}

The class \(\mathscr F\) is closely related to subadditivity of \(\Phi\)-entropy and has applications to concentration inequalities  \cite{chafai2004entropies}. For example, Raginsky uses it to obtain a unified proof of the tensorization of the strong data processing inequality constant \cite{raginsky2016strong}. The definition of $\mathscr{G}$ is new in the literature. For functions in $\mathscr{G}$, in a concurrent work \cite{AbinSubadditiveInformation}, the second author and his collaborators show that for every $\Phi\in\mathscr{G}$ and for any probability mass functions $p_X,q_X$ on $\cX$ and $p_Y,q_Y$ on $\cY$ such that $p_X\ll q_X$ and $p_Y\ll q_Y$, we have
\begin{align} \label{eq:sub-DPhi}
    \DPhi\big(p_X\otimes p_Y \,\big\|\, q_X\otimes q_Y\big)
    \le \DPhi(p_X\|q_X) + \DPhi(p_Y\|q_Y).
\end{align}

\begin{proposition}
    Suppose $\Phi\in \mathscr{F}$. 
    \begin{enumerate}[\rm (i)]
        \item (\cite[Theorem 4]{masiha2023f}) For any $X,Y$, 
        \begin{align*}
            \IPhi(X;Y) \leq (\ln 2) I(X;Y) \cdot \left[\lim_{t\to \infty} t \Phi''(t)\right].
        \end{align*}
        \item (Subadditivity of $\IPhi$ \cite[Theorem 8]{beigi2018phi}) Let $X_1,\dots,X_n$ be mutually independent random variables. Then for any random variable $U$, 
            \begin{align} \label{IPhi-subadditive}
            \IPhi(U;X^n) 
            \leq \sum_{i=1}^n \IPhi(U;X^n|X_{\widehat{i}}),
        \end{align}
        where $X_{\widehat{i}}=(X_1,\dots,X_{i-1},X_{i+1},\dots,X_n)$.
    \end{enumerate}
\end{proposition}

\begin{proposition}
Suppose that $\Phi\in\mathscr G$ and $\Phi(0)<\infty$. If $(X_1,Y_1)$ is independent of $(X_2,Y_2)$, then
\begin{align}
\IPhi(X_1X_2;Y_1Y_2)
\leq
\IPhi(X_1;Y_1)+\IPhi(X_2;Y_2).
\label{eq:IPhi-product-subadditive}
\end{align}
\end{proposition}

\begin{proof}
Write
\[
P_i\defeq p_{X_iY_i},
\qquad
Q_i\defeq p_{X_i}\otimes p_{Y_i},
\qquad i=1,2.
\]
The independence of $(X_1,Y_1)$ and $(X_2,Y_2)$ implies
\[
p_{X_1X_2Y_1Y_2}=P_1\otimes P_2
\]
and
\[
p_{X_1X_2}\otimes p_{Y_1Y_2}
=
Q_1\otimes Q_2.
\]
By the product subadditivity of $\Phi$-divergence for
$\Phi\in\mathscr G$,
\begin{align*}
\IPhi(X_1X_2;Y_1Y_2)
&=
\DPhi(P_1\otimes P_2\|Q_1\otimes Q_2)\\
&\leq
\DPhi(P_1\|Q_1)+\DPhi(P_2\|Q_2)\\
&=
\IPhi(X_1;Y_1)+\IPhi(X_2;Y_2)
\end{align*}
where the second line follows from \eqref{eq:sub-DPhi}. 
\end{proof}

Suppose $(X_i,Y_i)$ for $i\in [n]$ are mutually independent pairs. Then 
    \begin{align*}
        \IPhi(X^n;Y^n) \leq \sum_{i=1}^n \IPhi(X_i;Y_i). 
    \end{align*}

We will investigate the properties of $\Phi$ for the classes $\mathscr{F}$ and $\mathscr{G}$. 

\begin{lemma}[Lemma 1 \cite{masiha2023f}]
    \label{lem:class-F-properties}
    Suppose $\Phi\in \mathscr{F}$. Then $t\mapsto t\Phi''(t)$ is non-decreasing and then $t\mapsto t^2\Phi''(t)$ is convex. And 
    \begin{align*}
        \lim_{t\to \infty}
        \frac{\Phi(t)}{t\ln t}
        =\lim_{t\to \infty} t\Phi''(t)
        >0.
    \end{align*}
    
    Moreover, $t\mapsto \Phi''(t)$ is nonnegative and nonincreasing, and 
    $$
    \lim_{t\to \infty} \frac{\frac{1}{2}t^2}{\Phi(t)} 
    =\lim_{t\to \infty} \frac{1}{\Phi''(t)}. 
    $$
\end{lemma}
Lemma~\ref{lem:class-F-properties} implies that $\Phi(t)=\Omega(t\ln t)$ and $\Phi(t)=O(t^2)$.

We prove the following properties of $\Phi$ for $\Phi\in\mathscr{G}$. 
\begin{lemma}\label{lem:class-G-properties}
    Suppose $\Phi\in\mathscr G$ and its domain contains $(0,\infty)$.
    The function $t\mapsto t\Phi''(t)$ is nonincreasing, and
    \begin{align*}
        c_\Phi
        &\defeq\lim_{t\to\infty}t\Phi''(t)=\lim_{t\to\infty}\frac{\Phi'(t)}{1+\ln t}
        =
        \lim_{t\to\infty}\frac{\Phi(t)}{t\ln t}
        \in[0,\infty).
    \end{align*}
    Define
    \begin{align*}
        \Psi(t)\defeq\Phi(t)-c_\Phi t\ln t.
    \end{align*}
    Then $\Psi$ is convex, $t\mapsto t\Psi''(t)$ is nonincreasing, and
    \begin{align*}
        \lim_{t\to\infty}t\Psi''(t)&=0,\\
        \Psi(t)&=o(t\ln t),
    \end{align*}
    and either $\Psi$ is affine or $\Psi\in\mathscr G$.

    If, in addition, $\Phi(0)<\infty$, then the limit
    \begin{align*}
        \ell_\Phi\defeq
        \lim_{t\to\infty}\frac{\Psi(t)}{t}
        \in\mathbb R\cup\{+\infty\}
    \end{align*}
    exists.
\end{lemma}

The proof is in Appendix~\ref{app:proof-lem:class-G-properties}.

For $\Phi\in \mathscr{G}$, using the above notations, $\Phi(t)=O(t\ln t)$ and $\Psi(t)=o(t\ln t)$. Since $\Psi$ is convex, we can write
$$
\IPhi(X;Y)=c_\Phi(\ln 2) I(X;Y)+I_{\Psi}(X;Y).
$$

\begin{lemma} \label{lem:G-concave-X|Z}
    Suppose $\Phi\in \mathscr{G}$. Let $\nu_Z,\nu_{Y|XZ},\mu_{XYZ}$ be fixed. Then 
    \begin{align*}
        \nu_{X|Z} \mapsto 
        D_\Phi(\nu_{XZ}\|\mu_{XZ})
        -D_\Phi(\nu_{XYZ}\|\mu_{XYZ})  
    \end{align*}
    is convex on the set $\{\nu_{X|Z}:\nu_{XYZ}\ll\mu_{XYZ}\}$. 
\end{lemma}

\begin{proof}
    Note that 
\begin{align*}
    \DPhi(\nu_{XZ}\|\mu_{XZ})
    -\DPhi(\nu_{XYZ}\|\mu_{XYZ})
    &=
    \sum_{x,z} \mu(x,z) \Phi\left(\frac{\nu(x,z)}{\mu(x,z)}\right)
    -\sum_{x,y,z} \mu(x,y,z) \Phi\left(\frac{\nu(x,y,z)}{\mu(x,y,z)}\right)\\
    &=\sum_{x,z} \mu(x,z) 
    \left(
    \Phi\left(\frac{\nu(x,z)}{\mu(x,z)}\right)
    -\sum_y \mu(y|x,z) \Phi\left(\frac{\nu(z)\nu(x|z)\nu(y|x,z)}{\mu(x,y,z)}\right)
    \right). 
\end{align*}
Fixing $x,z$, we will prove the mapping 
\begin{align} \label{x|z-convex-lemma}
    \nu(x|z) \mapsto 
    \Phi\left(\frac{\nu(z)\nu(x|z)}{\mu(x,z)}\right)
    -\sum_y \mu(y|x,z) \Phi\left(\frac{\nu(z)\nu(x|z)\nu(y|x,z)}{\mu(x,y,z)}\right)
\end{align}
is convex. Differentiating twice with respect to $\nu(x|z)$, the second derivative is equal to 
\begin{align}
    \frac{\nu(z)^2}{\mu(x,z)^2}
    \Phi''\left(\frac{\nu(z)\nu(x|z)}{\mu(x,z)}\right)
    -\sum_y \mu(y|x,z)\frac{\nu(z)^2\nu(y|x,z)^2}{\mu(x,y,z)^2}
    \Phi''\left(\frac{\nu(z)\nu(x|z)\nu(y|x,z)}{\mu(x,y,z)}\right). \label{eq:map-2nd-derivative-lemma}
\end{align}
Since $\Phi\in \mathscr{G}$, $t\mapsto t^2\Phi''(t)$ is concave. Considering $t=\frac{\nu(y|x,z)}{\mu(x,y,z)}$, by Jensen's inequality, we have 
\begin{align*}
    &\sum_y \mu(y|x,z)\frac{\nu(z)^2\nu(y|x,z)^2}{\mu(x,y,z)^2}
    \Phi''\left(\frac{\nu(z)\nu(x|z)\nu(y|x,z)}{\mu(x,y,z)}\right)\\
    \leq& \nu(z)^2 
    \left(\sum_y \mu(y|x,z)\frac{\nu(y|x,z)}{\mu(x,y,z)}\right)^2 
    \Phi'' \left(
    \sum_y \mu(y|x,z) 
    \frac{\nu(z)\nu(x|z)\nu(y|x,z)}{\mu(x,y,z)}
    \right)\\
    =& \frac{\nu(z)^2}{\mu(x,z)^2}
    \Phi''\left(\frac{\nu(z)\nu(x|z)}{\mu(x,z)}\right). 
\end{align*}
Thus \eqref{eq:map-2nd-derivative-lemma} is nonnegative and then \eqref{x|z-convex-lemma} is convex. Since it holds for any $x,z$, 
\begin{align*}
    \nu_{X|Z} \mapsto 
    \DPhi(\nu_{XZ}\|\mu_{XZ})
    -\DPhi(\nu_{XYZ}\|\mu_{XYZ})
\end{align*}
is convex.
\end{proof}

By Lemma~\ref{lem:class-F-properties}, if $\Phi\in \mathscr{F}\cap \mathscr{G}$, $t\mapsto t^2\Phi''(t)$ has to be linear. Moreover, $\Phi(0)<\infty$. Thus the functions in $\mathscr{F}\cap \mathscr{G}$ are in the form 
\begin{align*}
    \Phi(t)=At\log t+Bt+C
\end{align*}
where $A> 0$.

\subsection{Supremum representations of conditional $\Phi$-mutual information}
\label{subsec:variational-cond-Phi-MI}

We derive a direct variational representation of conditional 
$\Phi$-mutual information. Each admissible
function in this representation yields a lower bound on
$I_\Phi(X;Y|Z)$.

We first recall the unconditional representation. Let
\[
    \Phi^*(t)
    \defeq
    \sup_{u\geq0}\{ut-\Phi(u)\},
    \qquad t\in\R,
\]
be the Fenchel conjugate of $\Phi$. For probability distributions
$P\ll Q$ on a common finite alphabet, Fenchel duality gives
\[
    D_\Phi(P\|Q)
    =
    \sup_T
    \left\{
        \E_P[T]-\E_Q[\Phi^*(T)]
    \right\}
    -\Phi(1),
\]
where the supremum is over all real-valued functions on that
alphabet. Taking $P=p_{XY}$ and $Q=p_X\otimes p_Y$ yields
\begin{equation}
    I_\Phi(X;Y)
    =
    \sup_T
    \left\{
        \E_{p_{XY}}[T(X,Y)]
        -
        \E_{p_X\otimes p_Y}[\Phi^*(T(X,Y))]
    \right\}
    -\Phi(1).
    \label{Phi-MI-sup-form}
\end{equation}
Restricting $T$ to a parametrized family produces a variational
lower bound. Such objectives motivate neural estimators of mutual
information \cite{belghazi2018mutual}. The lower-bound guarantee
applies to the population objective; replacing expectations by
sample averages need not preserve it.

Our conditional quantity is defined by
\[
    I_\Phi(X;Y|Z)
    =
    I_\Phi(X;YZ)-I_\Phi(X;Z).
\]
Applying \eqref{Phi-MI-sup-form} separately to the two terms does
not provide a lower bound on their difference. Moreover,
subtracting approximate values can lead to large relative errors
when the two terms are close. We therefore seek a direct
variational representation of $I_\Phi(X;Y|Z)$. Our approach
expresses it as a conditional Jensen gap and uses the convexity
of this gap when $\Phi\in\mathscr F$.

Set $P_{XYZ}\defeq p_{XYZ}$, $Q_{XYZ}\defeq p_X\otimes p_{YZ}$ 
and define the likelihood ratio
\[
    L(x,y,z)
    \defeq
    \frac{P(x,y,z)}{Q(x,y,z)}
    =
    \frac{p(x,y,z)}{p_X(x)p_{YZ}(y,z)}
\]
on the support of $Q$. Its values outside that support are
irrelevant. Since
\[
    \E_Q[L| X=x,Z=z]
    =
    \frac{p_{XZ}(x,z)}{p_X(x)p_Z(z)},
\]
we obtain
\begin{equation}
    I_\Phi(X;Y|Z)
    =
    \E_Q[\Phi(L)]
    -
    \E_Q\left[
        \Phi\left(\E_Q[L| X,Z]\right)
    \right].
    \label{eq:conditional-Phi-Jensen-gap}
\end{equation}

More generally, for a fixed distribution $Q$ on a finite product
alphabet and a nonnegative function $f=f(X,Y,Z)$, define
\begin{equation}
    \mathcal J_{\Phi,Q}(f;X,Z)
    \defeq
    \E_Q[\Phi(f)]
    -
    \E_Q\left[
        \Phi\left(\E_Q[f| X,Z]\right)
    \right].
    \label{eq:def-conditional-Jensen-gap}
\end{equation}
Thus
\[
    I_\Phi(X;Y|Z)
    =
    \mathcal J_{\Phi,Q}(L;X,Z).
\]
For $\Phi\in\mathscr F$ defined on $(0,\infty)$, we use its
finite continuous extension to zero.

The following two standard facts provide the ingredients for
the variational representation.

\begin{lemma}[{\cite[Lemma~32]{mahdi2018correlation}}]
\label{lem:conditional-Phi-entropy-convex}
Let $Q$ be a fixed probability distribution on a finite product
alphabet. Suppose that $\Phi\in\mathscr F$ and that its domain
contains $(0,\infty)$. Then $f\mapsto\mathcal J_{\Phi,Q}(f;X,Z)$ 
is convex on the set of nonnegative functions
$f:\cX\times\cY\times\cZ\to[0,\infty)$.
\end{lemma}

\begin{lemma}[{\cite[Proposition~16.10]{bauschke2017convex}}]
\label{lem:Fenchel-equality}
Suppose that $\Phi:(0,\infty)\to\R$ is convex and differentiable,
with 
$\Phi(0)\defeq \lim_{t\downarrow 0} \Phi(t)$. Then for every $u>0$, $\Phi^*(\Phi'(u))
    =
    u\Phi'(u)-\Phi(u)$. 
\end{lemma}

For a fixed probability distribution $Q_{XYZ}$ and a positive function $g=g(X,Y,Z)$, write 
\[
    g_{XZ}\defeq\E_Q[g|X,Z].
\]

\begin{theorem}
\label{thm:conditional-Phi-variational-form}
Let $Q$ be a fixed probability distribution and $\Phi\in \mathscr{F}$. Then, for every nonnegative function $f$,
\begin{equation}
    \mathcal J_{\Phi,Q}(f;X,Z)
    =
    \sup_{g>0}
    \left\{
        \E_Q\left[
            f\bigl(\Phi'(g)-\Phi'(g_{XZ})\bigr)
        \right]
        -
        \E_Q\left[
            \Phi^*(\Phi'(g))
            -
            \Phi^*(\Phi'(g_{XZ}))
        \right]
    \right\},
    \label{eq:conditional-Phi-single-critic-g}
\end{equation}
where $g_{XZ}=\E_Q[g| X,Z]$ and the supremum is over all positive functions
$g:\cX\times\cY\times\cZ\to(0,\infty)$.
If $f>0$, the supremum is attained at $g=f$.
For general nonnegative $f$, it is approached by
$g=f+\epsilon$ as $\epsilon\downarrow0$.
\end{theorem}

The representation follows by taking supporting hyperplanes of the convex functional $\cJ_{\Phi,Q}$ and applying 
Lemma~\ref{lem:Fenchel-equality}. The proof is given in 
Appendix~\ref{app:proof-thm:conditional-Phi-variational-form}.

Applying Theorem~\ref{thm:conditional-Phi-variational-form} to
$f=L=P/Q$ and using $\E_Q[Lh]=\E_P[h]$ gives the desired
representation.

\begin{corollary}[Variational representation of conditional
$\Phi$-mutual information]
\label{cor:conditional-Phi-MI-variational-form}
Suppose that $\Phi\in\mathscr F$ and that its domain contains
$(0,\infty)$. Let $P=p_{XYZ}$ and $Q=p_X\otimes p_{YZ}$. Then
\begin{equation}
    I_\Phi(X;Y|Z)
    =
    \sup_{g>0}
    \left\{
        \E_P\left[
            \Phi'(g)-\Phi'(g_{XZ})
        \right]
        -
        \E_Q\left[
            \Phi^*(\Phi'(g))
            -
            \Phi^*(\Phi'(g_{XZ}))
        \right]
    \right\},
    \label{eq:conditional-Phi-single-critic-g-MI}
\end{equation}
where $g_{XZ}=\E_Q[g| X,Z]$ and the supremum is over all
positive functions on $\cX\times\cY\times\cZ$.
\end{corollary}

Each positive $g$ therefore gives a lower bound on
$I_\Phi(X;Y|Z)$. Under the product distribution $Q$, the conditional
average appearing in the objective is
\[
    g_{XZ}(x,z)
    =
    \sum_y p(y|z)g(x,y,z).
\]

\begin{example}[Conditional Donsker--Varadhan representation]
\label{ex:conditional-DV-variational}
For $\Phi(t)=t\ln t$, we have $I_\Phi(X;Y|Z)=(\ln 2)I(X;Y|Z)$ and
\[
    \Phi'(t)=1+\ln t,
    \qquad
    \Phi^*(t)=e^{t-1}.
\]
Consequently,
\[
    \Phi'(g)-\Phi'(g_{XZ})
    =
    \ln\frac{g}{g_{XZ}},
    \qquad
    \Phi^*(\Phi'(g))=g.
\]
The second expectation in
\eqref{eq:conditional-Phi-single-critic-g-MI} vanishes because
$\E_Q[g-g_{XZ}]=0$. Hence
\begin{equation}
    (\ln 2) I(X;Y|Z)
    =
    \sup_{g>0}
    \E_P\left[
        \ln\frac{g}{g_{XZ}}
    \right].
    \label{eq:conditional-DV-g}
\end{equation}

Writing $g=e^T$ gives
\begin{equation}
    (\ln 2) I(X;Y|Z)
    =
    \sup_T
    \E_P\left[
        T(X,Y,Z)
        -
        \ln\E_Q\left[
            e^{T(X,Y,Z)}| X,Z
        \right]
    \right],
    \label{eq:conditional-DV-general}
\end{equation}
where the supremum is over all real-valued functions
$T:\cX\times\cY\times\cZ\to\R$. Here,
\[
    \E_Q[e^{T(X,Y,Z)}| X=x,Z=z]
    =
    \sum_y p(y|z)e^{T(x,y,z)}.
\]
If $p_{XYZ}$ has full support, an optimizer is
\[
    T(x,y,z)
    =
    \ln\frac{p(y|x,z)}{p(y|z)}.
\]
\end{example}

\section{Max-\texorpdfstring{$\Phi$}{Phi}-mutual information \texorpdfstring{$\wIPhi$}{max-IPhi}} \label{sec:max-Phi-MI}

\begin{definition}
    Define the \emph{max-$\Phi$-mutual information} as 
    \begin{align}\label{opt:IPhi}
        \wIPhi(X;Y) &\defeq
        \sup_{U}\left\{ \IPhi(U;X)+\IPhi(U;Y)-\IPhi(U;XY)\right\}
    \end{align}
    where the supremum is over all $p_{U|XY}$. 

    Then define the \emph{conditional max-$\Phi$-mutual information}: 
    \begin{align} \label{opt:cond-IPhi}
        \wIPhi(X;Y|Z) 
        &\defeq \sup_{U}\left\{ \IPhi(U;X|Z) +\IPhi(U;Y|Z) -\IPhi(U;XY|Z)\right\}
    \end{align}
    where the supremum is over all $p_{U|XYZ}$. 
\end{definition}
It is immediate that by the chain rule, the above definitions have the following equivalent forms.     The max-\(\Phi\)-mutual information satisfies
    \begin{align*}
        \wIPhi(X;Y) &= \sup_{U} \left\{\IPhi(U;X)+\IPhi(U;Y)-\IPhi(U;XY)\right\}\\
        &= \sup_{U} \left\{\IPhi(Y;U)-\IPhi(U;Y|X)\right\} \nonumber.
    \end{align*}
    The conditional max-\(\Phi\)-mutual information satisfies
    \begin{align*} 
        \wIPhi(X;Y|Z) 
        &= \sup_{U} \left\{\IPhi(U;X|Z) +\IPhi(U;Y|Z) -\IPhi(U;XY|Z)\right\} \\
        &= \sup_{U} \left\{\IPhi(U;Y|Z)-\IPhi(U;Y|XZ)\right\} \\
        &= \sup_{U} \left\{\IPhi(U;XZ)+\IPhi(U;YZ)-\IPhi(U;XYZ)-\IPhi(U;Z)\right\}.
    \end{align*}

When $\Phi(t)=t\log t$, the $\Phi$-mutual information reduces to
Shannon mutual information, and the conditional max-$\Phi$-mutual
information also reduces to conditional mutual information:
\[
\wIPhi(X;Y|Z)=I(X;Y|Z).
\]
Indeed, for every auxiliary random variable $U$, the chain rule gives
\begin{align*}
&I(U;X|Z)+I(U;Y|Z)-I(U;XY|Z)\\
&\qquad=
I(U;Y|Z)-I(U;Y|XZ)\\
&\qquad=
I(X;Y|Z)-I(X;Y|UZ)\\
&\qquad\leq I(X;Y|Z).
\end{align*}
Therefore, $\wIPhi(X;Y|Z)\leq I(X;Y|Z)$. On the other hand, taking
$U=X$ gives
\begin{align*}
&I(X;X|Z)+I(X;Y|Z)-I(X;XY|Z)
=
I(X;Y|Z),
\end{align*}
which proves the reverse inequality and hence the claimed equality.

The objective defining $\wIPhi(X;Y)$ coincides with the auxiliary-variable objective for the $I_\Phi$-ribbon at weights $(1,1)$. This connection also makes $\wIPhi$ useful for deriving information inequalities. In Section~\ref{sec:non-Shannon}, we establish a multipartite $\Phi$-generalization of the Zhang--Yeung inequality involving $I_\Phi$ and $\wIPhi$. We first study the optimization defining $\wIPhi$ and then establish its structural properties.

\subsection{Computation of $\wIPhi$}

In Appendix \ref{app:proof-optimizer}, we show that to evaluate \eqref{opt:IPhi}, it suffices to consider random variables $U$ with cardinality at most $|\cX||\cY|$. 
To evaluate \eqref{opt:cond-IPhi}, it suffices to consider random variables $U$ with cardinality at most $|\cX||\cY||\cZ|$ and thus the feasible region is compact. By the continuity and compactness, the supremum is always attained. Therefore, in the remaining parts, we will write the supremum as a maximum. 

We have the following theorem:
\begin{theorem} \label{thm:wIPhi-optimizer}
    Suppose $\Phi\in \mathscr{G}$ and $\Phi(0)<\infty$. Then the supremum of \eqref{opt:cond-IPhi} is attained by $p_{U|XYZ}$ satisfying $H(XY|UZ)=0$. In other words, $(X,Y)$ is a function of $(U,Z)$. 
\end{theorem}

The proof is in Appendix \ref{app:proof-optimizer}. The theorem yields the following closed-form expression for the unconditional quantity $\wIPhi(X;Y)$.

\begin{corollary}
\label{cor:G-closed-form}
    Suppose $\Phi\in \mathscr{G}$ and $\Phi(0)<\infty$. Then
    \begin{align*}
        \wIPhi(X;Y) =\IPhi(X;X)+\IPhi(Y;Y)-\IPhi(XY;XY).
    \end{align*}
\end{corollary}

\begin{proof}
Let $U$ be a maximizer in the definition of $\wIPhi(X;Y)$. By
Theorem \ref{thm:wIPhi-optimizer}, $U$ may be chosen so that
$H(XY|U)=0$. Hence $X$, $Y$, and $XY$ are all functions of $U$.
Item \ref{Phi-MI-self-reduction} of Proposition
\ref{prop:properties-Phi-MI} therefore gives
\begin{align*}
    \IPhi(U;X)&=\IPhi(X;X),\\
    \IPhi(U;Y)&=\IPhi(Y;Y),\\
    \IPhi(U;XY)&=\IPhi(XY;XY).
\end{align*}
Thus
\begin{align*}
    \wIPhi(X;Y)
    &=\IPhi(U;X)+\IPhi(U;Y)-\IPhi(U;XY)\\
    &=\IPhi(X;X)+\IPhi(Y;Y)-\IPhi(XY;XY).
\end{align*}
\end{proof}

\subsection{Basic structural properties}

\begin{proposition} \label{prop:max-Phi-MI}
Let \(X,Y,Z,W,T,U\) be finite random variables, and let
\(\Phi:\mathbb R_+\to \mathbb R\) be convex. Then the following properties hold.

\begin{enumerate}[\rm (i)]
    \item \textbf{Conditional symmetry and dominance over \(I_\Phi\).}
    We have
    \begin{align}
        \wIPhi(X;Y|Z)
        =
        \wIPhi(Y;X|Z),
        \label{max-Phi-symmetry}
    \end{align}
    and
    \begin{align}
        \wIPhi(X;Y|Z)
        \ge
        I_\Phi(X;Y|Z)
        \ge 0 .
        \label{max-Phi-LB}
    \end{align}

    \item \textbf{Conditional deterministic self-reduction.}
    If \(Y\) is a function of \((X,Z)\), then
    \begin{align}
        \wIPhi(X;Y|Z)
        =
        \wIPhi(Y;Y|Z).
        \label{max-Phi-deterministic}
    \end{align}

    \item \textbf{Self-information bounds.}
    We have
    \begin{align}
        \wIPhi(X;Y|Z)
        \le
        \wIPhi(X;X|Z),
        \label{max-Phi-MI-UB-self}
    \end{align}
    and
    \begin{align}
        \wIPhi(X;XY)
        =
        \wIPhi(X;X)
        =
        I_\Phi(X;X).
        \label{max-Phi-repetition}
    \end{align}
    Moreover, 
    \begin{align*}
        \wIPhi(X;Y|Z) 
        &\leq\min \{\IPhi(XZ;XZ), \IPhi(YZ;YZ)\}. 
    \end{align*}

    \item \textbf{Monotonicity under adjoining variables.}
    We have
    \begin{align}
        \wIPhi(XT;YZ|W)
        \ge
        \wIPhi(X;Y|W).
        \label{max-Phi-MI-more-RV}
    \end{align}

    \item \textbf{Subadditive chain rule.}
    We have
    \begin{subequations}\label{max-Phi-MI-subadditive}
\begin{align}
\wIPhi(X;YZ|W)
&\le
\wIPhi(X;Y|W)
+
\wIPhi(X;Z|YW),
\\
\wIPhi(XZ;Y|W)
&\le
\wIPhi(X;Y|W)
+
\wIPhi(Z;Y|XW).
\end{align}
\end{subequations}
    \item \textbf{$\wIPhi=0$ implies conditional independence.} 
    If, in addition, \(\Phi\) is strictly convex, then 
    \begin{align}
        \wIPhi(X;Y|Z)=0
        \quad\Longrightarrow\quad
        X\to Z\to Y
        \label{max-Phi-Markov}
    \end{align}
    is a Markov chain. In particular,
    \begin{align}
        \wIPhi(X;Y)=0
        \quad\Longrightarrow\quad
        X\independent Y .
        \label{max-Phi-independence}
    \end{align}
    The converse is not true in general, i.e., independence does not imply  $\wIPhi(X;Y)=0$.
    \item \textbf{Markov faithfulness for $\Phi \in \mathscr F$.}
    Suppose $\Phi\in \mathscr{F}$. Then 
    \begin{align}
        X\to Z\to Y \Longleftrightarrow 
        \wIPhi(X;Y|Z)=0. 
        \label{max-Phi-Markov-faithful}
    \end{align}
    In particular, 
    \begin{align}
        X\independent Y 
        \Longleftrightarrow 
        \wIPhi(X;Y)=0. 
        \label{max-Phi-independence-faithful}
    \end{align}
    Moreover, if $W_1\to X\to Y\to W_2$ is a Markov chain, then
    \begin{align}
        \wIPhi(X;Y)= \wIPhi(XW_1;YW_2). 
        \label{max-Phi-Markov-invariance}
    \end{align}
    \item \textbf{Subadditivity for $\Phi \in \mathscr F$.} Suppose $\Phi\in \mathscr F$. If \(X\to Z\to Y\) is a Markov chain, then
    \begin{align}
        \wIPhi(U;Y|XZ)\ge \wIPhi(U;Y|Z).
        \label{max-Phi-MI-F-class-Markov}
    \end{align}
    In particular, if \(X\) and \(Y\) are independent, then
    \begin{align}
        \wIPhi(U;Y|X)\ge \wIPhi(U;Y).
        \label{max-Phi-MI-F-class-independent}
    \end{align}
    \item \textbf{Information inequality.} We have 
    \begin{align} \label{wPhi-self-repetition}
        \wIPhi(X;Y)\leq \sup_U \left\{\wIPhi(U;X) +\wIPhi(U;Y) -\wIPhi(U;XY)\right\}.
    \end{align}
    Suppose in addition $\Phi\in \mathscr G$ and $\Phi(0)<\infty$. Then 
    \begin{align} \label{wPhi-self-repetition-G}
        \wIPhi(X;Y)=\sup_U \left\{\wIPhi(U;X) +\wIPhi(U;Y) -\wIPhi(U;XY)\right\}.
    \end{align}
    Thus, under the same assumptions, for every $P_{UXY}$ we have
    \begin{align} \label{wIPhi-info-ineq}
        \wIPhi(U;X) +\wIPhi(U;Y) -\wIPhi(U;XY) \leq \wIPhi(X;Y). 
    \end{align}
\end{enumerate}

\end{proposition}

The proof is in Appendix \ref{app:proof-prop:max-Phi-MI}.

\subsection{Private-randomness-assisted $\wIPhi$} 
\label{subsec:private-randomness}

A natural requirement for a dependence measure $\delta$ is invariance under adjoining local private randomness. To be precise, let $W_1\to X\to Y\to W_2$ be a Markov chain. Here $W_1$ is the private randomness owned by $X$ and $W_2$ is the private randomness owned by $Y$. 
One would then expect a dependence measure $\delta$ to satisfy
\[
\delta(W_1X;W_2Y)= \delta(X;Y).
\]
The max-$\Phi$-mutual information $\wIPhi$ does not satisfy this
property in general. In fact, adjoining independent private randomness
can strictly increase $\wIPhi$.

\begin{example} \label{eg:local-randomness}
Let $\Phi(t)=(\sqrt t-1)^2$ 
and \(W_1,W_2\) be independent and uniformly distributed on $\{0,1\}$. Take \(U=(W_1,W_2)\). One can verify that $\Phi\in \mathscr G\setminus \mathscr F$. By definition,
\[
\wIPhi(W_1;W_2)
\geq \IPhi(U;W_1)+\IPhi(U;W_2)-\IPhi(U;W_1W_2).
\]
By the formula \eqref{eq:self-uniform}, we can calculate 
\[
\IPhi(W_1;W_1)=\IPhi(W_2;W_2)
=\frac12\Phi(2)+\frac12\Phi(0)
=2-\sqrt2,
\]
and
\[
\IPhi(W_1W_2;W_1W_2)
=\frac14\Phi(4)+\frac34\Phi(0)
=1.
\]
Therefore
\[
\wIPhi(W_1;W_2)
\geq (2-\sqrt2)+(2-\sqrt2)-1
=3-2\sqrt2>0.
\]
Thus \(W_1\independent W_2\), but \(\wIPhi(W_1;W_2)>0\).
\end{example}
The above example shows that $\wIPhi$ may assign a positive value to independent random variables. Thus, in general $\wIPhi$ is not an independence measure. 

Motivated by this example, we define the largest value of $\wIPhi$
that can be obtained by supplying the two terminals with arbitrary
private randomness.

\begin{definition}
The \emph{private-randomness-assisted max-$\Phi$-mutual information}
is defined as
\[
\widehat I_\Phi(X;Y)
\defeq
\sup
\wIPhi(W_1X;W_2Y),
\]
where the supremum is over all joint distributions $p(w_1,w_2,x,y)
=
p(x,y)p(w_1|x)p(w_2|y)$. 
Equivalently, the supremum is over all Markov chains
\[
W_1\to X\to Y\to W_2.
\]
\end{definition}

\begin{proposition} \label{prop:properties-hat-IPhi}
The quantity $\widehat I_\Phi$ has the following properties.
\begin{enumerate}[\rm (i)]
    \item We have
    \[
    \widehat I_\Phi(X;Y)\geq \wIPhi(X;Y).
    \]
    Moreover, if $\Phi\in\mathscr{F}$, we have equality $\widehat I_\Phi(X;Y)=\wIPhi(X;Y)$.
    \item Suppose $W_1\to X\to Y\to W_2$ is a Markov chain. Then 
    \[
    \widehat I_\Phi(W_1X;W_2Y)
    =
    \widehat I_\Phi(X;Y).
    \]

    \item Let $\delta$ be any functional that is invariant under private randomness and satisfies
    \[
    \delta(X;Y)\geq \wIPhi(X;Y)
    \]
    for all $X,Y$. Then
    \[
    \delta(X;Y)\geq \widehat I_\Phi(X;Y).
    \]
\end{enumerate}
\end{proposition}

The proof is in Appendix \ref{app:proof-prop:properties-hat-IPhi}.

In particular, we study $\widehat I_\Phi(X;Y)$  for $\Phi\in\mathscr G$ since $\wIPhi$ has the closed form. By Lemma \ref{lem:class-G-properties}, the following limit exists and is finite,
\[
c_\Phi\defeq\lim_{t\to\infty}t\Phi''(t).
\]
Then we decompose 
\begin{align*}
    \Phi(t)=c_\Phi t\ln t+\Psi(t).
\end{align*}
Theorem \ref{thm:hat-IPhi} below also shows
that, provided the residual component $\Psi$ has finite asymptotic slope, its
$\widehat I_\Psi(X;Y)$ contributes only a constant independent of
the joint distribution of $(X,Y)$. Consequently, all nonconstant
dependence captured by $\widehat I_\Phi$ is carried by Shannon mutual
information.

\begin{theorem}
\label{thm:hat-IPhi}
Suppose $\Phi:(0,\infty)\to \R$, $\Phi\in\mathscr G$ and $\Phi(0)<\infty$. Then $c_\Phi \defeq \lim_{t\to\infty} t\Phi''(t)$ exists and is finite. Define
\[
\Psi(t)
\defeq
\Phi(t)-c_\Phi t\ln t.
\]
Then $\Psi$ is convex, and either $\Psi$ is affine or
$\Psi\in\mathscr G$. Moreover,
\[
\ell_\Phi
\defeq
\lim_{t\to\infty}\frac{\Psi(t)}{t}
\in\R\cup\{+\infty\}
\]
exists.

If $\ell_\Phi<\infty$, then
\[
\widehat I_\Phi(X;Y)
=
c_\Phi(\ln 2) I(X;Y)+K_\Phi,
\]
where
\[
K_\Phi
\defeq
\ell_\Phi+\Psi(0)-\Psi(1)\geq0.
\]
If $\ell_\Phi=+\infty$, then
\[
\widehat I_\Phi(X;Y)=+\infty
\]
for every pair of finite random variables $(X,Y)$.
\end{theorem}

The proof is in Appendix \ref{app:proof-thm:hat-IPhi}.
\begin{example}
    The following functions satisfy the assumptions $\Phi\in\mathscr G$ with $\Phi(0)<\infty$ and $\ell_\Phi<\infty$: $t\log t$, $(\sqrt{t}-1)^2$ and $t\log t-(1+t)\log \frac{1+t}{2}$. 
\end{example}

\begin{remark}
    Note that $\Phi\in \mathscr G$ and $\Phi(0)<\infty$ do not imply $\ell_{\Phi}<\infty$. Consider for \(t>0\),
\[
\Phi(t)\defeq
\int_{1}^{t}\frac{t-s}{s\log(e+s)}\,ds,
\]
and extend \(\Phi\) continuously to \(t=0\). Then
\[
\Phi''(t)=\frac{1}{t\log(e+t)}
\]
and hence
\[
t^2\Phi''(t)=\frac{t}{\log(e+t)},
\]
which is concave on \((0,\infty)\). Therefore
\(\Phi\in\mathscr G\). Moreover,
\[
c_\Phi=\lim_{t\to\infty}t\Phi''(t)=0,
\]
so \(\Psi=\Phi\). Since
\[
\frac{\Phi(t)}{t}
=
\int_1^t
\left(1-\frac{s}{t}\right)
\frac{ds}{s\log(e+s)},
\]
we have
\[
\ell_\Phi
=
\lim_{t\to\infty}\frac{\Psi(t)}{t}
=+\infty.
\]
Furthermore, \(\Phi(0)<\infty\).
\end{remark}

\subsection{Comparison between \texorpdfstring{$\IPhi$}{IPhi} and \texorpdfstring{$\wIPhi$}{max-IPhi}} 
\label{sec:comparison}

The preceding sections introduced two dependence measures associated with a
convex function \(\Phi\). The first one, \(I_\Phi\), is directly induced by
\(\Phi\)-divergence and \(\Phi\)-entropy. It has a natural entropy structure and
an exact chain rule, but its conditional version does not retain all Shannon-type
identities. The second one, \(\wIPhi\), is a variational envelope of
\(I_\Phi\). It restores conditional symmetry and conditional deterministic self-reduction, but it is generally harder to compute and satisfies only a subadditive
chain rule.

Table \ref{tab:IPhi-vs-wIPhi} summarizes the main differences.

\begin{table}[H]
\centering
\footnotesize
\renewcommand{\arraystretch}{1.35}
\setlength{\tabcolsep}{4pt}
\setlength{\arrayrulewidth}{0.4pt}

\begin{tabular}{|p{0.38\textwidth}|c|c|c|c|}
\hline
\textbf{Property}
&
\textbf{\(I_\Phi\)}
&
\textbf{\(\wIPhi\)}
&
\textbf{\(I_\Phi,\Phi\in\mathscr F\)}
&
\textbf{\(\wIPhi,\Phi\in\mathscr F\)}
\\
\hline

Symmetry
&
\(\surd\)
&
\(\surd\)
&
\(\surd\)
&
\(\surd\)
\\
\hline

Conditional symmetry
&
&
\(\surd\)
&
&
\(\surd\)
\\
\hline

Markov faithfulness for strictly convex $\Phi$
&
\(\surd\)
&
&
\(\surd\)
&
\(\surd\)
\\
\hline

Deterministic self-reduction
&
\(\surd\)
&
\(\surd\)
&
\(\surd\)
&
\(\surd\)
\\
\hline

Conditional deterministic self-reduction
&

&
\(\surd\)
&

&
\(\surd\)
\\
\hline

Idempotence
&
\(\surd\)
&
\(\surd\)
&
\(\surd\)
&
\(\surd\)
\\
\hline

Repetition invariance
&
\(\surd\)
&
\(\surd\)
&
\(\surd\)
&
\(\surd\)
\\
\hline

Self-information bound
&
\(\surd\)
&
\(\surd\)
&
\(\surd\)
&
\(\surd\)
\\
\hline

Conditional monotonicity in the second argument
&
\(\surd\)
&
\(\surd\)
&
\(\surd\)
&
\(\surd\)
\\
\hline

Conditional monotonicity for both arguments
&
&
\(\surd\)
&
&
\(\surd\)
\\
\hline

Subadditivity
&
&

&
\(\surd\)
&
\(\surd\)
\\
\hline

Chain rule on the second component
&
\(\surd\)
&
&
\(\surd\)
&
\\
\hline

\end{tabular}
\vspace{8pt}
\caption{Comparison of structural properties of \(I_\Phi\) and \(\wIPhi\).}
\label{tab:IPhi-vs-wIPhi}
\end{table}

\section{$\Phi$-ribbons from independence structures} \label{sec:characterization-ribbon}

In this section, we show a universal inner bound for $\Phi$-ribbons for all random variables with the independence structure. Moreover, this inner bound is tight for the HC ribbon.

\subsection{$\Phi$-ribbons}

The hypercontractivity (HC) ribbon was first defined by Ahlswede and G\'acs \cite{ahlswede1976spreading}. The following equivalent form in terms of mutual information is proven by Nair \cite{nair2014equivalent}: the HC ribbon of $(X_1,\dots,X_n)$, denoted as $\fR(X_1;\dots;X_n)$, is defined as the set of all $(\lambda_1,\dots,\lambda_n)\in [0,1]^n$, such that for any $p_{U|X^n}$, the following inequality holds: 
\begin{align*}
    \sum_{i=1}^n \lambda_i I(U;X_i) \leq I(U;X^n).
\end{align*} 
Beigi and the second author generalize the HC ribbon to the $\Phi$-ribbon and the $\IPhi$-ribbon \cite{beigi2018phi}, stated in Definition~\ref{def:Phi-ribbon}. These two ribbons are the same if $\HPhi$ is homogeneous, i.e., there exists $\kappa(\cdot)$ such that, for all $c>0$ and $f(X)$, $\HPhi(cf)=\kappa(c) 
\HPhi(f)$ \cite{raginsky2016strong}.

\begin{definition}[\cite{beigi2018phi}] \label{def:Phi-ribbon}
    Let $\Phi\in \mathscr{F}$. For finite random variables $(X_1,\dots,X_n)$, the \emph{$\Phi$-ribbon}, denoted as $\fR_{\Phi}(X_1;\dots;X_n)$, is defined as the set of all $(\lambda_1,\dots,\lambda_n) \in [0,1]^n$ such that for any function $f(X^n)$ with $f(\cX^n)\subseteq \dom(\Phi)$, 
    \begin{align*}  
        \sum_{i=1}^n \lambda_i \HPhi(\E[f|X_i])  
        &\leq \HPhi(f).
    \end{align*}
    The \emph{$\IPhi$-ribbon}, denoted as $\fR_{\IPhi}(X_1;\dots;X_n)$, is defined as the set of all $(\lambda_1,\dots,\lambda_n) \in [0,1]^n$ such that for any distribution $p_{U|X^n}$, 
    \begin{align*}  
        \sum_{i=1}^n \lambda_i \IPhi(U;X_i)
        &\leq \IPhi(U;X^n).
    \end{align*}
\end{definition}

When we impose that $f\geq 0$ and $\E[f]=1$ for the $\Phi$-ribbon, the above two ribbons are the same \cite[Theorem 14]{beigi2018phi}.

Note that we can recover the HC ribbon when $\Phi(t)=t\log t$. The $\Phi$-entropy and the $\Phi$-ribbon have applications in the concentration inequalities, Markov chain mixing times, etc., as shown in \cite{raginsky2016strong}.

\begin{proposition}
\label{prop:phi-ribbon-properties}
    For $\Phi\in \mathscr{F}$, we have 
    \begin{enumerate}[\rm (i)]
        \item (Example 10 \cite{beigi2018phi}) $\{(\lambda_1,\dots,\lambda_n)\in[0,1]^n:\sum_{i=1}^n \lambda_i \leq 1\}$ is always a subset of the $\Phi$-ribbon for any random variables $(X_1,\dots,X_n)$. Moreover, when $X_1=\dots=X_n$ and these random variables are nonconstant, the above set is exactly the $\Phi$-ribbon.
        \item (Proposition 11 \cite{beigi2018phi}) If $X_1,\dots,X_n$ are mutually independent, the $\Phi$-ribbon is exactly $[0,1]^n$. 
    \end{enumerate}
\end{proposition}

\subsection{General independence structures}

Given a hypergraph $\cH=(\cV,\cE)$ with $\cV=[n]$ and $\cE\subseteq \cP([n])$, we say that $\cH$ specifies the independence structure of the random variables $(X_1,\dots,X_n)$ if 
\begin{align*}
    \cE
    =
    \left\{S\in \cP([n]): (X_i)_{i\in S} \textrm{ are mutually independent}
    \right\}.
\end{align*}
Conversely, every independence structure can be represented by hypergraphs. Note that $\{i\}\in \cE$ for $i\in [n]$. Moreover, $\cE$ is downward closed: if $S\in \cE$, then for any $T\subseteq S$, $T\in \cE$. When $\cE=\cP([n])$, we say that $\cH$ is complete.

\begin{theorem} \label{thm:Phi-ribbon-general-hypergraph}
    Suppose $\Phi\in \mathscr{F}$. Let $\cH=([n],\cE)$ be the hypergraph that specifies the independence structure of $X_1,\dots,X_n$. Define 
    $$J_{\cE}\defeq\{\vec{0},\vec{e}_1,\dots,\vec{e}_n\}\cup \{\vec{e}_{S}\}_{S\in \cE}$$
    where $\vec{e}_i\in\mathbb{R}^n$ stands for the $i$-th standard basis vector and $\vec{e}_S=\sum_{i\in S}\vec{e}_i$. Then 
    $$\Conv(J_{\cE})\subseteq \fR_{\Phi}(X_1;\dots;X_n).$$ 
\end{theorem}
\begin{proof}
    Let $f=f(X^n)$ be a function of $X^n$. 
    For each $i\in [n]$, we have
    \begin{align} \label{ineq:proof-i}
        \HPhi(\E[f|X_i]) \leq \HPhi(f),
    \end{align}
    which follows from the chain rule and the nonnegativity of the $\Phi$-entropy. 
    For each $S\in \cE$, $(X_i)_{i\in S}$ are mutually independent. Thus by Proposition~\ref {prop:phi-ribbon-properties}~(ii), we have 
    \begin{align} \label{ineq:proof-S}
        \sum_{i\in S} \HPhi(\E[f|X_i]) \leq \HPhi(\E[f|X_S])\leq \HPhi(f).
    \end{align}
    Taking the convex combination of \eqref{ineq:proof-i} and  \eqref{ineq:proof-S} for each $i$ and $S$, we obtain the desired inequality. 
\end{proof}

For the HC ribbon, this inner bound is optimal among bounds determined solely by the prescribed independence structure.

\begin{theorem}
\label{thm:tlogt-arbitrary-independence-converse}
Let $\cH=([n],\cE)$ be a hypergraph that is not complete where $\cE$ contains all singletons and is downward closed. Define 
$$J_{\cE}\defeq\{\vec{0},\vec{e}_1,\dots,\vec{e}_n\}\cup \{\vec{e}_{S}\}_{S\in \cE}$$
If $\vec\lambda=(\lambda_1,\dots,\lambda_n)\in [0,1]^n\setminus \Conv(J_{\cE})$, 
then there exist finite random variables $X_1,\ldots,X_n$ satisfying the independence structure of $\cH$, but 
\[
    \vec\lambda\notin \fR(X_1;\ldots;X_n).
\]
\end{theorem}

The proof is in Appendix \ref{app:proof-tlogt-arbitrary-independence-converse}. The assumption that $\cH$ is not complete is to exclude the trivial case. If $\cH$ is complete, $\Conv(J_{\cE})=[0,1]^n$. Thus there is nothing to prove.

\section{Ribbons from non-Shannon type inequalities} \label{sec:non-Shannon}

Beyond the bounds obtained directly from independent subsets, non-Shannon type inequalities yield additional ribbon bounds. Then we provide a multipartite generalization and an $\IPhi$-ribbon generalization. 

The Zhang--Yeung inequality states that for any four random variables $Z$, $U$, $X$, and $Y$ (\cite{zhang1998characterization}, \cite[Theorem 14.7]{yeung2012first}):
\begin{align*}
    I(Z; U) \leq &2I(Z; U|X)+I(Z; U|Y )+I(X; Y)+I(X; Z|U)+I(X; U|Z)
\end{align*}
and 
\begin{align*}
    2I(X;Y) \leq &I(Z;U) + I(Z;XY) + 3I(X;Y|Z) + I(X;Y|U).
\end{align*}
Moreover, a generalization of the Zhang--Yeung inequality to a five-random-variable inequality was given in \cite[Theorem 2]{makarychev2002new}: 
\begin{align*}
    I(Z;U) \leq I(Z;U|X)+I(Z; U|Y)+I(X;Y)+I(Z;V|U)+I(Z; U|V)+I(U;V|Z).
\end{align*}

Renaming the variables $X\to S$,
$Y\to T$,
$Z\to X$,
$U\to Y$,
$V\to J$, the 5-variable non-Shannon type inequality can be rewritten in the following form: 
\begin{align*}
    \frac{2}{3}I(J;X)
    +\frac{2}{3}I(J;Y)-I(J;XY)\leq 
    \frac{1}{3}
    \bigg(I(S;T)+I(X;Y|S) +I(X;Y|T)\bigg).
\end{align*}
Zhang and Yang \cite{zhang2002new} define $(S,T)$ to be an independent perfect covering of $(X,Y)$ if $S\independent T$, $X\independent Y|S$ and $X\independent Y|T$. They also provide an example in which $(X,Y)$ are dependent but admit an independent perfect covering. The existence of the independent perfect coverings implies that $(2/3,2/3)$ is in the HC ribbon of $(X,Y)$. 

\subsection{Generalization of the Zhang--Yeung inequality}
Zhang and Yang also provide a multipartite generalization of the Zhang--Yeung inequality \cite[Theorem 3]{zhang2002new}: 
\begin{align*}
    n I(J;X;Y)
    -I(J;XY)
    \leq \sum_{i=1}^n I(X;Y|T_i)
    +\sum_{i=1}^n H(T_i)-H(T^n),
\end{align*}
where $I(J;X;Y)=I(J;X)-I(J;X|Y)$. We will give another generalized version. 

\begin{theorem}
    For any random variables $U,X_1,\dots,X_n,S,T$, we have
    \begin{align*}
        \sum_{i=1}^n \gamma_i I(U;X_i)
        -I(U;X^n)
        \leq 
        \frac{1}{3}
        \Bigg(
        \sum_{i<j}
        \bigg(
        I(X_i;X_j|S)
        +I(X_i;X_j|T)
        \bigg)
        +I(S;T)
        \Bigg)
    \end{align*}
    if $\vec\gamma=(\gamma_1,\dots,\gamma_n)\in[0,2/3]^n$ and $\sum_{i=1}^n\gamma_i\le 4/3$. 

    Thus, if there exist independent $S,T$ such that $(X_1,\dots,X_n)$ are pairwise independent given $S$ and given $T$, then $\{(\gamma_1,\dots,\gamma_n)\in [0,2/3]^n:\sum_{i=1}^n \gamma_i\leq \frac{4}{3}\}$ is a subset of the HC ribbon for $X_1,\dots,X_n$. 
\end{theorem}

The proof follows the same idea as Theorem \ref{thm:multipartite-Phi-ZY} so we omit it here.

Then we can derive the $\Phi$-mutual information generalization of the Zhang--Yeung inequality. 

\begin{theorem}[Multipartite $\Phi$-Zhang--Yeung inequality] \label{thm:multipartite-Phi-ZY}
Suppose $\Phi:\R_+\to \R$ is convex and $n\geq 2$. If $\vec\gamma\in[0,2/3]^n$ and $\sum_{i=1}^n\gamma_i\le 4/3$, then
\[
\sum_{i=1}^n\gamma_i \IPhi(J;X_i)-\IPhi(J;X^n)
\le
\frac13
\left[
\sum_{i<j}
\big(
\wIPhi(X_i;X_j|S)+\wIPhi(X_i;X_j|T)
\big)
+
\wIPhi(S;T)
\right],
\]
for any $J,X^n,S,T$. 
\end{theorem}
\begin{proof}
    Since the expression depends only on the marginal distributions of $(J,X^n)$ and $(X^n,S,T)$, without loss of generality assume that $J\rightarrow X^n\rightarrow (S,T)$. 
    It suffices to prove the result when $\sum_{i=1}^n\gamma_i=4/3$, because the $\Phi$-mutual information is nonnegative. Define $\lambda_i\defeq \frac{3}{2}\gamma_i$ for each $i\in[n]$. 
    Then $\vec\lambda\in[0,1]^n$ and $\sum_{i=1}^n\lambda_i=2$. Hence $\vec\lambda$ is a convex combination of the vectors $\vec e_i+\vec e_j$, $i<j$; write
    \[
        \vec\lambda=\sum_{i<j}\omega_{ij}(\vec e_i+\vec e_j),
        \qquad
        \omega_{ij}\geq 0,
        \qquad
        \sum_{i<j}\omega_{ij}=1.
    \]

    For every $i<j$, we have 
    \begin{align*}
        &\IPhi(J;X_i|S)+\IPhi(J;X_j|S)-\IPhi(J;X^n|S)\\
        \leq&
        \IPhi(J;X_i|S)+\IPhi(J;X_j|S)-\IPhi(J;X_iX_j|S)\\
        \leq& \wIPhi(X_i;X_j|S)
    \end{align*}
    where the first inequality follows from the chain rule and the nonnegativity of $\IPhi$, and the second inequality follows from the definition of $\wIPhi$. 

    Taking the convex combination of these inequalities, we obtain
    \begin{align*}
        \sum_{i=1}^n\lambda_i\IPhi(J;X_i|S)-\IPhi(J;X^n|S)
        \leq \sum_{i<j}\wIPhi(X_i;X_j|S).
    \end{align*}
    Using the chain rule, $\sum_{i=1}^n\lambda_i=2$, and the nonnegativity of the $\Phi$-mutual information, this implies 
    \begin{align}
        \sum_{i=1}^n\lambda_i\IPhi(J;X_i)
        -\IPhi(J;S)-\IPhi(J;X^nST)
        \leq \sum_{i<j}\wIPhi(X_i;X_j|S).
        \label{ineq:multipartite-Phi-ZY-S}
    \end{align}
    Similarly,
    \begin{align}
        \sum_{i=1}^n\lambda_i\IPhi(J;X_i)
        -\IPhi(J;T)-\IPhi(J;X^nST)
        \leq \sum_{i<j}\wIPhi(X_i;X_j|T).
        \label{ineq:multipartite-Phi-ZY-T}
    \end{align}
    Finally, the definition of $\wIPhi(S;T)$ and the nonnegativity of the $\Phi$-mutual information yield
    \begin{align}
        \IPhi(J;S)+\IPhi(J;T)-\IPhi(J;X^nST)
        \leq \wIPhi(S;T).
        \label{ineq:multipartite-Phi-ZY-ST}
    \end{align}
    Summing \eqref{ineq:multipartite-Phi-ZY-S}, \eqref{ineq:multipartite-Phi-ZY-T}, and \eqref{ineq:multipartite-Phi-ZY-ST}, and then dividing by $3$, gives
    \begin{align*}
        \sum_{i=1}^n\frac{2\lambda_i}{3}\IPhi(J;X_i)
        -\IPhi(J;X^nST)
        \leq \frac{1}{3}\left[
        \sum_{i<j}\big(
        \wIPhi(X_i;X_j|S)+\wIPhi(X_i;X_j|T)
        \big)+\wIPhi(S;T)
        \right].
    \end{align*}
    Since $2\lambda_i/3=\gamma_i$ and the Markov chain gives $\IPhi(J;X^nST)=\IPhi(J;X^n)$, this is the desired inequality.
\end{proof}

\begin{corollary}
Suppose $\Phi \in \mathscr F$. 
Let $X_1,\dots,X_n,S,T$ be random variables such that
\[
S\independent T,
\qquad
X_i\independent X_j|S,
\qquad
X_i\independent X_j|T,
\quad \forall i<j.
\]
If $\vec\gamma\in[0,2/3]^n$ and $\sum_{i=1}^n\gamma_i\le 4/3$, then
\[
\vec\gamma\in\fR_{\IPhi}(X_1;\dots;X_n).
\]
\end{corollary}

\section{$\Phi$-strong data processing inequality constants}
\label{sec:SDPI}

\begin{definition}
Suppose $\Phi:\cI\to \R$ is convex where $\cI\subseteq \R_+$. Define
$\eta_{\Phi}(X;Y)$ to be the smallest (the infimum of)
$\lambda \geq 0$ such that, for any function $f=f(X)$
(whose range is in the domain of $\Phi$), we have
\[
\lambda H_{\Phi}(f)
\geq
H_{\Phi}\bigl(\mathbb{E}[f | Y]\bigr).
\]
We call $\eta_{\Phi}(X;Y)$ the \emph{$\Phi$-strong data processing
inequality constant} ($\Phi$-SDPI constant). When $X$ is constant or $\Phi$ is affine, we define $\eta_{\Phi}(X;Y)=0$. 
\end{definition}

By the chain rule and the nonnegativity of $\Phi$-entropy, 
\begin{align*}
    \HPhi(f)\geq \HPhi \left(\E[f|Y]\right). 
\end{align*}
Thus, $\eta_{\Phi}(X;Y)$ is well defined and $\eta_{\Phi}(X;Y)\in [0,1]$. 

Equivalently, we have 
\begin{align*}
    \eta_{\Phi}(X;Y)=\sup \frac{\HPhi(\E[f|Y])}{\HPhi(f)}
\end{align*}
where the supremum is over all $f(X)$ such that $f(\cX)\subseteq \dom(\Phi)$ and $\HPhi(f)>0$. When $\Phi$ is affine or $X$ is constant, we define $\eta_{\Phi}(X;Y)=0$. 

For $\Phi\in \mathscr{F}$, the $\Phi$-SDPI constant was first studied by Raginsky \cite{raginsky2016strong} where his version has the constraint $\E[f]=1$. The above version is proposed by Beigi and the second author \cite{beigi2018phi}.

In \cite[Theorem 20]{beigi2018phi}, it is shown that when $\Phi\in \mathscr F$ and the domain of $\Phi$ is a compact interval, we have the identity 
\begin{align*}
    \eta_{\Phi}(X;Y) &= \inf\frac{1-\lambda_1}{\lambda_2}
\end{align*}
where the infimum is over all $(\lambda_1,\lambda_2)\in \fR_{\Phi} (X;Y)$ with $\lambda_2\neq 0$. In concurrent work, we show that this characterization remains valid without the compactness assumption, and that it can be extended to the matrix setting.

\subsection{Z-channel source}
We say that $(X,Y)$ is a Z-channel source if the joint distribution is given by 
$$p_{XY}=\begin{pmatrix}
    1-s & 0 \\
    s\delta & s(1-\delta)
\end{pmatrix}$$ 
where $s,\delta\in(0,1)$. Here, we discuss the $\Phi$-SDPI constant only in the scalar case.

\begin{theorem} \label{thm:Z-channel}
Let $(X,Y)$ be a Z-channel source with
\[
p_{XY}
=
\begin{pmatrix}
1-s & 0\\
s\delta & s(1-\delta)
\end{pmatrix},
\qquad s,\delta\in(0,1).
\]
Suppose that $\Phi:(0,\infty)\to\mathbb R$ is strictly convex and three times differentiable such that $\Phi''(x)>0$ for all $x>0$. 

For $m>0$ and $0<v<m/s$, define
\begin{align*}
    F_m(v)\defeq 
    \begin{cases}
        \frac{
        (1-s(1-\delta))\Phi\left(\frac{m-s(1-\delta)v}{1-s(1-\delta)}\right)
        +s(1-\delta)\Phi(v)-\Phi(m)
        }{
        (1-s)\Phi\left(\frac{m-sv}{1-s}\right)
        +s\Phi(v)-\Phi(m)
        },\quad v\neq m\\
        \frac{(1-s)(1-\delta)}{1-s(1-\delta)},\quad v=m.
    \end{cases}
\end{align*}

Assume that $F_m(v)$ is nonincreasing in $(0,m/s)$ for all $m>0$. 

If $\Phi(0)<\infty$, the optimization defining
$\eta_\Phi(X;Y)$ can be restricted to functions satisfying
$f(1)=0$.

If $\Phi(0)=\infty$, then
\[
\eta_\Phi(X;Y)
=
\lim_{v\downarrow0}F_m(v)
=
1-\delta.
\]
\end{theorem}

The proof is in Appendix \ref{app:proof-Z-channel}.

\begin{theorem} \label{thm:sufficient-Phi-Z-channel}
    Let $\Phi:(0,\infty)\to\R$ be strictly convex and three times differentiable with $\Phi(0)\defeq \lim_{t\downarrow 0}\Phi(t)\in \R\cup \{+\infty\}$. Each of the following conditions guarantees that the function $F_m(v)$ in Theorem \ref{thm:Z-channel} is nonincreasing:
    \begin{enumerate}[\rm (i)]
        \item $\Phi'''(t)\leq 0$ for every $t>0$, and
        \begin{align} \label{sup-condition1}
            &\bigg(\Phi(x)+\Phi'(x)(y-x)-\Phi(y)\bigg)\bigg(\Phi'(y)-\Phi'(x)\bigg)+\Phi''(x)(y-x)^2 \bigg(\Phi'(y)-\frac{\Phi(x)-\Phi(y)}{x-y}\bigg)\geq 0
        \end{align}
        for every $x,y>0$ with $x\neq y$.
        \item For all $x,y>0$ such that $\Phi'''(x)\neq 0$ and $\Phi'''(y)\neq 0$,
        \begin{align} 
            x+\frac{\Phi''(x)}{\Phi'''(x)}
            \geq y+3\frac{\Phi''(y)}{\Phi'''(y)}.
            \label{sup-condition2}
        \end{align}
        \item For any $x>0$, 
        \begin{align} \label{sup-condition3}
            1\leq -\frac{x\Phi'''(x)}{\Phi''(x)}\leq 3. 
        \end{align}
    \end{enumerate}
\end{theorem}
The proof is in Appendix \ref{apx:sufficient-Phi-Z-channel}.

\begin{example}
    The following functions satisfy the condition in Theorem \ref{thm:sufficient-Phi-Z-channel} (iii): $t\log(t),1/t,-\log(t)$. Define 
    \begin{align*}
        B_{\Phi}(t)&=-\frac{t\Phi'''(t)}{\Phi''(t)}. 
    \end{align*}
    Observe the following:
    \begin{align*}
        B_{t\log t}(t) &=1, \\
        B_{1/t}(t) &= 3, \\
        B_{-\log t}(t) &=2.
    \end{align*}
\end{example}

\section{Future work}

\textbf{Private randomness assisted information.}
We saw that for class $\mathscr{G}$, $\widetilde{I}_\Phi$ captures both the dependence and the randomness in the individual variables $X$ and $Y$. We have shown that $\widehat{I}_\Phi(X;Y)$ is either an affine function of Shannon’s mutual information or identically infinite. Moreover, for $\Phi\in \mathscr F$, $\widehat{I}_\Phi(X;Y)=\wIPhi(X;Y)$. What happens to $\widehat I_\Phi$ when $\Phi$ does not belong to $\mathscr{G}\cup \mathscr{F}$?

\textbf{Other notions of $\Phi$-mutual information.} The definition  
\[
I_\Phi(X;Y)
=
D_\Phi(p_{XY}\|p_X\otimes p_Y)
\]
that we use in this paper is not unique. See Table \ref{tab:Phi-MI-definitions} for other definitions. It is interesting to investigate the properties of $\wIPhi$ and $\IPhi$ based on the other definitions of $I_\Phi(X;Y)$.

\section*{Acknowledgement}

Theorems~\ref{thm:conditional-Phi-variational-form}, \ref{thm:hat-IPhi}, \ref{thm:tlogt-arbitrary-independence-converse} and Proposition~\ref{prop:properties-hat-IPhi} are developed with the assistance of ChatGPT 5.6 Sol. The authors have verified AI-generated content. The authors also use ChatGPT 5.6 Sol for proofreading.

\bibliographystyle{IEEEtran}
\bibliography{refs}

\appendices
\section{Proof of Lemma \ref{lem:class-G-properties}}
\label{app:proof-lem:class-G-properties}

Let
\begin{align*}
    g(t)\defeq t^2\Phi''(t).
\end{align*}
By the definition of $\mathscr G$, $g$ is concave and nonnegative.
For any $0<s<t$ and $0<\epsilon<s$, the concavity of $g$ gives
\begin{align*}
    g(s)
    &\geq
    \frac{t-s}{t-\epsilon}g(\epsilon)
    +\frac{s-\epsilon}{t-\epsilon}g(t)\\
    &\geq
    \frac{s-\epsilon}{t-\epsilon}g(t).
\end{align*}
Letting $\epsilon\to 0^+$, we obtain
\begin{align*}
    \frac{g(s)}{s}\geq \frac{g(t)}{t}.
\end{align*}
Therefore, $t\mapsto t\Phi''(t)$ is nonincreasing and nonnegative.
It follows that
\begin{align*}
    c_\Phi\defeq\lim_{t\to\infty}t\Phi''(t)
\end{align*}
exists and is finite. Moreover,
\begin{align*}
    0\leq c_\Phi\leq t\Phi''(t),
    \qquad t>0.
\end{align*}

Since $1+\ln t\to\infty$, L'H\^{o}pital's rule (\cite[Theorem 5.13]{rudin2021principles}) yields
\begin{align*}
    \lim_{t\to\infty}\frac{\Phi'(t)}{1+\ln t}
    &=
    \lim_{t\to\infty}
    \frac{\Phi''(t)}{1/t}\\
    &=c_\Phi.
\end{align*}
Applying L'H\^{o}pital's rule once more gives
\begin{align*}
    \lim_{t\to\infty}\frac{\Phi(t)}{t\ln t}
    &=
    \lim_{t\to\infty}
    \frac{\Phi'(t)}{1+\ln t}\\
    &=c_\Phi.
\end{align*}
Hence, for every $t>0$,
\begin{align*}
    \Psi''(t)
    =\Phi''(t)-\frac{c_\Phi}{t}
    =\frac{t\Phi''(t)-c_\Phi}{t}
    \geq 0,
\end{align*}
so $\Psi$ is convex. Furthermore,
\begin{align*}
    t\Psi''(t)=t\Phi''(t)-c_\Phi
\end{align*}
is nonincreasing and converges to zero. Moreover,
\begin{align*}
    \lim_{t\to\infty}\frac{\Psi(t)}{t\ln t}
    &=
    \lim_{t\to\infty}\frac{\Phi(t)}{t\ln t}-c_\Phi\\
    &=0.
\end{align*}
Equivalently, $\Psi(t)=o(t\ln t)$.
Finally,
\begin{align*}
    t^2\Psi''(t)
    =t^2\Phi''(t)-c_\Phi t,
\end{align*}
which is concave since $t^2\Phi''(t)$ is concave. Therefore, $\Psi$
satisfies the defining curvature condition of $\mathscr G$. Thus, either
$\Psi$ is affine or $\Psi\in\mathscr G$.

Finally, suppose $\Phi(0)<\infty$. Then $\Psi(0)<\infty$. For
$0<s<t$, the convexity of $\Psi$ gives
\begin{align*}
    \frac{\Psi(s)-\Psi(0)}{s}
    \leq
    \frac{\Psi(t)-\Psi(0)}{t}.
\end{align*}
Hence $t\mapsto(\Psi(t)-\Psi(0))/t$ is nondecreasing, so its limit
exists in $\mathbb R\cup\{+\infty\}$. Since $\Psi(0)/t\to0$, the same
is true of
\begin{align*}
    \ell_\Phi
    =\lim_{t\to\infty}\frac{\Psi(t)}{t}.
\end{align*}

\section{Proof of Proposition \ref{prop:properties-Phi-MI}} \label{app:proof-prop:properties-Phi-MI}

\begin{enumerate}[\rm (i)]
    \item \textbf{Deterministic self-reduction.}
    See \cite[Theorem 5]{zvarova1974measures} for the proof.

    \item \textbf{Markov faithfulness.}
    Suppose $X\to Z\to Y$ is a Markov chain. Then 
    \begin{align*}
        \IPhi(X;YZ)
        &=\sum_{x,y,z} p(x) p(y,z )
        \Phi \left(\frac{p(x,y,z)}{p(x)p(y,z)}\right) -\Phi(1)\\
        &=\sum_{x,y,z} p(x) p(y,z)
        \Phi \left(\frac{p(x|y,z)}{p(x)}\right) -\Phi(1)\\
        &=\sum_{x,y,z} p(x) p(y,z)
        \Phi \left(\frac{p(x|z)}{p(x)}\right) -\Phi(1)\\
        &=\sum_{x,z} p(x) p(z)
        \Phi \left(\frac{p(x|z)}{p(x)}\right) -\Phi(1)\\
        &=\IPhi(X;Z)
    \end{align*}
    where the third line follows from the Markov condition $X\to Z\to Y$. 

    Conversely, assume that $\Phi$ is strictly convex and $\IPhi(X;Y|Z)=0$. Let $f_x(y,z)=\frac{p(x,y,z)}{p(x)p(y,z)}$. Then
    \begin{align*}
        \IPhi(X;Y|Z)
        &=\sum_{x}p(x) \HPhi(f_x(Y,Z)|Z).
    \end{align*}
    Since $\IPhi(X;Y|Z)=0$, we have $\HPhi(f_x(Y,Z)|Z)=0$ for any $x$ with $p(x)>0$. By Jensen's inequality,
    \begin{align*}
        \HPhi(f_x(Y,Z)|Z)
        &=\E[\Phi(f_x(Y,Z))]-\E[\Phi(\E[f_x(Y,Z)|Z])]\geq 0.
    \end{align*}
    Since $\Phi$ is strictly convex, $f_x(Y,z)$ is constant for any $z$ with $p(z)>0$. That is, for any $y,z$, $f_x(y,z)=\E[f_x(Y,Z)|Z=z]$. Note that
    \begin{align*}
        \E[f_x(Y,Z)|Z=z]
        &=\sum_{y} p(y|z) \frac{p(x,y,z)}{p(x)p(y,z)}\\
        &=\frac{p(x|z)}{p(x)}.
    \end{align*}
    Then $\frac{p(x,y,z)}{p(x)p(y,z)}=\frac{p(x|z)}{p(x)}$. Rearranging the terms, we obtain $p(x|y,z)=p(x|z)$. Thus $X\to Z\to Y$ is a Markov chain.

    \item \textbf{Idempotence.}
    The unconditional quantity $\IPhi(A;B)$ is symmetric in $A$ and $B$ by definition. Since $X$ is a deterministic function of both $(X,Y,Z)$ and $(X,Z)$, the deterministic-transformation property proved in (i) gives
    \begin{align*}
        \IPhi(X;XYZ)
        &=\IPhi(XYZ;X)
        =\IPhi(X;X),\\
        \IPhi(X;XZ)
        &=\IPhi(XZ;X)
        =\IPhi(X;X).
    \end{align*}
    Hence $\IPhi(X;XYZ)=\IPhi(X;XZ)$. Subtracting $\IPhi(X;Z)$ from both sides gives
    \[
        \IPhi(X;XY|Z)=\IPhi(X;X|Z).
    \]

    \item \textbf{Permutation invariance.}
    Let
    \[
    A\defeq f(X),
    \qquad
    B\defeq g(Y),
    \qquad
    C\defeq h(Z).
    \]
    Since \(f\), \(g\), and \(h\) are one-to-one, the mappings
    \[
    x\mapsto f(x),
    \qquad
    (y,z)\mapsto (g(y),h(z))
    \]
    are one-to-one on the corresponding supports. Therefore, for every
    \(x\in\cX\), \(y\in\cY\), and \(z\in\cZ\),
    \begin{align*}
        p_A(f(x))
        &=p_X(x),\\
        p_{BC}(g(y),h(z))
        &=p_{YZ}(y,z),\\
        p_{ABC}(f(x),g(y),h(z))
        &=p_{XYZ}(x,y,z).
    \end{align*}
    Using the definition of the \(\Phi\)-mutual information, we obtain
    \begin{align*}
        \IPhi(f(X);g(Y)h(Z))
        &=
        \sum_{\substack{a\in f(\cX)\\
                        b\in g(\cY),\,c\in h(\cZ)}}
        p_A(a)p_{BC}(b,c)
        \Phi\left(
            \frac{p_{ABC}(a,b,c)}
                {p_A(a)p_{BC}(b,c)}
        \right)
        -\Phi(1)\\
        &=
        \sum_{x,y,z}
        p_X(x)p_{YZ}(y,z)
        \Phi\left(
            \frac{p_{XYZ}(x,y,z)}
                {p_X(x)p_{YZ}(y,z)}
        \right)
        -\Phi(1)\\
        &=
        \IPhi(X;YZ).
    \end{align*}
    Similarly,
    \begin{align*}
        \IPhi(f(X);h(Z))
        &=
        \sum_{\substack{a\in f(\cX)\\c\in h(\cZ)}}
        p_A(a)p_C(c)
        \Phi\left(
            \frac{p_{AC}(a,c)}
                {p_A(a)p_C(c)}
        \right)
        -\Phi(1)\\
        &=
        \sum_{x,z}
        p_X(x)p_Z(z)
        \Phi\left(
            \frac{p_{XZ}(x,z)}
                {p_X(x)p_Z(z)}
        \right)
        -\Phi(1)\\
        &=
        \IPhi(X;Z).
    \end{align*}
    Hence, by the definition of conditional \(\Phi\)-mutual information,
    \begin{align*}
        \IPhi(f(X);g(Y)|h(Z))
        &=
        \IPhi(f(X);g(Y)h(Z))
        -
        \IPhi(f(X);h(Z))\\
        &=
        \IPhi(X;YZ)-\IPhi(X;Z)\\
        &=
        \IPhi(X;Y|Z).
    \end{align*}
    This proves \eqref{permutation-invariance}.

    \item \textbf{Self-information bound.} See \cite[Theorem 3]{zvarova1974measures} for the proof. 

    \item \textbf{Conditional absorption on the second argument.}  By the chain rule, 
    \begin{align*}
        \IPhi(X;YZ|Z)
        &=\IPhi(X;YZZ)-\IPhi(X;Z)\\
        &=\IPhi(X;YZ)-\IPhi(X;Z)\\
        &=\IPhi(X;Y|Z)
    \end{align*}
    where the second line follows from permutation invariance because
    the map $(Y,Z)\mapsto(Y,Z,Z)$ is one-to-one.

    \item \textbf{Data processing on the second argument.}
    By the data processing property of $\Phi$-divergence applied to the projection $(X,Y,Z)\mapsto (X,Z)$,
    \[
        \IPhi(W;XZ)\leq \IPhi(W;XYZ).
    \]
    Subtracting $\IPhi(W;Z)$ from both sides gives
    \[
        \IPhi(W;X|Z)\leq \IPhi(W;XY|Z).
    \]
\end{enumerate}

\section{Proof of Proposition \ref{prop:IPhi-counterexample}} \label{app:proof-prop:IPhi-counterexample}

    \begin{enumerate}[\rm (i)]
        \item Take
    \[
    \Phi(t)=(t-1)^2.
    \]
    Let \(X=Y\sim\operatorname{Bern}(1/2)\), and let
    \(Z\sim\operatorname{Bern}(1/2)\) be independent of \(X\).
    Then \(XZ=YZ\) is uniform on four points. By \eqref{eq:self-uniform}, we have
    \[
    \IPhi(XZ;YZ)=3,
    \qquad
    \IPhi(XZ;Z)=\IPhi(Z;Z)=1,
    \]
    and therefore
    \[
    \IPhi(XZ;Y|Z)=\IPhi(XZ;YZ)-\IPhi(XZ;Z)=3-1=2.
    \]
    On the other hand,
    \[
    \IPhi(X;YZ)=\IPhi(X;X)=1,
    \qquad
    \IPhi(X;Z)=0,
    \]
    so
    \[
    \IPhi(X;Y|Z)=\IPhi(X;YZ)-\IPhi(X;Z)=1.
    \]
    Thus
    \[
    \IPhi(XZ;Y|Z)=2\neq1=\IPhi(X;Y|Z).
    \]
    \item Take
\[
\Phi(t)=(\sqrt{t}-1)^2.
\]
Let \(Y,Z\sim\Bern(1/2)\) be independent, and set
\[
X=Y,
\qquad
W=Z.
\]
Then \(YZ\) is uniformly distributed with support size \(4\).
By \eqref{eq:self-uniform}, we have
\begin{align*}
    \IPhi(Y;YZ)
    &=
    \IPhi(Y;Y)
    =
    2-\sqrt{2},
\end{align*}
\begin{align*}
    \IPhi(YZ;YZ)
    =
    1,
\end{align*}
and
\begin{align*}
    \IPhi(Z;Z)
    =
    2-\sqrt{2}.
\end{align*}
Since \(Y\independent Z\),
\begin{align*}
    \IPhi(Y;Z)
    =
    0.
\end{align*}
Therefore,
\begin{align*}
    \IPhi(X;Y|Z)
    &=
    \IPhi(Y;YZ)-\IPhi(Y;Z)\\
    &=
    2-\sqrt{2}.
\end{align*}
On the other hand,
\begin{align*}
    \IPhi(XW;Y|Z)
    &=
    \IPhi(YZ;YZ)-\IPhi(YZ;Z)\\
    &=
    \IPhi(YZ;YZ)-\IPhi(Z;Z)\\
    &=
    1-(2-\sqrt{2})\\
    &=
    \sqrt{2}-1.
\end{align*}
Hence,
\[
    \IPhi(X;Y|Z)
    =
    2-\sqrt{2}
    >
    \sqrt{2}-1
    =
    \IPhi(XW;Y|Z).
\]

However, if \(Z\independent(X,Y,W)\), then
\begin{align*}
    \IPhi(X;Y|Z)
    &=
    \IPhi(X;Y),
\end{align*}
and
\begin{align*}
    \IPhi(XW;Y|Z)
    &=
    \IPhi(XW;Y).
\end{align*}
In this case, by the symmetry of the unconditional \(\Phi\)-mutual
information and its data processing inequality,
\begin{align*}
    \IPhi(X;Y)
    =
    \IPhi(Y;X)
    \leq
    \IPhi(Y;XW)
    =
    \IPhi(XW;Y).
\end{align*}
Thus, the failure of monotonicity in the first argument above is
intrinsically a conditional phenomenon.
    
\end{enumerate}

\section{Proof of Theorem \ref{thm:CDFI-forces-MI}} \label{app:proof-thm:CDFI-forces-MI}

(i)
Suppose the conditional deterministic self-reduction holds. Consider $(YZ,Z,X)$. Then $Z$ is a function of $(YZ,X)$. Thus 
\begin{align*}
    \IPhi(YZ;Z|X)=\IPhi(Z;Z|X). 
\end{align*}
By the chain rule, we have 
\begin{align*}
    \IPhi(XZ;YZ)-\IPhi(X;YZ)
    &=\IPhi(Z;XZ)-\IPhi(X;Z)\\
    &=\IPhi(Z;Z)-\IPhi(X;Z).
\end{align*}
Thus, 
\begin{align*}
    \IPhi(X;Y|Z)
    &=\IPhi(X;YZ)-\IPhi(X;Z)\\
    &=\IPhi(XZ;YZ)-\IPhi(Z;Z). 
\end{align*}
Note that this form is symmetric in $X$ and $Y$. Thus we have the conditional symmetry:
\begin{align*}
    \IPhi(X;Y|Z)
    &=\IPhi(Y;X|Z).
\end{align*}

(ii) Suppose $\IPhi$ is conditionally symmetric, i.e. 
\begin{align*}
    \IPhi(X;Y|Z)
    &=\IPhi(Y;X|Z).
\end{align*}
Take $X=ST$, $Y=S$ and $Z=T$. Then $Y=f(X,Z)$. Then  
\begin{align*}
    \IPhi(ST;S|T)
    &=\IPhi(S;ST|T)\\
    &=\IPhi(S;S|T). 
\end{align*}
Since $\IPhi(ST;S|T)=\IPhi(ST;ST)-\IPhi(ST;T)$ and $\IPhi(S;S|T)=\IPhi(S;ST)-\IPhi(S;T)$, we obtain 
\begin{align*}
    \IPhi(ST;ST)-\IPhi(ST;T)
    =\IPhi(S;ST)-\IPhi(S;T).
\end{align*}
Since $\IPhi(ST;T)=\IPhi(T;T)$ and $\IPhi(S;ST)=\IPhi(S;S)$, rearranging the equation, we obtain 
\begin{align} \label{IPhi-S-T}
    \IPhi(S;T)=\IPhi(S;S)+\IPhi(T;T)-\IPhi(ST;ST). 
\end{align}

Since $\Phi(0)<\infty$, $\IPhi(S;S)<\infty$ for all $S$. 
By \eqref{IPhi-S-T}, for $S\independent T$, we have 
\begin{equation}
    \IPhi(ST;ST)
    =
    \IPhi(S;S)+\IPhi(T;T).
    \label{eq:self-info-product-additive}
\end{equation}

Let
\(S\sim P=(p_i)_{i=1}^n\) and \(T\sim Q=(q_j)_{j=1}^m\) be fully-supported and independent. Then 
\[
    \IPhi(S;S)
    =
    \sum_{i=1}^n p_i^2 \Phi\left(\frac1{p_i}\right)
    +
    \sum_{\substack{1\le i,j\le n\\ i\neq j}}p_i p_j \Phi(0)
    -
    \Phi(1).
\]
Equivalently,
\begin{equation}
    \IPhi(S;S)
    =
    \sum_{i=1}^n h(p_i)+C,
    \label{eq:self-info-trace-form}
\end{equation}
where
\[
    h(p)
    \defeq
    p^2\left(\Phi\left(\frac1p\right)-\Phi(0)\right),
    \qquad
    C\defeq \Phi(0)-\Phi(1).
\]

We now use the product additivity \eqref{eq:self-info-product-additive}. 
By \eqref{eq:self-info-product-additive} and \eqref{eq:self-info-trace-form}, we have 
\begin{equation}
    \sum_{i=1}^n\sum_{j=1}^m h(p_iq_j)+C
    =
    \sum_{i=1}^n h(p_i)+C+\sum_{j=1}^m h(q_j)+C.
    \label{eq:self-h-product-additive}
\end{equation}
Fix \(Q\), and define
\[
    g_Q(x)
    \defeq
    \sum_{j=1}^m h(xq_j)-h(x).
\]
Then \eqref{eq:self-h-product-additive} implies that, for every probability vector
\(P=(p_i)_{i=1}^n\),
\[
    \sum_{i=1}^n g_Q(p_i)=\IPhi(T;T).
\]
Comparing the two probability vectors
\[
    P=(x,y,1-x-y)
    \quad\text{and}\quad
    P=(x+y,1-x-y),
\]
we obtain 
\begin{align} \label{eq:gQ-linear}
    g_Q(x)+g_Q(y)=g_Q(x+y)
\end{align}
for \(x,y\in (0,1)\) such that \(x+y<1\). Since $\Phi$ is twice continuously differentiable, \(g_Q\) is continuous on $(0,1)$. Take $x_0\in (0,1)$.  Note that 
\begin{align*}
    \lim_{\epsilon\downarrow 0}
    g_Q(\epsilon)
    &\overset{\rm (a)}{=}\lim_{\epsilon\downarrow 0}
    g_Q(x_0+\epsilon)-g_Q(x_0)\\
    &\overset{\rm (b)}{=}0
\end{align*}
where (a) follows from \eqref{eq:gQ-linear} and (b) follows from the continuity at $x_0$. Thus $g_Q$ extends continuously to $[0,1]$.  Thus the solution to the Cauchy functional equation \eqref{eq:gQ-linear} must be linear \cite[Theorem 1]{wright2019gleason}, i.e., 
\[
    g_Q(x)=a_Q x
\]
for some constant \(a_Q\). Differentiating twice gives
\begin{equation}
    \sum_{j=1}^m q_j^2 h''(xq_j)=h''(x).
    \label{eq:self-h-second-functional}
\end{equation}

Taking \(Q=(q,1-q)\), \eqref{eq:self-h-second-functional} gives
\[
    q^2h''(xq)+(1-q)^2h''(x(1-q))=h''(x).
\]
Define
\[
    F(x)\defeq x^2h''(x).
\]
Multiplying the previous identity by \(x^2\), we obtain
\[
    F(xq)+F(x(1-q))=F(x).
\]
Equivalently, 
\begin{align} \label{eq:F-linear}
    F(u)+F(v)=F(u+v)
\end{align}
for all \(u,v>0\) with \(u+v<1\). A similar argument for the Cauchy functional equation \eqref{eq:F-linear} shows that 
\[
    F(x)=a x
\]
for some constant \(a\). Hence
\[
    h''(x)=\frac{a}{x}.
\]
Integrating twice gives
\[
    h(x)=a x\ln x+b x+d
\]
for some constants \(b,d\) where $x\in[0,1]$. Substituting this form into
\eqref{eq:self-h-product-additive} gives
\[
    b+nm\,d+C=2b+(n+m)d+2C
\]
for all alphabet sizes \(n,m\). Hence \(d=0\) and \(b+C=0\). Therefore
\[
    \IPhi(S;S)=a\sum_{i=1}^n p_i\ln p_i.
\]
Since \(\IPhi(S;S)\geq 0\), let \(c=-a\ln(2)\) so that \(c\geq 0\), and then 
\begin{equation}
    \IPhi(S;S)=cH(S).
    \label{eq:self-info-Shannon}
\end{equation}

Substituting \eqref{eq:self-info-Shannon} into \eqref{IPhi-S-T}, we get
\[
    \IPhi(S;T)
    =
    \IPhi(S;S)+\IPhi(T;T)-\IPhi(ST;ST)
    =
    cH(S)+cH(T)-cH(ST)
    =
    cI(S;T)
\]
for every joint distribution \(p_{ST}\). Therefore, by the strict convexity of $\Phi$, $c>0$ and then $\IPhi$ is a positive multiple of Shannon's mutual information.

\section{Proof of Theorem \ref{thm:self-information-optimality}}
\label{app:proof-thm:self-information-optimality}

By \eqref{eq:self-information-subquadratic}, $h_\Phi$ is continuous on
$[0,1]$. Hence the objective in \eqref{opt:self-information} is
continuous on the compact probability simplex, so a maximizer exists.

We next show that every maximizer lies in the interior. For $p>0$,
direct differentiation gives
\begin{align}
    h_\Phi'(p)
    =
    2p\left[
        \Phi\left(\frac{1}{p}\right)-\Phi(0)
    \right]
    -
    \Phi'\left(\frac{1}{p}\right).
    \label{eq:self-information-h-prime}
\end{align}
Setting $t=1/p$, we obtain
\begin{align*}
    h_\Phi'(p)
    =
    2G_\Phi(t)-\Phi'(t).
\end{align*}
By convexity of $\Phi$,
\begin{align*}
    \Phi'(t)
    \geq
    \frac{\Phi(t)-\Phi(0)}{t}
    =
    G_\Phi(t).
\end{align*}
Therefore,
\begin{align}
    h_\Phi'(p)
    \leq
    G_\Phi\left(\frac{1}{p}\right)
    \leq
    s_\Phi.
    \label{eq:self-information-h-prime-upper}
\end{align}
We claim that 
\begin{align}
    h_\Phi'(p)
    <
    s_\Phi,
    \qquad
    p>0.
    \label{eq:self-information-h-prime-strict}
\end{align}

If $s_\Phi=+\infty$, this is immediate since $h_\Phi'(p)$ is finite
for every fixed $p>0$. Suppose instead that $s_\Phi<\infty$ and that
equality holds in \eqref{eq:self-information-h-prime-upper} for some
$p>0$. Let $t=1/p$. Since
\begin{align*}
    h_\Phi'(p)
    \leq
    G_\Phi(t)
    \leq
    s_\Phi,
\end{align*}
equality $h_\Phi'(p)=s_\Phi$ implies
\begin{align}
    \Phi'(t)
    =
    G_\Phi(t)
    =
    s_\Phi.
    \label{eq:self-information-equality-case}
\end{align}

The equality
\begin{align*}
    \Phi'(t)
    =
    \frac{\Phi(t)-\Phi(0)}{t}
\end{align*}
implies, by convexity, that $\Phi$ is affine on $[0,t]$. Indeed, for
every $u\in[0,t]$, convexity gives
\begin{align*}
    \Phi(u)
    &\leq
    \left(1-\frac{u}{t}\right)\Phi(0)
    +
    \frac{u}{t}\Phi(t),
\end{align*}
while the supporting-line inequality at $t$ gives
\begin{align*}
    \Phi(u)
    &\geq
    \Phi(t)+\Phi'(t)(u-t).
\end{align*}
By \eqref{eq:self-information-equality-case}, the two right-hand sides
are equal, and hence $\Phi$ agrees with the affine function joining
$(0,\Phi(0))$ and $(t,\Phi(t))$ throughout $[0,t]$.

Moreover, since $G_\Phi$ is nondecreasing and
\begin{align*}
    G_\Phi(t)
    =
    s_\Phi
    =
    \lim_{u\to\infty}G_\Phi(u),
\end{align*}
we have
\begin{align*}
    G_\Phi(u)
    =
    s_\Phi,
    \qquad
    u\geq t.
\end{align*}
Thus
\begin{align*}
    \Phi(u)
    =
    \Phi(0)+s_\Phi u,
    \qquad
    u\geq t.
\end{align*}
Consequently, $\Phi$ is affine on $[0,\infty)$, contradicting the
assumption. This proves \eqref{eq:self-information-h-prime-strict}.

Suppose now, toward a contradiction, that $p_X^\ast$ is a boundary
maximizer. Then there exist $x_0,x_1\in\cX$ such that
\begin{align*}
    p_X^\ast(x_0)
    =
    a>0,
    \qquad
    p_X^\ast(x_1)
    =
    0.
\end{align*}
For $0<\epsilon<a$, define $p_X^{(\epsilon)}$ by
\begin{align*}
    p_X^{(\epsilon)}(x_0)
    &=
    a-\epsilon,\\
    p_X^{(\epsilon)}(x_1)
    &=
    \epsilon,\\
    p_X^{(\epsilon)}(x)
    &=
    p_X^\ast(x),
    \qquad
    x\notin\{x_0,x_1\}.
\end{align*}
Given $Y\sim p_Y$, denote $\IPhi(p_Y;p_Y)=\IPhi(Y;Y)$. 
Then
\begin{align}
    &
    \frac{
        I_\Phi(p_X^{(\epsilon)};p_X^{(\epsilon)})
        -
        I_\Phi(p_X^\ast;p_X^\ast)
    }{\epsilon}
    \nonumber\\
    &\qquad=
    \frac{h_\Phi(a-\epsilon)-h_\Phi(a)}{\epsilon}
    +
    \frac{h_\Phi(\epsilon)}{\epsilon}.
    \label{eq:self-information-boundary-perturbation}
\end{align}
By \eqref{eq:self-information-h-G},
\begin{align*}
    \frac{h_\Phi(\epsilon)}{\epsilon}
    =
    G_\Phi\left(\frac{1}{\epsilon}\right).
\end{align*}
Therefore, combined with \eqref{eq:self-information-h-prime-strict}, we have 
\begin{align*}
    \lim_{\epsilon\downarrow0}
    \frac{
        I_\Phi(p_X^{(\epsilon)};p_X^{(\epsilon)})
        -
        I_\Phi(p_X^\ast;p_X^\ast)
    }{\epsilon}
    =
    s_\Phi-h_\Phi'(a)
    >
    0,
\end{align*}
where the limit is $+\infty$ when $s_\Phi=+\infty$. Hence, for all
sufficiently small $\epsilon>0$,
\begin{align*}
    I_\Phi(p_X^{(\epsilon)};p_X^{(\epsilon)})
    >
    I_\Phi(p_X^\ast;p_X^\ast),
\end{align*}
contradicting the maximality of $p_X^\ast$. Thus every maximizer lies
in the interior of the probability simplex.

Since all nonnegativity constraints are inactive, the KKT condition for
\begin{align*}
    \sum_{x\in\cX}p(x)=1
\end{align*}
gives
\begin{align*}
    h_\Phi'(p_X^\ast(x))
    =
    \lambda,
    \qquad
    \forall x\in\cX,
\end{align*}
for some $\lambda\in\R$. This proves
\eqref{eq:self-information-interior-KKT}, while
\eqref{eq:self-information-interior-KKT-explicit} follows from
\eqref{eq:self-information-h-prime}.

\section{Maximum $\IPhi$ coupling}
\label{app:max-IPhi-coupling}

Let $\Pi(P_X;P_Y)$ denote the set of all couplings of $X$ and $Y$ with marginals $P_X,P_Y$, i.e., given $P_X,P_Y$, for any $Q_{XY}\in \Pi(P_X;P_Y)$, $Q_X=P_X$, $Q_Y=P_Y$. When the marginal distributions are clear from context, we also write $\Pi(X;Y)=\Pi(P_X;P_Y)$. Consider the following maximization problem about $\IPhi$ over couplings, 
\begin{align} 
    \max\quad  &\IPhi(X;Y) \label{maximize-Phi-MI} \\
    \operatorname{s.t.} \quad &p_{XY}\in \Pi(P_X;P_Y).  \nonumber
\end{align}
This problem is defined in \cite{zvarova1974measures} and studied for certain binary alphabet distributions. 
When $\IPhi$ is Shannon's mutual information ($\Phi(t)=t\log t$), the above problem reduces to the minimum entropy coupling problem (cf. \cite{kovavcevic2015entropy}), i.e., 
\begin{align*}
    \min\quad &  H(Q)\\
    \operatorname{s.t.} \quad & Q\in \Pi(P_X;P_Y). 
\end{align*}
Since $I(X;Y)=H(P_X)+H(P_Y)-H(Q_{XY})$, the above minimization is equivalent to the maximization of $I(X;Y)$. The minimum entropy coupling is shown to be NP-hard in \cite{kovavcevic2015entropy} and \cite{vidyasagar2012metric}. 

The optimization \eqref{maximize-Phi-MI} is equivalent to
\begin{align*}
    \max\quad & \DPhi (Q_{XY}\|P_{X}\otimes P_{Y})\\
    \operatorname{s.t.} \quad &Q_{XY}\in \Pi(P_X;P_Y). 
\end{align*}
Since $\DPhi$ is convex in $Q_{XY}$, one maximizer is an extreme point in $\Pi(P_X;P_Y)$. If $\Phi$ is strictly convex and $\Phi(0)<\infty$, the maximizer must be an extreme point in $\Pi(P_X;P_Y)$. 

Zv\'arov\'a gives the explicit maximum for special marginal distributions. 
\begin{proposition}[\cite{zvarova1974measures}]
    \label{prop:bern-unif-max}
    Suppose $\Phi:\R_+\to \R$. 
    Let $P_X= \Bern(b)$ and $P_Y=\Unif([m])$ where $b\in (0,1)$ and $m$ is a positive integer. Then the maximum of $\IPhi(X;Y)$ over $\Pi(P_X;P_Y)$ is equal to 
    \begin{align*}
        k\varphi \left(\frac{1}{m}\right) +\varphi \left(b-\frac{k}{m}\right) +(m-k-1) \varphi(0)-\Phi(1)
    \end{align*}
    where 
    \begin{align*}
        \varphi(t)
        = \frac{b}{m} \Phi \left(\frac{mt}{b}\right) +\frac{1-b}{m} \Phi \left(\frac{1-mt}{1-b}\right)
    \end{align*}
    and $k=\lfloor mb\rfloor$. 
\end{proposition}

In general, the coupling problem \eqref{maximize-Phi-MI} is hard to compute as maximizing Shannon's mutual information over couplings is already known to be NP-hard. Therefore, we seek to get upper bounds. Here, we develop an upper bound using the ideas in \cite{liese2006divergences}. By restricting to the supports of $p_X$ and $p_Y$, we assume all the marginal distributions in this section to be fully supported. 

For convex $\Phi$, Liese \cite{liese2006divergences} shows the following spectral representation of $\Phi$-divergence, 
\begin{align} \label{eq:hockey-stick-representation}
    D_\Phi (Q\|P)
    =
    \int_{(0,\infty)}
    D_{\Phi_\alpha}(Q\|P)
    \nu_\Phi(d\alpha),
\end{align}
where $\Phi_\alpha(t)=(t-\alpha)_+ \defeq \max \{0,t-\alpha\}$ and $\nu_{\Phi}$ is the curvature measure corresponding to $\Phi$ (cf. Fact~\ref{fact:curv-Phi}). Then 
\begin{align} \label{eq:spec-IPhi}
    \IPhi(X;Y)
    &= \DPhi(p_{XY}\| p_X \otimes p_Y)\nonumber \\
    &=\int_{(0,\infty)}
    D_{\Phi_\alpha}(p_{XY}\|p_X \otimes p_Y)
    \nu_\Phi(d\alpha).
\end{align}
Thus, we obtain the following upper bound, 
\begin{align*}
    \max_{Q_{XY}\in \Pi(X;Y)} \DPhi (Q_{XY}\|P_{X}\otimes P_{Y})
    &=\max_{Q_{XY}\in \Pi(X;Y)}
    \int_{(0,\infty)}
    D_{\Phi_\alpha} (Q_{XY}\|P_{X}\otimes P_{Y})
    \nu_\Phi(d\alpha)\\
    &\leq \int_{(0,\infty)}
    \left(\max_{Q_{XY}\in \Pi(X;Y)}
    D_{\Phi_\alpha} (Q_{XY}\|P_{X}\otimes P_{Y})\right)\nu_\Phi(d\alpha).
\end{align*}

\begin{fact}[Theorem 1 \cite{liese2006divergences}]
    \label{fact:curv-Phi}
    Suppose $\Phi:(0,\infty)\to \R$ is convex and $a,b\in (0,\infty)$. Then 
    \begin{align*}
        \Phi(b)
        =\Phi(a)+\Phi_{+}'(a) (b-a) 
        +\int_{(0,\infty)}
        \left[(b-t)_+ -(a-t)_+ -(b-a) \1_{\{t\leq a\}}\right] \nu_{\Phi}(dt),
    \end{align*}
    where 
    \begin{align*}
        \Phi_{+}'(a)
        \defeq \lim_{\epsilon\downarrow 0} \frac{\Phi(a+\epsilon)-\Phi(a)}{\epsilon}
    \end{align*}
    and $\nu_{\Phi}(dt)$ is a nonnegative measure corresponding to $\Phi$, called the curvature measure.  
\end{fact}

To compute the upper bound, we need to compute $\max_{Q_{XY}\in \Pi(X;Y)}
    D_{\Phi_\alpha} (Q_{XY}\|P_{X}\otimes P_{Y})$. 

Fix \(\pi\in\Pi(p_X;q_Y)\), and write
\[
    \mu\defeq p\otimes q,
    \qquad
    L_\pi(x,y)
    \defeq
    \frac{\pi(x,y)}{p(x)q(y)}.
\]

Since \(\pi\) and \(\mu\) are probability distributions,
\begin{align}
    \E_\mu[L_\pi]
    &=
    \sum_{x,y}p(x)q(y)
    \frac{\pi(x,y)}{p(x)q(y)}
    =
    \sum_{x,y}\pi(x,y)
    =
    1.
\label{eq:L-pi-mean-one}
\end{align}

Define 
\[
    \kappa(p,q)
    \defeq
    \max_{\pi\in\Pi(p;q)}
    \max_{(x,y)\in\mathcal X\times\mathcal Y}
    L_\pi(x,y).
\]
Thus every \(\pi\in\Pi(p;q)\) satisfies
\[
    0\le L_\pi(x,y)\le\kappa(p,q).
\]
Moreover, \eqref{eq:L-pi-mean-one} implies that \(\kappa(p,q)\ge 1\).

Observe that 
\begin{align} \label{eq:kappa-pq}
    \kappa(p,q)
    = \min\left\{
        \max_{x\in \cX} \frac{1}{p(x)}, 
        \max_{y\in \cY} \frac{1}{q(y)}
    \right\}.
\end{align}

For \(t\ge0\), define
\begin{align*}
    \mathsf B_{p,q}(t)
    &\defeq
    \max_{\pi\in\Pi(p;q)}
    \E_\mu\!\left[(L_\pi-t)_+\right]\\
    &=\max_{\pi\in\Pi(p;q)}
    \sum_{x,y}p(x)q(y)
    \bigl(L_\pi(x,y)-t\bigr)_+, 
\end{align*}
where
\[
    (a)_+\defeq\max\{a,0\}.
\]
In particular, since $a\mapsto \max\{a,0\}$ is convex, by Jensen's inequality, we have 
\begin{align} \label{eq:Bpq-hinge}
    \mathsf B_{p,q}(t)
    \ge
    (1-t)_+.
\end{align}

Let 
\begin{align*}
    \sJ_{\Phi}(p_X)\defeq \IPhi(X;X).
\end{align*}
When the distribution $p_X$ is clear from context, we also write it as $\sJ_{\Phi}(X)$. 

\begin{lemma}
    Let \(p\) and \(q\) be probability distributions on $\cX$ and $\cY$, and let $\Phi_\alpha(u)\defeq (u-\alpha)_+$ for $\alpha\geq 0$.
Then
\[
    \max_{\pi\in\Pi(p;q)}
    D_{\Phi_\alpha}\!\left(
        \pi\,\|\,p\otimes q
    \right)
    =
    \mathsf B_{p,q}(\alpha)-(1-\alpha)_+.
\]
\end{lemma}
\begin{proof}

By the definition of
\(\mathsf B_{p,q}\),
\[
\begin{aligned}
    \max_{\pi\in\Pi(p;q)}
    D_{\Phi_\alpha}\!\left(
        \pi\,\|\,p\otimes q
    \right)
    &=
    \max_{\pi\in\Pi(p;q)}
    \left\{
        \E_{p\otimes q}
        \left[
            \left(
                \frac{\pi(X,Y)}{p(X)q(Y)}-\alpha
            \right)_+
        \right]
        -(1-\alpha)_+
    \right\} \\
    &=
    \mathsf B_{p,q}(\alpha)-(1-\alpha)_+.
\end{aligned}
\]

\end{proof}

\begin{theorem}
\label{thm:hinge-divergence-bound}
Let \(p\) and \(q\) be probability distributions with full support on $\cX$ and $\cY$, respectively, and let
\(\Phi:[0,\infty)\to\R\) be convex with curvature measure
\(\nu_\Phi\). For \(\alpha\ge0\), define the hinge generator $\Phi_\alpha(u)\defeq (u-\alpha)_+$. 
Then we have 
\[
\begin{aligned}
    \max_{\pi\in\Pi(p;q)}
    D_\Phi\!\left(
        \pi\,\|\,p\otimes q
    \right)
    &\le
    \int_{(0,\kappa(p,q)]}
    \max_{\pi\in\Pi(p;q)}
    D_{\Phi_\alpha}\!\left(
        \pi\,\|\,p\otimes q
    \right)
    \nu_\Phi(d\alpha)\\
    &=
    \int_{(0,\kappa(p,q)]}
    \left[
        \mathsf B_{p,q}(\alpha)-(1-\alpha)_+
    \right]
    \nu_\Phi(d\alpha).
\end{aligned}
\]
Moreover, 
\begin{align*}
    \int_{(0,\kappa(p,q)]}
    \left[
        \mathsf B_{p,q}(\alpha)-(1-\alpha)_+
    \right]
    \nu_\Phi(d\alpha)
    \leq \min \left\{\sJ_\Phi(p), \sJ_\Phi(q)\right\}. 
\end{align*}
\end{theorem}
\begin{proof}

By definition, \(\kappa(p,q)\ge 1\).
Consequently, for \(\alpha>\kappa(p,q)\), both
\((L_\pi-\alpha)_+\) and \((1-\alpha)_+\) vanish. By
\eqref{eq:hockey-stick-representation},
\begin{align*}
    \DPhi\!\left(
        \pi\,\|\,p\otimes q
    \right)
    &=
    \int_{(0,\kappa(p,q)]}
    \left[
        \E_\mu[(L_\pi-\alpha)_+]
        -(1-\alpha)_+
    \right]
    \nu_\Phi(d\alpha).
\end{align*}
By the definition of \(\mathsf B_{p,q}(\alpha)\),
\[
    \E_\mu[(L_\pi-\alpha)_+]
    \le
    \mathsf B_{p,q}(\alpha).
\]
Since \(\nu_\Phi\) is a nonnegative measure, it follows that
\[
    \DPhi\!\left(
        \pi\,\|\,p\otimes q
    \right)
    \le
    \int_{(0,\kappa(p,q)]}
    \left[
        \mathsf B_{p,q}(\alpha)-(1-\alpha)_+
    \right]
    \nu_\Phi(d\alpha).
\]
Taking the maximum over \(\pi\in\Pi(p;q)\) proves the first inequality. 

For the second inequality, it suffices to prove the upper bound by
$\sJ_{\Phi}(p)$, since the proof for $\sJ_{\Phi}(q)$ is analogous. For
$\alpha\ge0$, let
\begin{align*}
    \sS_p(\alpha)
    \defeq
    \sum_x p(x) (1-\alpha p(x))_+.
\end{align*}
We claim that 
\begin{align} \label{eq:Bpq-Sp}
    \sB_{p,q}(\alpha)
    \leq \sS_p(\alpha).
\end{align}
Indeed, since $0\leq \pi(x,y)/q(y)\leq 1$, the convexity of
$u\mapsto(u-\alpha)_+$ gives
\begin{align*}
    \left(\frac{\pi(x,y)}{p(x)q(y)}-\alpha\right)_+
    &\leq \frac{\pi(x,y)}{q(y)}
    \left(\frac{1}{p(x)}-\alpha\right)_+
    + \left(1-\frac{\pi(x,y)}{q(y)} \right) (0-\alpha)_+\\
    &=\frac{\pi(x,y)}{q(y)}
    \left(\frac{1}{p(x)}-\alpha\right)_+.
\end{align*}
Multiplying by $p(x)q(y)$ and summing over $x,y$, we obtain
\begin{align*}
    \sum_{x,y}p(x)q(y)\left(L_\pi(x,y)-\alpha\right)_+
    &\leq
    \sum_{x,y}p(x)\pi(x,y)
    \left(\frac{1}{p(x)}-\alpha\right)_+\\
    &=\sum_x p(x)^2
    \left(\frac{1}{p(x)}-\alpha\right)_+\\
    &=\sum_x p(x)(1-\alpha p(x))_+
    =\sS_p(\alpha).
\end{align*}
Taking the maximum over $\pi\in \Pi(p;q)$ proves
\eqref{eq:Bpq-Sp}. Let
$K_p\defeq\max\{1/p(x):x\in\cX\}$. Applying
\eqref{eq:hockey-stick-representation} to the diagonal coupling of a
random variable with law $p$ gives
\begin{align*}
    \sJ_{\Phi}(p)
    &= \int_{(0,K_p]}
    \left[\sS_p(\alpha)-(1-\alpha)_+\right]
    \nu_{\Phi}(d\alpha).
\end{align*}
By \eqref{eq:kappa-pq}, $\kappa(p,q)\leq K_p$. Moreover,
\eqref{eq:Bpq-hinge} and \eqref{eq:Bpq-Sp} imply
$\sS_p(\alpha)-(1-\alpha)_+\geq0$. Hence
\begin{align*}
    &\int_{(0,\kappa(p,q)]}
    \left[\sB_{p,q}(\alpha)-(1-\alpha)_+\right]
    \nu_\Phi(d\alpha)\\
    &\quad\leq
    \int_{(0,\kappa(p,q)]}
    \left[\sS_p(\alpha)-(1-\alpha)_+\right]
    \nu_\Phi(d\alpha)\\
    &\quad\leq
    \int_{(0,K_p]}
    \left[\sS_p(\alpha)-(1-\alpha)_+\right]
    \nu_\Phi(d\alpha)\\
    &=\sJ_{\Phi}(p). 
\end{align*}

\end{proof}

\section{Proof of Theorem \ref{thm:conditional-Phi-variational-form}}
\label{app:proof-thm:conditional-Phi-variational-form}

For convenience, write $\cJ(f)
    \defeq
    \mathcal J_{\Phi,Q}(f;X,Z)$. 
For a positive function \(g=g(X,Y,Z)\), define $g_{XZ}
    \defeq
    \E_Q[g|XZ]$. 

First we prove \eqref{eq:conditional-Phi-single-critic-g} holds for all positive $f$. 

By Lemma \ref{lem:conditional-Phi-entropy-convex}, the mapping $f\mapsto\cJ(f)$ is convex. Its directional derivative at a positive function \(g\) in
the direction \(h\) is
\begin{align*}
    \frac{d}{dt}\mathcal J(g+th)\Big|_{t=0}
    &=
    \E_Q[\Phi'(g)h]
    -
    \E_Q\left[
        \Phi'(g_{XZ})\E_Q[h|XZ]
    \right]\\
    &=
    \E_Q\left[
        \left(
            \Phi'(g)-\Phi'(g_{XZ})
        \right)h
    \right].
\end{align*}
Let $U_g\defeq \Phi'(g)-\Phi'(g_{XZ})$. 
By the convexity of the mapping $f\mapsto\cJ(f)$, we have 
\begin{align}
    \cJ(f)
    \geq
    \cJ(g)
    +
    \E_Q[U_g(f-g)].
    \label{eq:conditional-Phi-supporting-hyperplane}
\end{align}
By the definition of \(U_g\), 
\begin{align*}
    \cJ(g)-\E_Q[U_g g]
    &=
    \E_Q[\Phi(g)]
    -
    \E_Q[\Phi(g_{XZ})]
    -
    \E_Q[g\Phi'(g)]
    +
    \E_Q[g\Phi'(g_{XZ})]\\
    &=\E_Q[\Phi(g)]
    -
    \E_Q[\Phi(g_{XZ})]
    -
    \E_Q[g\Phi'(g)]
    +\E_Q[g_{XZ}\Phi'(g_{XZ})]\\
    &=-\E_Q\left[
        g\Phi'(g)-\Phi(g)
    \right]
    +
    \E_Q\left[
        g_{XZ}\Phi'(g_{XZ})-\Phi(g_{XZ})
    \right]
\end{align*}
where the second line follows because \(\Phi'(g_{XZ})\) is a function of \((X,Z)\).

By Lemma \ref{lem:Fenchel-equality}, $\Phi^*(\Phi'(u))=u\Phi'(u)-\Phi(u)$. Then we obtain
\begin{align*}
    \cJ(g)-\E_Q[U_g g]
    &=
    -\E_Q\left[
        \Phi^*(\Phi'(g))
        -
        \Phi^*(\Phi'(g_{XZ}))
    \right].
\end{align*}
By \eqref{eq:conditional-Phi-supporting-hyperplane}, we have 
\begin{align*}
    \cJ(f)
    \geq
    &\E_Q\left[
        f\left(
            \Phi'(g)-\Phi'(g_{XZ})
        \right)
    \right]
    -
    \E_Q\left[
        \Phi^*(\Phi'(g))
        -
        \Phi^*(\Phi'(g_{XZ}))
    \right].
\end{align*}
Taking the supremum over \(g>0\) gives
\begin{align*}
    \cJ(f)
    &\geq
    \sup_{g>0}
    \left\{
        \E_Q\left[
            f\left(
                \Phi'(g)-\Phi'(g_{XZ})
            \right)
        \right]
        -
        \E_Q\left[
            \Phi^*(\Phi'(g))
            -
            \Phi^*(\Phi'(g_{XZ}))
        \right]
    \right\}.
\end{align*}
On the other hand, choosing \(g=f\) gives equality in
\eqref{eq:conditional-Phi-supporting-hyperplane}. Therefore,
\begin{align*}
    \cJ(f)
    &=
    \sup_{g>0}
    \left\{
        \E_Q\left[
            f\left(
                \Phi'(g)-\Phi'(g_{XZ})
            \right)
        \right]
        -
        \E_Q\left[
            \Phi^*(\Phi'(g))
            -
            \Phi^*(\Phi'(g_{XZ}))
        \right]
    \right\}.
\end{align*}

It remains to consider nonnegative $f$. By
Lemma~\ref{lem:class-F-properties}, $t\mapsto t\Phi''(t)$ is nondecreasing and hence 
\[
    0<\Phi''(t)\leq \frac{\Phi''(1)}{t},
    \qquad 0<t\leq 1.
\]
After integration, we have $\Phi'(t)=O(1+|\log t|)$ as $t\downarrow0$.
Consequently, $\Phi$ has a finite continuous extension to zero,
and $t\Phi'(t)\to0$. Since the alphabet is finite,
$\mathcal J(f+\epsilon)\to\cJ(f)$. 
For each fixed $g>0$, apply
\eqref{eq:conditional-Phi-supporting-hyperplane}
to $f+\delta$ and let $\delta\downarrow0$. This yields
\[
    \cJ(g)+\E_Q[U_g(f-g)]
    \leq \cJ(f).
\]
Then 
\begin{align*} 
    \sup_{g>0}\left\{\cJ(g)+\E_Q[U_g(f-g)]\right\}
    \leq \cJ(f).
\end{align*}
For the reverse inequality, set
$g^{(\epsilon)}=f+\epsilon$. Denote  
$f_{XZ}\defeq\E_Q[f|XZ]$ and $g_{XZ}^{(\epsilon)}\defeq\E_Q[g^{(\epsilon)}|XZ]$. Then $g^{(\epsilon)}>0$ and 
$g_{XZ}^{(\epsilon)}=f_{XZ}+\epsilon$. Moreover, 
\begin{align*}
    \cJ(g^{(\epsilon)})
      +\E_Q[U_{g^{(\epsilon)}}(f-g^{(\epsilon)})]
    &=
      \mathcal J(f+\epsilon)
      -\epsilon\E_Q\left[
          \Phi'(f+\epsilon)
          -\Phi'(f_{XZ}+\epsilon)
      \right]\\
    &\to \cJ(f),
\end{align*}
as $\epsilon\to 0$ where the convergence follows from the fact that $\epsilon\Phi'(a+\epsilon)\to0$
for every $a\geq0$. It finishes the proof of 
\eqref{eq:conditional-Phi-single-critic-g}
for all nonnegative $f$.

\section{Proof of Theorem \ref{thm:wIPhi-optimizer}} \label{app:proof-optimizer}

Consider
\begin{align*}
    \wIPhi(X;Y|Z)
    &= \sup_{U} \left\{\IPhi(U;XZ)+\IPhi(U;YZ)-\IPhi(U;XYZ)-\IPhi(U;Z)\right\}.
\end{align*}
Note that
\begin{align*}
    \IPhi(U;XZ)+\IPhi(U;YZ)-\IPhi(U;XYZ)-\IPhi(U;Z)
    &=\sum_{u} p(u) g(p_{XYZ|u})
\end{align*}
where 
\begin{align*}
    g(q_{XYZ})=\DPhi(q_{XZ}\|p_{XZ})
    +\DPhi(q_{YZ}\|p_{YZ}) 
    -\DPhi(q_{XYZ}\|p_{XYZ}) 
    -\DPhi(q_{Z}\|p_{Z}).
\end{align*}
Then $\wIPhi(X;Y|Z)$ is the upper concave envelope of $g(q_{XYZ})$ at $p_{XYZ}$. By Carath\'eodory's theorem, it suffices to consider the random variable $U$ with cardinality at most $|\cX||\cY||\cZ|$. 

We have the following two claims:
\begin{enumerate}[\rm (i)]
    \item Fixing $q_Z$ and $q_{Y|XZ}$, $g(q_{XYZ})$ is convex in $q_{X|Z}$. \label{g-convex-1}
    \item Fixing $q_Z$ and $q_{X|YZ}$, $g(q_{XYZ})$ is convex in $q_{Y|Z}$. \label{g-convex-2}
\end{enumerate}
Indeed, fixing $q_Z$ and $q_{Y|XZ}$, by Lemma~\ref{lem:G-concave-X|Z}, since $\Phi\in \mathscr{G}$, 
\begin{align*}
    q_{X|Z}\mapsto 
    \DPhi(q_{XZ}\|p_{XZ})
    -\DPhi(q_{XYZ}\|p_{XYZ})
\end{align*}
is convex. Since $\Phi$ is convex, $q_{X|Z}\mapsto \DPhi(q_{YZ}\|p_{YZ})$ is convex. Note that $\DPhi(q_{Z}\|p_{Z})$ is constant. Thus, their sum, $q_{X|Z}\mapsto g(q_{XYZ})$ is convex. The proof of the second claim follows similarly. 

To compute 
the upper concave envelope of 
$g(q_{XYZ})$, we need to maximize
\begin{align}
    \sum_{i}\omega_i g(q^{(i)}_{XYZ})\label{objf}
\end{align}
subject to
$$\sum_{i}\omega_i q^{(i)}_{XYZ}=p_{XYZ}.$$
Let $U$ be the maximizer of \eqref{objf} with the joint distribution 
\begin{align*}
    p_{XYZU}(x,y,z,i) =\omega_i q^{(i)}_{XYZ}(x,y,z). 
\end{align*}
For each $i$, we can write  
\begin{align*}
    q_{XYZ}^{(i)}
    &=q_{Z}^{(i)} q_{X|Z}^{(i)} q_{Y|XZ}^{(i)}\\
    &=q_{Z}^{(i)} \left(\sum_j \nu_{ij} q_{X|Z}^{(i,j)}\right) q_{Y|XZ}^{(i)}
\end{align*}
where $X$ is a deterministic function of $Z$ under $q_{X|Z}^{(i,j)}$. 
By Jensen's inequality, Item \ref{g-convex-1} implies that 
\begin{align*}
    g(q^{(i)}_{XYZ})
    \leq \sum_{j} \nu_{ij} g \left(q_{Z}^{(i)}q_{X|Z}^{(i,j)}q_{Y|XZ}^{(i)}\right).
\end{align*}
Then 
\begin{align*}
    \sum_{i}\omega_i g(q^{(i)}_{XYZ})
    &\leq 
    \sum_{i}\omega_i
    \left(\sum_{j} \nu_{ij} g\left(q_{Z}^{(i)}q_{X|Z}^{(i,j)}q_{Y|XZ}^{(i)}\right)\right)\\
    &=\sum_{i,j} \omega_i\nu_{ij} g\left(q_{Z}^{(i)}q_{X|Z}^{(i,j)}q_{Y|XZ}^{(i)}\right). 
\end{align*}
Note that 
\begin{align*}
    \sum_{i,j} \omega_i\nu_{ij} q_{Z}^{(i)}q_{X|Z}^{(i,j)}q_{Y|XZ}^{(i)}
    &=\sum_{i}\omega_i q_{Z}^{(i)}q_{X|Z}^{(i)}q_{Y|XZ}^{(i)}=p_{XYZ}.
\end{align*}
Take $U'$ with joint distribution 
\begin{align*}
    p_{XYZU'}(x,y,z,(i,j)) =\omega_i\nu_{ij}q_Z^{(i)}(z)q_{X|Z}^{(i,j)}(x|z)q_{Y|XZ}^{(i)}(y|x,z).
\end{align*}
Then $U'$ satisfies the optimization constraint and has objective function value at least as large as $U$. Thus $U'$ is also a maximizer. Note that $p_{X|ZU'}(x|z,(i,j))=q_{X|Z}^{(i,j)}(x|z)$ where $X$ is a deterministic function of $Z$. Therefore, the maximum in \eqref{objf} is always achieved by a joint distribution under which $X$ is a function of $(Z,U')$. 

Starting from this optimizer $U'$, apply the same procedure using Item~\ref{g-convex-2} separately to each $p_{XYZ|U'=u'}$. Thus the maximum is attained by an auxiliary variable for which
$(X,Y)$ is a deterministic function of $(Z,U'')$.
It finishes the proof.

\section{Proof of Proposition \ref{prop:max-Phi-MI}} \label{app:proof-prop:max-Phi-MI}

\begin{enumerate}[(i)]
    \item \textbf{Conditional symmetry and dominance over \(I_\Phi\).}
    Conditional symmetry follows from the definition
    \[
        \wIPhi(X;Y|Z)=\max_U \left[
        \IPhi(U;X|Z)+\IPhi(U;Y|Z)-\IPhi(U;XY|Z)\right],
    \]
    since the expression is unchanged when $X$ and $Y$ are interchanged.
    Taking the special choice $U=X$, we have 
    \begin{align*}
        \IPhi(X;X|Z) +\IPhi(X;Y|Z) -\IPhi(X;XY|Z)
        &=\IPhi(X;Y|Z),
    \end{align*}
    where the equality follows from \eqref{Phi-MI-redundant-RV}. Thus $\wIPhi(X;Y|Z)\geq \IPhi(X;Y|Z)$. The nonnegativity of $\IPhi(X;Y|Z)$ gives the second inequality.

    \item \textbf{Conditional deterministic self-reduction.}
    Since $Y$ is a function of $(X,Z)$, we have 
    \begin{align*}
        \wIPhi(X;Y|Z)
        &=\max_U \left[\IPhi(U;Y|Z) -\IPhi(U;Y|XZ)\right]\\
        &=\max_U \left[\IPhi(U;Y|Z)\right]
    \end{align*}
    where the last equality follows from \eqref{Phi-MI-Markov}, since
    $U\to XZ\to Y$ is a Markov chain. Moreover, note that 
    \begin{align*}
        \wIPhi(Y;Y|Z)
        &=\max_U \left[\IPhi(U;Y|Z) -\IPhi(U;Y|YZ)\right]\\
        &=\max_U \left[\IPhi(U;Y|Z)\right].
    \end{align*}
    The last equality again follows from \eqref{Phi-MI-Markov}.
    Thus $\wIPhi(X;Y|Z)=\wIPhi(Y;Y|Z)$.

    \item \textbf{Self-information bounds.}
    By conditional symmetry and the equivalent form in \eqref{opt:cond-IPhi}, 
    \begin{align*}
        \wIPhi(X;Y|Z)
        &=\max_U \left[\IPhi(U;X|Z)-\IPhi(U;X|YZ)\right]\\
        &\leq \max_U \left[\IPhi(U;X|Z)\right]\\
        &= \max_U \left[2\IPhi(U;X|Z)-\IPhi(U;XX|Z)\right]\\
        &=\wIPhi(X;X|Z),
    \end{align*}
    where the inequality follows from the nonnegativity of conditional $\Phi$-mutual information and the third line follows from \eqref{Phi-MI-redundant-RV-2}. 
    Next, note that 
    \begin{align*}
        \wIPhi(X;XY)&=\max_{U}\left[\IPhi(U;XY)+\IPhi(U;X)-\IPhi(U;XXY)\right] \\
        &=\max_{U}\left[\IPhi(U;X)\right],
    \end{align*}
    where the second equality follows from repetition invariance applied to the repeated copy of $X$ in the second argument, 
    and 
    \begin{align*}
        \wIPhi(X;X)&=\max_U\left[\IPhi(U;X)+\IPhi(U;X)-\IPhi(U;XX)\right]\\
        &=\max_U\left[\IPhi(U;X)\right]
    \end{align*}
    where the second equality follows from \eqref{Phi-MI-redundant-RV-2}. Then $\wIPhi(X;XY)=\wIPhi(X;X)$. By \eqref{Phi-MI-UB-self}, for any $U$, $\IPhi(U;X)\leq \IPhi(X;X)$. Taking the maximum on the left hand side, we have 
    \begin{align*}
        \wIPhi(X;X)
        &=\max_U \left[\IPhi(U;X)\right] \\
        &\leq \IPhi(X;X). 
    \end{align*}
    Moreover, by \eqref{max-Phi-LB}, $\wIPhi(X;X)\geq \IPhi(X;X)$. Thus $\wIPhi(X;X)=\IPhi(X;X)$, and $\wIPhi(X;XY)=\wIPhi(X;X)=\IPhi(X;X)$.

    Recall the equivalent definition of $\wIPhi$: 
    \begin{align*} 
        \wIPhi(X;Y|Z) 
        &= \sup_{U} \left\{\IPhi(U;XZ)+\IPhi(U;YZ)-\IPhi(U;XYZ)-\IPhi(U;Z)\right\}.
    \end{align*}
    Then 
    \begin{align*}
        \wIPhi(X;Y|Z) 
        &= \sup_{U} \left\{\IPhi(U;XZ)+\IPhi(U;YZ)-\IPhi(U;XYZ)-\IPhi(U;Z)\right\}\\
        &\leq \sup_{U}\IPhi(U;XZ)\\
        &= \IPhi(XZ;XZ)
    \end{align*}
    where the last line follows from the self information bound \eqref{Phi-MI-UB-self}. Similarly, 
    \begin{align*}
        \wIPhi(X;Y|Z) 
        &\leq \IPhi(YZ;YZ). 
    \end{align*}
    Thus, 
    \begin{align*}
        \wIPhi(X;Y|Z) 
        &\leq\min \{\IPhi(XZ;XZ), \IPhi(YZ;YZ)\}. 
    \end{align*}

    \item \textbf{Monotonicity under adjoining variables.} 
    We first show 
    \begin{align*}
        \wIPhi(X;Y|W)\leq \wIPhi(X;YZ|W). 
    \end{align*}
    Suppose $p_{U|XYW}$ is the maximizer of $\wIPhi(X;Y|W)$. Given $p_{XYZW}$, let $p_{UXYZW}=p_{XYW}p_{U|XYW}p_{Z|XYW}$. Then 
    \begin{align*}
        \wIPhi(X;Y|W)
        &=\IPhi(U;X|W)+\IPhi(U;Y|W)-\IPhi(U;XY|W)\\
        &=\IPhi(U;X|W)+\IPhi(U;Y|W)-\IPhi(U;XYZ|W)\\
        &\leq \IPhi(U;X|W)+\IPhi(U;YZ|W)-\IPhi(U;XYZ|W)\\
        &\leq \wIPhi(X;YZ|W),
    \end{align*}
    where the second line follows from the chain rule on the second coordinate
$$\IPhi(U;XYZ|W)=\IPhi(U;XY|W)+\IPhi(U;Z|XYW)$$
    and the fact that
    the Markov chain $U\to XYW\to Z$ implies $\IPhi(U;Z|XYW)=0$ (Markov faithfulness property),    
    and the third line follows from \eqref{Phi-MI-more-RV}.  
    Applying this inequality once, then applying it again after using
    conditional symmetry, gives
    \begin{align*}
        \wIPhi(X;Y|W)
        \leq \wIPhi(X;YZ|W)
        =\wIPhi(YZ;X|W)
        \leq\wIPhi(YZ;XT|W)
        =\wIPhi(XT;YZ|W),
    \end{align*}
    which proves \eqref{max-Phi-MI-more-RV}.

    \item \textbf{Subadditive chain rule.}
    By definition, 
    \begin{align*}
        \wIPhi(X;YZ|W)
        &=\max_U \left[\IPhi(U;X|W)+\IPhi(U;YZ|W)-\IPhi(U;XYZ|W)\right]\\
        &=\max_U \left[\IPhi(U;X|W)+\IPhi(U;Y|W)+\IPhi(U;Z|YW)-\IPhi(U;XY|W)-\IPhi(U;Z|XYW)\right]\\
        &\leq \max_U \left[\IPhi(U;X|W)+\IPhi(U;Y|W)-\IPhi(U;XY|W)\right]\\
        &\quad +\max_U \left[\IPhi(U;Z|YW)-\IPhi(U;Z|XYW)\right]\\
        &=\wIPhi(X;Y|W)+\wIPhi(X;Z|YW). 
    \end{align*}
Due to the conditional symmetry of $\wIPhi$, the second inequality follows similarly.
    \item \textbf{$\wIPhi=0$ implies conditional independence.}
    By \eqref{max-Phi-LB}, $\IPhi(X;Y|Z)\leq \wIPhi(X;Y|Z)=0$. By \eqref{Phi-MI-Markov2}, since $\Phi$ is strictly convex, $X\to Z\to Y$ is a Markov chain. Setting $Z$ to be constant gives the independence statement.

    \item \textbf{Markov faithfulness for $\Phi \in \mathscr F$.}
    Since $\Phi\in\mathscr F$ is strictly convex, the implication
    $\wIPhi(X;Y|Z)=0\Rightarrow X\to Z\to Y$ follows from
    \eqref{max-Phi-Markov}; taking $Z$ to be constant gives
    $\wIPhi(X;Y)=0\Rightarrow X\independent Y$.
    Conversely, suppose $X\to Z\to Y$ is a Markov chain. By the chain rule,
    \begin{align*}
        \wIPhi(X;Y|Z) 
        &= \max_{U} \left[\IPhi(U;X|Z)+\IPhi(U;Y|Z)-\IPhi(U;XY|Z)\right] \\
        &= \max_{U} \left[\IPhi(U;Y|Z)-\IPhi(U;Y|XZ)\right].
    \end{align*}
    By the conditional $\Phi$-entropy subadditivity
    \cite[Lemma 7(b)]{beigi2018phi}, the Markov condition implies that
    $\IPhi(U;Y|Z)-\IPhi(U;Y|XZ)\leq 0$ for every $U$. Hence
    $\wIPhi(X;Y|Z)\leq 0$, and \eqref{max-Phi-LB} gives
    $\wIPhi(X;Y|Z)=0$. Taking $Z$ to be constant gives the independence
    equivalence.

    To prove the last statement, it suffices to prove that if $X\to Y\to Z$, then $\wIPhi(X;YZ)=\wIPhi(X;Y)$. By \eqref{max-Phi-MI-more-RV}, $\wIPhi(X;YZ)\geq \wIPhi(X;Y)$. For the other direction, let $V$ be the maximizer of $\wIPhi(X;YZ)$. Then
    \begin{align*}
        \wIPhi(X;YZ)
        &=\IPhi(V;X)-\IPhi(V;X|YZ)\\
        &\overset{\rm (a)}{\leq} \IPhi(V;X)-\IPhi(V;X|Y)\\
        &\overset{\rm (b)}{\leq} \wIPhi(X;Y),
    \end{align*}
    where (a) follows from the conditional $\Phi$-entropy subadditivity
    \cite[Lemma 7(b)]{beigi2018phi} and the Markov chain
    $X\to Y\to Z$, and (b) follows from the definition of $\wIPhi$.
    Thus $\wIPhi(X;YZ)=\wIPhi(X;Y)$. 
    \item \textbf{Subadditivity for $\Phi \in \mathscr F$.}
    We have 
    \begin{align*}
        \wIPhi(U;Y|Z)
        &\leq \wIPhi(UX;Y|Z)\\
        &\leq \wIPhi(X;Y|Z)+\wIPhi(U;Y|XZ)\\
        &=\wIPhi(U;Y|XZ)
    \end{align*}
    where the first line follows from \eqref{max-Phi-MI-more-RV}, the second line follows from \eqref{max-Phi-MI-subadditive}, and the third line follows from \eqref{max-Phi-Markov-faithful}. Taking $Z$ to be constant, we obtain \eqref{max-Phi-MI-F-class-independent}. 

    \item \textbf{Information inequality.}
    For an arbitrary convex $\Phi$, take the auxiliary random variable in
    the right-hand side of \eqref{wPhi-self-repetition} to be $U=X$. Then
    \begin{align*}
        \wIPhi(U;X)+\wIPhi(U;Y)-\wIPhi(U;XY)
        &=\wIPhi(X;X)+\wIPhi(X;Y)-\wIPhi(X;XY)\\
        &=\wIPhi(X;Y),
    \end{align*}
    where the second equality follows from \eqref{max-Phi-repetition}.
    Taking the supremum over $U$ proves \eqref{wPhi-self-repetition}.

    Next, suppose $\Phi\in\mathscr G$ and $\Phi(0)<\infty$. By
    Corollary~\ref{cor:G-closed-form}, for every joint distribution
    $P_{UXY}$,
    \begin{align*}
        &\wIPhi(U;X)+\wIPhi(U;Y)-\wIPhi(U;XY)\\
        &\quad
        =\Big(\IPhi(U;U)+\IPhi(X;X)-\IPhi(UX;UX)\Big)\\
        &\qquad
        +\Big(\IPhi(U;U)+\IPhi(Y;Y)-\IPhi(UY;UY)\Big)\\
        &\qquad
        -\Big(\IPhi(U;U)+\IPhi(XY;XY)-\IPhi(UXY;UXY)\Big)\\
        &\quad
        =\wIPhi(X;Y)+\IPhi(U;U)-\wIPhi(UX;UY)\\
        &\quad
        \overset{\rm (a)}{\leq}
        \wIPhi(X;Y)+\IPhi(U;U)-\IPhi(UX;UY)\\
        &\quad
        \overset{\rm (b)}{\leq}
        \wIPhi(X;Y).
    \end{align*}
    Here the second equality uses the permutation invariance of $\IPhi$,
    (a) follows from \eqref{max-Phi-LB}, and (b) follows from
    \eqref{Phi-MI-more-RV}. Taking the
    supremum over $U$ and combining it with 
    \eqref{wPhi-self-repetition} proves
    \eqref{wPhi-self-repetition-G}. 

    Finally, note that \eqref{wIPhi-info-ineq} follows directly from \eqref{wPhi-self-repetition-G}. 
\end{enumerate}

\section{Proof of Proposition \ref{prop:properties-hat-IPhi}} \label{app:proof-prop:properties-hat-IPhi}

\begin{enumerate}[\rm (i)]
    \item Taking the identity local maps gives
    $\widehat I_\Phi(X;Y)\geq \wIPhi(X;Y)$. If $\Phi\in\mathscr F$, then
    \eqref{max-Phi-Markov-invariance} gives
    \[
        \wIPhi(W_1X;W_2Y)=\wIPhi(X;Y)
    \]
    for every $W_1\to X\to Y\to W_2$. Taking the supremum over
    $W_1,W_2$ gives $\widehat I_\Phi(X;Y)=\wIPhi(X;Y)$.

    \item Fix a Markov chain $W_1\to X\to Y\to W_2$. Consider any
    pair $(V_1,V_2)$ admissible in the definition of
    $\widehat I_\Phi(W_1X;W_2Y)$. Its joint distribution factors as
    \[
        p(x,y)p(w_1|x)p(w_2|y)
        p(v_1|w_1,x)p(v_2|w_2,y).
    \]
    Hence $(V_1,W_1)\to X\to Y\to(V_2,W_2)$ is a Markov chain, so
    \[
        \wIPhi(V_1W_1X;V_2W_2Y)
        \leq \widehat I_\Phi(X;Y).
    \]
    Taking the supremum over $(V_1,V_2)$ gives
    \[
        \widehat I_\Phi(W_1X;W_2Y)
        \leq \widehat I_\Phi(X;Y).
    \]

    Conversely, let $A\to X\to Y\to B$ be any Markov chain. Adjoin
    $(A,B)$ conditionally independently of $(W_1,W_2)$ given $(X,Y)$.
    Then $A\to W_1X\to W_2Y\to B$ is a Markov chain, and
    \eqref{max-Phi-MI-more-RV} gives
    \[
        \wIPhi(AW_1X;BW_2Y)
        \geq \wIPhi(AX;BY).
    \]
    Taking the supremum over $(A,B)$ gives the reverse inequality.
    Therefore,
    \[
        \widehat I_\Phi(W_1X;W_2Y)
        =\widehat I_\Phi(X;Y).
    \]

    \item Let $\delta$ be invariant under private randomness and suppose
    $\delta(A;B)\geq\wIPhi(A;B)$ for all $A,B$. For every Markov chain
    $W_1\to X\to Y\to W_2$, we have
    \[
        \delta(X;Y)
        =\delta(W_1X;W_2Y)
        \geq\wIPhi(W_1X;W_2Y).
    \]
    Taking the supremum over $(W_1,W_2)$ gives
    $\delta(X;Y)\geq\widehat I_\Phi(X;Y)$.
\end{enumerate}

\section{Proof of Theorem \ref{thm:hat-IPhi}} \label{app:proof-thm:hat-IPhi}

Since $\Phi\in \mathscr{G}$, by Lemma \ref{lem:class-G-properties}, $t\Phi''(t)$ is nonincreasing, and
$c_{\Phi}\defeq \lim_{t\to\infty} t\Phi''(t)$ exists and is finite. Moreover, $\Psi(t)$ is convex and either $\Psi$ is affine or $\Psi\in \mathscr G$. And the limit
$\ell_\Phi\defeq \lim_{t\to \infty} \frac{\Psi(t)}{t}$ exists but can be
infinite. If $\Psi(t)$ is affine, we have 
\begin{align*}
    \IPhi(X;Y)&=c_{\Phi}(\ln 2) I(X;Y)+I_{\Psi}(X;Y)\\
    &=c_{\Phi}(\ln 2) I(X;Y).
\end{align*}
Then 
\begin{align*}
    \widehat I_{\Phi}(X;Y)=c_{\Phi}(\ln 2) I(X;Y).
\end{align*}
Note that $\ell_{\Phi}=\Psi(1)-\Psi(0)<\infty$ and then $K_{\Phi}=0$. Thus, in the following, we assume $\Psi\in \mathscr{G}$. 

We now discuss the dichotomy. 

(i) By the assumption, $\ell_\Phi<\infty$. Since the map
\[
t\mapsto \frac{\Psi(t)-\Psi(0)}{t}
\]
is nondecreasing, we have
\begin{align}
    \Psi(t)-\Psi(0)\leq \ell_\Phi t,
\qquad t>0.\label{PsiDiff}
\end{align}
Taking \(t=1\) gives
\[
K_\Phi\defeq\ell_\Phi+\Psi(0)-\Psi(1)\geq 0.
\]

We next separate the Shannon mutual information component. By the definition of $I_\Phi$, we have 
\[
I_\Phi(A;B)=c_\Phi(\ln 2) I(A;B)+I_\Psi(A;B).
\]
Applying Corollary~\ref{cor:G-closed-form} to $\Phi$ and $\Psi$ therefore gives
\[
\wIPhi(A;B)
=
c_\Phi(\ln 2) I(A;B)+\widetilde I_\Psi(A;B).
\]
Moreover, the Markov chain $W_1\to X\to Y\to W_2$ implies
\[
I(W_1X;W_2Y)=I(X;Y).
\]
Therefore,
\[
\widehat I_\Phi(X;Y)
=
c_\Phi(\ln 2) I(X;Y)+\widehat I_\Psi(X;Y).
\]
It remains to show that $\widehat I_\Psi(X;Y)=K_\Phi$.

For any finite random variable \(A\),
\begin{align}
I_\Psi(A;A)\nonumber
&=
\sum_a p(a)^2\Psi\left(\frac{1}{p(a)}\right)
+
\left(1-\sum_a p(a)^2\right)\Psi(0)
-\Psi(1)\\\nonumber
&=
\sum_a p(a)^2
\left(
\Psi\left(\frac{1}{p(a)}\right)-\Psi(0)
\right)
+\Psi(0)-\Psi(1)\\\label{eqnineqd}
&\leq
\ell_\Phi\sum_a p(a)^2\frac{1}{p(a)}
+\Psi(0)-\Psi(1)\\\nonumber
&=K_\Phi,
\end{align}
where the inequality in \eqref{eqnineqd} follows from \eqref{PsiDiff}.

By \eqref{max-Phi-MI-UB-self} and \eqref{max-Phi-repetition},
\[
\widetilde I_\Psi(A;B)
\leq
\widetilde I_\Psi(A;A)
=
I_\Psi(A;A)
\leq K_\Phi.
\]
Applying this to \(A=W_1X\) and \(B=W_2Y\), and taking the supremum,
gives
\[
\widehat I_\Psi(X;Y)\leq K_\Phi.
\]

For the reverse inequality, let \(R_n\) and \(T_n\) be independent
and uniform on \([n]\), and independent of \((X,Y)\). Then
\[
R_n\to X\to Y\to T_n,
\]
and \eqref{max-Phi-MI-more-RV} gives
\[
\widehat I_\Psi(X;Y)
\geq
\widetilde I_\Psi(R_nX;T_nY)
\geq
\widetilde I_\Psi(R_n;T_n).
\]
Since $\Psi\in\mathscr G$ and $\Psi(0)<\infty$,
Corollary~\ref{cor:G-closed-form} yields
\[
\widetilde I_\Psi(R_n;T_n)
=
I_\Psi(R_n;R_n)
+
I_\Psi(T_n;T_n)
-
I_\Psi(R_nT_n;R_nT_n).
\]
Furthermore,
\[
I_\Psi(R_n;R_n)
=
\frac{\Psi(n)}{n}
+
\left(1-\frac1n\right)\Psi(0)
-\Psi(1),
\]
and
\[
I_\Psi(R_nT_n;R_nT_n)
=
\frac{\Psi(n^2)}{n^2}
+
\left(1-\frac1{n^2}\right)\Psi(0)
-\Psi(1).
\]
Since
\[
\lim_{n\to\infty}\frac{\Psi(n)}{n}
=
\lim_{n\to\infty}\frac{\Psi(n^2)}{n^2}
=
\ell_\Phi,
\]
we obtain
\[
\lim_{n\to\infty}\widetilde I_\Psi(R_n;T_n)
=
\ell_\Phi+\Psi(0)-\Psi(1)
=
K_\Phi.
\]
Therefore,
\[
\widehat I_\Psi(X;Y)\geq K_\Phi.
\]
Combining the two bounds gives
\[
\widehat I_\Psi(X;Y)=K_\Phi,
\]
and hence
\[
\widehat I_\Phi(X;Y)
=
c_\Phi(\ln 2) I(X;Y)+K_\Phi.
\]

(ii) By Lemma \ref{lem:class-G-properties}, we have shown that \(\Psi\in\mathscr G\), and $\Psi(t)=o(t\ln t)$.

Let \(R_n,T_n\) be independent and uniform on \([n]\), and
independent of \((X,Y)\). Then
\[
R_n\to X\to Y\to T_n.
\]
By monotonicity of \(\wIPhi\) under enlarging its arguments,
\[
\widehat I_\Phi(X;Y)
\geq
\wIPhi(R_nX;T_nY)
\geq
\wIPhi(R_n;T_n).
\]
Since \(R_n\independent T_n\), the Shannon mutual information component vanishes, and hence
\[
\wIPhi(R_n;T_n)=\widetilde I_\Psi(R_n;T_n).
\]

Define
\[
a_n
\defeq
I_\Psi(R_n;R_n)
=
\frac{\Psi(n)}{n}
+
\left(1-\frac1n\right)\Psi(0)
-\Psi(1).
\]
Since $\Psi\in\mathscr{G}$, Corollary~\ref{cor:G-closed-form} gives
\[
\widetilde I_\Psi(R_n;T_n)
=
2a_n-a_{n^2}.
\]
Since \(\ell_\Phi=+\infty\), we have
\[
a_n\to+\infty.
\]
We claim that
\[
\sup_{n\geq 1}\left\{2a_n-a_{n^2}\right\}=+\infty.
\]
Since $\Psi(t)=o(t\ln t)$,
\[
a_n=o(\log n).
\]

Suppose, toward a contradiction, that there exists \(M<\infty\)
such that
\[
2a_n-a_{n^2}\leq M
\]
for every \(n\). Then
\begin{align} \label{ineq:a_n}
    a_{n^2}\geq 2a_n-M.
\end{align}
Since $\lim_{n\rightarrow\infty}a_n=\infty$, one can find \(m\) such that \(a_m>M\). Iterating \eqref{ineq:a_n} on $m$, we have 
\[
a_{m^{2^k}}
\geq
2^k a_m-(2^k-1)M
=
2^k(a_m-M)+M.
\]
Therefore,
\[
\frac{a_{m^{2^k}}}{\log(m^{2^k})}
\geq
\frac{a_m-M}{\log m}
+
\frac{M}{2^k\log m}.
\]
Taking \(k\to\infty\), we obtain
\[
\liminf_{k\to\infty}
\frac{a_{m^{2^k}}}{\log(m^{2^k})}
\geq
\frac{a_m-M}{\log m}
>0,
\]
contradicting \(a_n=o(\log n)\). Thus
\[
\sup_{n\geq1}\widetilde I_\Psi(R_n;T_n)=+\infty.
\]

Finally,
\[
\widehat I_\Phi(X;Y)
\geq
\sup_{n\geq1}\widetilde I_\Psi(R_n;T_n)
=+\infty.
\]
Therefore,
\[
\widehat I_\Phi(X;Y)=+\infty.
\]

\section{Proof of Theorem \ref{thm:tlogt-arbitrary-independence-converse}}
\label{app:proof-tlogt-arbitrary-independence-converse}

The following lemma is a consequence of Reed--Solomon codes \cite{reed1960polynomial,huffman2003fundamentals}. The proof is by a probabilistic argument. We remark that there is an explicit construction by the Vandermonde matrix. 

\begin{lemma}\label{lem:ext-lnr-map}
    Let $\cE\subseteq 2^{[n]}$ be nonempty. 
    Let $c_i$, $i\in [n]$ be nonnegative integers and $r\defeq \max_{S\in \cE} \left\{\sum_{i\in S} c_i\right\}$. For a sufficiently large prime number $q$, there exist linear maps $\{L_i:\F_q^r\to \F_q^{c_i}\}_{i\in[n]}$ such that for any $S\in \cE$, the vertical concatenation of $L_i$ for $i\in S$ has full rank, i.e.,
    \[
    L_S:\mathbb F_q^r\to \prod_{i\in S}\mathbb F_q^{c_i},
    \qquad
    L_S(z)=(L_i (z))_{i\in S},
    \]
    has rank
    \[
        \rank(L_S)=\sum_{i\in S}c_i=\min\left(\sum_{i\in S}c_i,r\right).
    \]
\end{lemma}
\begin{proof}
    Consider each $L_i$ as a $c_i\times r$ matrix in $\F_q$. Let all entries of $L_i$ be independently and uniformly distributed over $\F_q$. Given $S\in \cE$, let $t=\sum_{i\in S}c_i$. By definition $t\leq r$. Then 
    \begin{align*}
        \Pr \left[\rank(L_S)<\sum_{i\in S}c_i\right]
        &=1-\prod_{j=0}^{t-1}\left(1-\frac{1}{q^{r-j}}\right)\\
        &\leq 1- \left(1-\frac{1}{q^{r-t+1}}\right)^{t}\\
        &\leq \frac{t}{q^{r-t+1}}\\
        &\leq \frac{r}{q}.
    \end{align*}
    By the union bound, 
    \begin{align*}
        \Pr \left[\exists S\in \cE\,\rank(L_S)<\sum_{i\in S}c_i\right]
        &\leq \frac{|\cE|r}{q}.
    \end{align*}
    Take $q>|\cE|r$. Then there exist $\{L_i:\F_q^r\to \F_q^{c_i}\}_{i\in[n]}$ that satisfy the requirements. 
\end{proof}

\begin{proof}[Proof of Theorem \ref{thm:tlogt-arbitrary-independence-converse}]
    Since $\vec\lambda\notin \Conv(J_{\cE})$ and $\Conv(J_{\cE})$ is a compact
    convex set, by the separating hyperplane theorem, there exists 
    $\vec a\in\mathbb R^n$ such that 
    \begin{align}
        \sum_{i=1}^n a_i\lambda_i
        >
        \max_{S\in\cE}\sum_{i\in S}a_i .
    \end{align}
    We first show that we can replace $a_i$ with nonnegative integers. 
    Let $a_i^+=\max\{a_i,0\}$. Since $\vec \lambda$ is nonnegative, 
    \begin{align*}
        \sum_{i=1}^n a_i^+\lambda_i
        \geq
        \sum_{i=1}^n a_i\lambda_i.
    \end{align*}
We claim that
 \begin{align*}
        \max_{S\in\cE}\sum_{i\in S}a_i^+
        =
        \max_{S\in\cE}\sum_{i\in S}a_i.
    \end{align*}
    
    Since $a_i^+\geq a_i$ for all $i\in [n]$, 
    \begin{align*}
        \max_{S\in\cE}\sum_{i\in S}a_i^+
        \geq
        \max_{S\in\cE}\sum_{i\in S}a_i.
    \end{align*}
    On the other hand, let $T\in \cE$ be a maximizing subset for
    $\max_{S\in\cE}\sum_{i\in S}a_i^+$. Take
    $T'=\{i\in T:a_i\geq 0\}$. By the downward closure of $\cE$, $T'\in \cE$. Then
    \begin{align*}
        \max_{S\in\cE}\sum_{i\in S}a_i^+
        &=\sum_{i\in T}a_i^+\\
        &= \sum_{i\in T'}a_i \\
        &\leq \max_{S\in\cE}\sum_{i\in S}a_i.
    \end{align*}
    Thus
    \[
        \max_{S\in\cE}\sum_{i\in S}a_i^+
        =
        \max_{S\in\cE}\sum_{i\in S}a_i .
    \]
    Combining this with the separating inequality, we obtain
    \begin{align*}
        \sum_{i=1}^n a_i^+\lambda_i
        > 
        \max_{S\in\cE}\sum_{i\in S}a_i^+.
    \end{align*}
    Since the inequality is strict, there exist nonnegative rational numbers $\{b_i\}_{i\in[n]}$ such that 
    \begin{align*}
        \sum_{i=1}^n b_i\lambda_i
        > 
        \max_{S\in\cE}\sum_{i\in S}b_i.
    \end{align*}
    After multiplying by a common denominator, there exist nonnegative integers $\{c_i\}_{i\in[n]}$ such that 
    \begin{align*}
        \sum_{i=1}^n c_i\lambda_i
        > 
        \max_{S\in\cE}\sum_{i\in S}c_i.
    \end{align*}
    Let $r\defeq \max_{S\in\cE}\sum_{i\in S}c_i$. Then
    \[
        \sum_{i=1}^n c_i\lambda_i>r.
    \]

    Now we construct random variables with exactly the independence structure $([n],\cE)$. 
    By Lemma \ref{lem:ext-lnr-map}, given $\cE$ and $c_1,\dots,c_n$, there exist a prime $q$ and linear maps 
    \[
        L_i:\mathbb F_q^r\to \mathbb F_q^{c_i}, 
        \qquad i\in[n],
    \]
    such that for every $S\in\cE$, the combined map
    \[
        L_S:\mathbb F_q^r\to \prod_{i\in S}\mathbb F_q^{c_i},
        \qquad
        L_S(z)=(L_i (z))_{i\in S},
    \]
    has rank
    \[
        \rank(L_S)=\sum_{i\in S}c_i.
    \]

    Let $U$ be uniform on $\mathbb F_q^r$, and define $Y_i=L_i(U)$ for $i\in [n]$. Note that $L_i$ is a surjective map and hence each $Y_i$ is uniform on $\F_q^{c_i}$. Since $L_S$ has rank $\sum_{i\in S}c_i$, it guarantees that for any $S\in \cE$, $Y_S$ are mutually independent. We further construct random variables $X^n$ based on $Y^n$ such that $X_S$ are not mutually independent for $S\notin \cE$. Let 
    \begin{align*}
        \cM=\{T\notin \cE: \forall S\subsetneq T, S\in \cE\}. 
    \end{align*}
    Then $\cM$ consists of the minimal forbidden sets with respect to  $\cE$.  
    For each \(T\in\mathcal M\), let \((B_i^T)_{i\in T}\) be uniformly distributed on \(\{b\in\mathbb F_q^T:\sum_{i\in T}b_i=0\}\). Choose these vectors independently across \(T\), with their entire collection independent of \(U\). Then for any $T'\subsetneq T$, $(B_i^T)_{i\in T'}$ are mutually independent. Define $C_i=(B_i^T:T\in \cM,i\in T)$. 
    For $S\in \cE$, $T\cap S\subsetneq T$ for every $T\in \cM$ and thus $C_S$ are mutually independent. For $S\notin \cE$, there exists $T\in \cM$ such that $T\subseteq S$ and thus $C_S$ are not mutually independent. 
    
    Define $X_i=(Y_i,C_i)$. Then for every $S\in \cE$, $X_S$ are mutually independent.  For every $S\notin \cE$, $X_S$ are not mutually independent since $C_S$ are not. 

    For each $i\in [n]$, by the independence of $C_i$ and $U$,  
    \begin{align*}
        I(U;X_i)&=I(U;Y_iC_i)\\
        &=I(U;Y_i)\\
        &=H(Y_i)\\
        &=c_i\log q.
    \end{align*}
    Moreover, 
    \[
        I(U;X^n)\leq H(U)=r\log q.
    \]
    Therefore
    \[
    \begin{aligned}
        \sum_{i=1}^n \lambda_i I(U;X_i)-I(U;X^n)
        &\ge
        \left(\sum_{i=1}^n c_i\lambda_i-r\right)\log q\\
        &>0.
    \end{aligned}
    \]
    This completes the proof. 
    
\end{proof}

\section{Proof of Theorem \ref{thm:Z-channel}}
\label{app:proof-Z-channel}

Consider nonconstant functions \(f(X)\) such that $0<\HPhi(f)<\infty$. 
Write
\[
u\defeq f(0),
\qquad
v\defeq f(1),
\qquad
m\defeq \E[f]=(1-s)u+sv.
\]
Since \(f\) is nonnegative and nonconstant, \(m>0\). Moreover,
\[
u=\frac{m-sv}{1-s}.
\]

The distribution of \(Y\) is given by
\[
\Pr[Y=0]=1-s(1-\delta),
\qquad
\Pr[Y=1]=s(1-\delta).
\]
Furthermore,
\begin{align*}
\E[f|Y=0]
&=
\frac{(1-s)u+s\delta v}{1-s(1-\delta)}\\
&=
\frac{m-s(1-\delta)v}{1-s(1-\delta)}.
\end{align*}
Since \(Y=1\) implies \(X=1\), we also have
\[
\E[f|Y=1]=v.
\]
Therefore,
\begin{align}
\HPhi(\E[f|Y])
&=
\bigl(1-s(1-\delta)\bigr)
\Phi\left(
\frac{m-s(1-\delta)v}{1-s(1-\delta)}
\right)
\nonumber\\
&\quad
+s(1-\delta)\Phi(v)-\Phi(m),
\label{eq:Z-numerator}
\\
\HPhi(f)
&=
(1-s)\Phi\left(
\frac{m-sv}{1-s}
\right)
+s\Phi(v)-\Phi(m).
\label{eq:Z-denominator}
\end{align}
Consequently,
\[
\frac{\HPhi(\E[f|Y])}{\HPhi(f)}
=
F_m(v).
\]

Suppose first that \(\Phi(0)<\infty\). For every fixed \(m>0\),
the function \(F_m\) extends continuously to \(v=0\). Since \(F_m\)
is nonincreasing,
\[
F_m(v)\leq F_m(0)
\]
for every admissible \(v\).

Define
\[
f_m^*(1)\defeq0,
\qquad
f_m^*(0)\defeq\frac{m}{1-s}.
\]
Then
\[
\E[f_m^*]=m,
\]
and, by \eqref{eq:Z-numerator} and
\eqref{eq:Z-denominator},
\[
F_m(0)
=
\frac{\HPhi(\E[f_m^*|Y])}
     {\HPhi(f_m^*)}.
\]
Thus every admissible function \(f\) can be replaced, without
decreasing the SDPI ratio, by a function satisfying \(f(1)=0\).
Therefore,
\[
\eta_\Phi(X;Y)
=
\sup_{\substack{
f:\{0,1\}\to[0,\infty)\\
f(1)=0,\ \HPhi(f)>0
}}
\frac{\HPhi(\E[f|Y])}{\HPhi(f)}.
\]

Now suppose that
\[
\Phi(0)
\defeq
\lim_{t\downarrow0}\Phi(t)
=
+\infty.
\]
Any function satisfying \(0<\HPhi(f)<\infty\) must then have
\(u,v>0\). Hence
\[
0<v<\frac{m}{s}.
\]
Since \(F_m\) is nonincreasing,
\[
F_m(v)
\leq
\lim_{\epsilon\downarrow0}F_m(\epsilon).
\]

As \(\epsilon\downarrow0\), the arguments
\[
\frac{m-s(1-\delta)\epsilon}{1-s(1-\delta)}
\qquad\text{and}\qquad
\frac{m-s\epsilon}{1-s}
\]
converge to the positive numbers
\[
\frac{m}{1-s(1-\delta)}
\qquad\text{and}\qquad
\frac{m}{1-s},
\]
respectively. Therefore,
\begin{align*}
&
\bigl(1-s(1-\delta)\bigr)
\Phi\left(
\frac{m-s(1-\delta)\epsilon}
     {1-s(1-\delta)}
\right)
+s(1-\delta)\Phi(\epsilon)-\Phi(m)
\\
&\qquad
=
s(1-\delta)\Phi(\epsilon)+O(1),
\end{align*}
whereas
\[
(1-s)\Phi\left(
\frac{m-s\epsilon}{1-s}
\right)
+s\Phi(\epsilon)-\Phi(m)
=
s\Phi(\epsilon)+O(1).
\]
Since \(\Phi(\epsilon)\to+\infty\), it follows that
\[
\lim_{\epsilon\downarrow0}F_m(\epsilon)
=
\frac{s(1-\delta)}{s}
=
1-\delta.
\]
Hence
\[
\eta_\Phi(X;Y)\leq1-\delta.
\]

For the reverse inequality, fix \(m>0\) and, for
\(0<\epsilon<m\), define
\[
f_\epsilon(1)\defeq\epsilon,
\qquad
f_\epsilon(0)
\defeq
\frac{m-s\epsilon}{1-s}.
\]
Then \(f_\epsilon\) is strictly positive and
\[
\E[f_\epsilon]=m.
\]
Moreover,
\[
\frac{\HPhi(\E[f_\epsilon|Y])}
     {\HPhi(f_\epsilon)}
=
F_m(\epsilon)
\to
1-\delta
\]
as \(\epsilon\downarrow0\). Thus
\[
\eta_\Phi(X;Y)\geq1-\delta.
\]
Combining the upper and lower bounds gives
\[
\eta_\Phi(X;Y)
=
\lim_{v\downarrow0}F_m(v)
=
1-\delta.
\]

\section{Proof of Theorem \ref{thm:sufficient-Phi-Z-channel}}
\label{apx:sufficient-Phi-Z-channel}

First suppose that condition (i) holds. We show directly that $F_m$ is nonincreasing.

Let $u=f(0)$ and $v=f(1)$ for some $u,v\geq 0$. Then $m\defeq \E[f]=(1-s)u+sv$. Then we can rewrite $u=\frac{m-sv}{1-s}$. 

For fixed $m>0$, $F_m(v)$ being nonincreasing on $v\in \left(0,\frac{m}{s}\right)$ is equivalent to
\begin{align} \label{ineq:derivative-condition}
    \frac{s(1-\delta)\Phi'(v)-s(1-\delta)\Phi'\left(\frac{m-s(1-\delta)v}{1-s(1-\delta)}\right)}{(1-s(1-\delta))\Phi\left(\frac{m-s(1-\delta)v}{1-s(1-\delta)}\right)+s(1-\delta)\Phi(v)-\Phi(m)}
    \le
    \frac{s\Phi'(v)-s\Phi'\left(\frac{m-sv}{1-s}\right)}{(1-s)\Phi\left(\frac{m-sv}{1-s}\right)+s\Phi(v)-\Phi(m)}. 
\end{align}
It is implied by 
\begin{align*}
    R(t)= \frac{\Phi'(v)-\Phi'\left(\frac{m-svt}{1-st}\right)}{(\frac{1}{st}-1)\Phi\left(\frac{m-svt}{1-st}\right)+\Phi(v)-\frac{1}{st}\Phi(m)}
\end{align*}
being nondecreasing 
for $t\in (0,1)$. Let $D(t)$ denote the above denominator of $R(t)$. Note that we can write $R'(t)=C(t)/D(t)^2$ where  
\begin{align*}
    C(t)
    =& -\frac{1}{st^2}\left(\Phi'(v)-\Phi'\left(\frac{m-svt}{1-st}\right)\right)\left(-\Phi\left(\frac{m-svt}{1-st}\right)+\frac{(m-v)st}{1-st}\Phi'\left(\frac{m-svt}{1-st}\right)+\Phi(m)\right)  \\
    & - \frac{s(m-v)}{(1-st)^2} \Phi''\left(\frac{m-svt}{1-st}\right) \left(\Phi(v)+\frac{1-st}{st}\Phi\left(\frac{m-svt}{1-st}\right)-\frac{1}{st}\Phi(m)\right).
\end{align*}
The condition that $R(t)$ is nondecreasing is equivalent to $C(t)\geq 0$, $\forall t\in (0,1)$. 
Define $x_1=v$, $x_2=\frac{m-svt}{1-st}$. Then we can compute $s$ from $x_1$ and $x_2$ as follows: $$s=\frac{m-x_2}{t(x_1-x_2)}.$$
Then after a change of variables we obtain
\begin{align*}
    C(t) &= \frac{x_1-x_2}{t(m-x_2)} \Bigg(\bigg(\Phi(x_2)+\Phi'(x_2)(m-x_2)-\Phi(m)\bigg) \bigg(\Phi'(x_1)-\Phi'(x_2)\bigg)  \nonumber\\
    &\qquad\qquad\qquad +\Phi''(x_2)(m-x_2)^2 \bigg(\frac{\Phi(x_1)-\Phi(m)}{x_1-m} +\frac{\Phi(x_2)-\Phi(m)}{m-x_2} \bigg) \Bigg).
\end{align*} 
Since $\frac{x_1-x_2}{t(m-x_2)}=1/(st^2)\geq 0$, it suffices to show that 
\begin{align*}
    w_{m,x_2}(x_1)&\defeq\bigg(\Phi(x_2)+\Phi'(x_2)(m-x_2)-\Phi(m)\bigg) \bigg(\Phi'(x_1)-\Phi'(x_2)\bigg)  \\
    &\quad +\Phi''(x_2)(m-x_2)^2 \bigg(\frac{\Phi(x_1)-\Phi(m)}{x_1-m} +\frac{\Phi(x_2)-\Phi(m)}{m-x_2} \bigg) \\
    &\geq 0. 
\end{align*}
Taking the partial derivative with respect to $x_1$, we obtain 
\begin{align*}
    w'_{m,x_2}(x_1) &=\bigg(\Phi(x_2)+\Phi'(x_2)(m-x_2)-\Phi(m)\bigg) \Phi''(x_1) +\Phi''(x_2) \frac{(m-x_2)^2}{(x_1-m)^2} \bigg(\Phi'(x_1)(x_1-m)-\Phi(x_1)+\Phi(m)\bigg).
\end{align*}

Since $m=stx_1+(1-st)x_2$, the following two cases are possible: $x_1\geq m \geq x_2$ or 
$x_2\geq m \geq x_1$. We will show that in both cases, $w_{m,x_2}(x_1)\geq w_{m,x_2}(m)$.

First consider the case of  $x_1\geq m \geq x_2$: by Taylor's expansion we obtain 
\begin{align*}
    \Phi(m) &= \Phi(x_1)+\Phi'(x_1)(m-x_1)+\frac{1}{2}\Phi''(x_1)(m-x_1)^2+\frac{1}{6}\Phi'''(\tilde{x}_1)(m-x_1)^3
\end{align*}
for some $\tilde{x}_1\in [m,x_1]$. And 
\begin{align*}
    \Phi(m) &= \Phi(x_2)+\Phi'(x_2)(m-x_2)+\frac{1}{2}\Phi''(x_2)(m-x_2)^2+\frac{1}{6}\Phi'''(\tilde{x}_2)(m-x_2)^3
\end{align*}
for some $\tilde{x}_2\in [x_2,m]$. Since $\Phi'''\leq 0$, we have 
\begin{align*}
    \Phi'(x_1)(x_1-m)+\Phi(m)-\Phi(x_1) &\geq \frac{1}{2}\Phi''(x_1)(m-x_1)^2 \\
    \Phi(x_2)+\Phi'(x_2)(m-x_2)-\Phi(m)+\frac{1}{2}\Phi''(x_2)(m-x_2)^2 &\geq 0.
\end{align*}
Note that 
\begin{align*}
    w'_{m,x_2}(x_1) &=\bigg(\Phi(x_2)+\Phi'(x_2)(m-x_2)-\Phi(m)\bigg) \Phi''(x_1) +\Phi''(x_2) \frac{(m-x_2)^2}{(x_1-m)^2} \bigg(\Phi'(x_1)(x_1-m)-\Phi(x_1)+\Phi(m)\bigg).
\end{align*}
Thus, 
\begin{align*}
    w'_{m,x_2}(x_1) &\geq \bigg(\Phi(x_2)+\Phi'(x_2)(m-x_2)-\Phi(m)\bigg) \Phi''(x_1) +\Phi''(x_2) \frac{(m-x_2)^2}{(x_1-m)^2}  \bigg(\frac{1}{2}\Phi''(x_1)(m-x_1)^2\bigg)\\
    &= \bigg(\Phi(x_2)+\Phi'(x_2)(m-x_2)-\Phi(m)+\frac{1}{2}\Phi''(x_2)(m-x_2)^2\bigg)\Phi''(x_1)\\
    &\geq 0.
\end{align*}
This shows that $w_{m,x_2}(x_1)\geq w_{m,x_2}(m)$. 

Next, consider the case of $x_2\geq m \geq x_1$. A similar argument shows that $w'_{m,x_2}(x_1)\leq 0$ in this case for $x_1\leq m\leq x_2$. Thus, we also obtain $w_{m,x_2}(x_1)\geq w_{m,x_2}(m)$ in the second case.

Then it remains to prove $w_{m,x_2}(m)\geq 0$. Applying condition \eqref{sup-condition1} with $x=x_2$, $y=m$, we have 
\begin{align*}
    \bigg(\Phi(x_2)+\Phi'(x_2)(m-x_2)-\Phi(m)\bigg)\bigg(\Phi'(m)-\Phi'(x_2)\bigg)+\Phi''(x_2)(m-x_2)^2 \bigg(\Phi'(m)-\frac{\Phi(x_2)-\Phi(m)}{x_2-m}\bigg)\geq 0.
\end{align*}
This is exactly $w_{m,x_2}(m)\geq 0$, which proves the result under condition (i).

Suppose Condition (ii) holds. It suffices to show that Condition (ii) implies (i). Define
\begin{align*}
    Q_x(y)=\bigg(\Phi(x)+\Phi'(x)(y-x)-\Phi(y)\bigg)\bigg(\Phi'(y)-\Phi'(x)\bigg)+\Phi''(x)(y-x)^2 \bigg(\Phi'(y)-\frac{\Phi(x)-\Phi(y)}{x-y}\bigg).
\end{align*}
The partial derivative with respect to $y$ equals
\begin{align*}
    Q_x'(y) &= -\bigg(\Phi'(x)-\Phi'(y)\bigg)^2+\Phi''(y)\bigg(\Phi(x)-\Phi(y)+\Phi'(x)(y-x)\bigg)\\
    &\qquad +\Phi''(x)\bigg(\Phi(x)-\Phi(y)+\Phi'(y)(y-x)+\Phi''(y)(y-x)^2\bigg).
\end{align*}
Since $Q_x(x)=Q'_x(x)=0$, it suffices to show that $Q_x$ is convex in $y$, or equivalently that
\begin{align*}
    Q_x''(y) ={}& 3\Phi''(y)\bigg(\Phi'(x)-\Phi'(y)+\Phi''(x)(y-x)\bigg)\\
    &+\Phi'''(y)\bigg(\Phi(x)-\Phi(y)+\Phi'(x)(y-x)+\Phi''(x)(y-x)^2\bigg)\geq 0.
\end{align*}
Differentiating $Q_x''(y)$ with respect to $x$ gives
\begin{align*}
    (x-y)\Phi'''(x)\Phi'''(y)
    \bigg(-3\frac{\Phi''(y)}{\Phi'''(y)}
    +\frac{\Phi''(x)}{\Phi'''(x)}+x-y\bigg).
\end{align*}

Note that Condition (ii) implies that $\Phi'''\leq 0$ wherever it is nonzero: indeed, setting $x=y$ in \eqref{sup-condition2} gives $\Phi''(x)/\Phi'''(x)\leq 0$ and $\Phi''(x)\geq 0$ by the convexity. It also gives
\begin{align*}
    -3\frac{\Phi''(y)}{\Phi'''(y)}
    +\frac{\Phi''(x)}{\Phi'''(x)}+x-y\geq 0.
\end{align*}
Consequently, $Q_x''(y)$, viewed as a function of $x$, is increasing for $x\geq y$ and decreasing for $x\leq y$. Since $Q_y''(y)=0$, we conclude that $Q_x''(y)\geq 0$ whenever the displayed ratios are defined; the remaining cases follow by continuity. Thus $Q_x(y)\geq 0$, so condition (i) holds and the conclusion follows from the first part of the proof.

Suppose Condition (iii) holds. It suffices to show that Condition (iii) implies (ii). Assume that
\[
1\leq -\frac{t\Phi'''(t)}{\Phi''(t)}\leq 3,
\qquad t>0,
\]
and define
\[
\kappa(t)\defeq
-\frac{t\Phi'''(t)}{\Phi''(t)}.
\]
Then
\[
1\leq \kappa(t)\leq 3
\]
and
\[
\frac{\Phi''(t)}{\Phi'''(t)}
=
-\frac{t}{\kappa(t)}.
\]
Consequently,
\[
x+\frac{\Phi''(x)}{\Phi'''(x)}
=
x\left(1-\frac{1}{\kappa(x)}\right)
\geq 0,
\]
where the inequality follows from \(\kappa(x)\geq 1\). On the other hand,
\[
y+3\frac{\Phi''(y)}{\Phi'''(y)}
=
y\left(1-\frac{3}{\kappa(y)}\right)
\leq 0,
\]
where the inequality follows from \(\kappa(y)\leq 3\). Therefore,
\[
x+\frac{\Phi''(x)}{\Phi'''(x)}
\geq
y+3\frac{\Phi''(y)}{\Phi'''(y)}
\]
for every \(x,y>0\). Hence condition {\rm (iii)} implies
condition {\rm (ii)}.

\end{document}

%% file: defns.tex
\usepackage{amsmath,amsthm,amssymb,mathrsfs,mathtools,bbm}

\usepackage[shortlabels]{enumitem}
\usepackage{color,xcolor,framed,tcolorbox,graphicx}

\theoremstyle{plain}
    \newtheorem{theorem}{Theorem}
    \newtheorem{proposition}{Proposition} 
    \newtheorem{corollary}[theorem]{Corollary}
    
    \newtheorem{lemma}[theorem]{Lemma}

    \newtheorem{fact}{Fact}
\theoremstyle{definition}
    \newtheorem{definition}{Definition}
    \newtheorem{example}{\bf Example}
    \newtheorem{remark}{Remark}

\newcommand{\Conv}{\operatorname{Conv}}

\renewcommand{\vec}[1]{\boldsymbol{#1}}
\newcommand{\Bern}{\operatorname{Bern}}

\newcommand{\HPhi}{H_{\Phi}}

\newcommand{\DPhi}{D_{\Phi}}

\newcommand{\IPhi}{I_{\Phi}}
\newcommand{\wIPhi}{\widetilde{I}_{\Phi}}

\newcommand{\supp}{\operatorname{supp}}
\newcommand{\dom}{\operatorname{dom}}
\newcommand{\Unif}{\operatorname{Unif}}

\newcommand{\E}{\mathbb{E}}
\newcommand{\F}{\mathbb{F}}

\newcommand{\R}{\mathbb{R}}

\newcommand{\fR}{\mathfrak{R}}

\newcommand\independent{\protect\mathpalette{\protect\independenT}{\perp}}
\def\independenT#1#2{\mathrel{\rlap{$#1#2$}\mkern2mu{#1#2}}}
\newcommand{\defeq}{\triangleq}

\newcommand{\Var}{\operatorname{Var}}

\newcommand{\1}{\mathbf{1}}
\makeatletter
\def\rv{\@ifnextchar\bgroup{\rv@i}{\rv@ii}}
\def\rv@i#1{\mathbf{#1}} 
\def\rv@ii#1{\mathbf{#1}} 
\makeatother

\makeatletter
\def\mt{\@ifnextchar\bgroup{\mt@i}{\mt@ii}}
\def\mt@i#1{\boldsymbol{#1}}
\def\mt@ii#1{\boldsymbol{#1}}
\makeatother

\newcommand{\cA}{\mathcal{A}}
\newcommand{\cB}{\mathcal{B}}
\newcommand{\cC}{\mathcal{C}}

\newcommand{\cE}{\mathcal{E}}

\newcommand{\cH}{\mathcal{H}}
\newcommand{\cI}{\mathcal{I}}
\newcommand{\cJ}{\mathcal{J}}

\newcommand{\cM}{\mathcal{M}}

\newcommand{\cP}{\mathcal{P}}

\newcommand{\cV}{\mathcal{V}}

\newcommand{\cX}{\mathcal{X}}
\newcommand{\cY}{\mathcal{Y}}
\newcommand{\cZ}{\mathcal{Z}}

\newcommand{\sB}{\mathsf{B}}
\newcommand{\sJ}{\mathsf{J}}
\newcommand{\sS}{\mathsf{S}}


%% file: refs.bib
@article{ziv1973functionals,
  title={On functionals satisfying a data-processing theorem},
  author={Ziv, Jacob and Zakai, Moshe},
  journal={IEEE Transactions on Information Theory},
  volume={19},
  number={3},
  pages={275--283},
  year={1973},
  publisher={IEEE}
}

@book{cohen1998comparisons,
  title={Comparisons of stochastic matrices with applications in information theory, statistics, economics and population},
  author={Cohen, JOEL and Kempermann, Johannes HB and Zbaganu, Gheorghe},
  year={1998},
  publisher={Springer Science \& Business Media}
}

@incollection{zakai1975generalization,
  title={A generalization of the rate-distortion theory and applications},
  author={Zakai, Moshe and Ziv, Jacob},
  booktitle={Information Theory New Trends and Open Problems},
  pages={87--123},
  year={1975},
  publisher={Springer}
}

@article{tridenski2015ziv,
  title={The Ziv--Zakai--R{\'e}nyi bound for joint source-channel coding},
  author={Tridenski, Sergey and Zamir, Ram and Ingber, Amir},
  journal={IEEE Transactions on Information Theory},
  volume={61},
  number={8},
  pages={4293--4315},
  year={2015},
  publisher={IEEE}
}

@inproceedings{hsu2018generalizing,
  title={Generalizing bottleneck problems},
  author={Hsu, Hsiang and Asoodeh, Shahab and Salamatian, Salman and Calmon, Flavio P},
  booktitle={2018 IEEE International Symposium on Information Theory (ISIT)},
  pages={531--535},
  year={2018},
  organization={IEEE}
}

@article{cohen1993relative,
  title={Relative entropy under mappings by stochastic matrices},
  author={Cohen, Joel E and Iwasa, Yoh and Rautu, Gh and Ruskai, Mary Beth and Seneta, Eugene and Zbaganu, Gh},
  journal={Linear algebra and its applications},
  volume={179},
  pages={211--235},
  year={1993},
  publisher={Elsevier}
}

@inproceedings{polyanskiy2010arimoto,
  title={Arimoto channel coding converse and R{\'e}nyi divergence},
  author={Polyanskiy, Yury and Verd{\'u}, Sergio},
  booktitle={2010 48th Annual Allerton Conference on Communication, Control, and Computing (Allerton)},
  pages={1327--1333},
  year={2010},
  organization={IEEE}
}

@article{mojahedian2019correlation,
  title={A Correlation Measure Based on Vector-Valued $ L\_p $-Norms},
  author={Mojahedian, Mohammad Mahdi and Beigi, Salman and Gohari, Amin and Yassaee, Mohammad Hossein and Aref, Mohammad Reza},
  journal={IEEE Transactions on Information Theory},
  volume={65},
  number={12},
  pages={7985--8004},
  year={2019},
  publisher={IEEE}
}

@article{merhav2011data,
  title={Data processing theorems and the second law of thermodynamics},
  author={Merhav, Neri},
  journal={IEEE transactions on information theory},
  volume={57},
  number={8},
  pages={4926--4939},
  year={2011},
  publisher={IEEE}
}

@article{AbinSubadditiveInformation,
  title   = {A Class of Subadditive Information Measures and their Applications},
  author  = {Abin, Hamidreza and Zinati, Mahdi and Gohari, Amin and Yassaee, Mohammad Hossein and Mojahedian, Mohammad Mahdi},
  journal = {arXiv},
  year    = {},
}

@article{beigi2018phi,
  title={Phi-entropic measures of correlation},
  author={Beigi, Salman and Gohari, Amin},
  journal={IEEE Transactions on Information Theory},
  volume={64},
  number={4},
  pages={2193--2211},
  year={2018},
  publisher={IEEE}
}

@article{raginsky2016strong,
  title={Strong data processing inequalities and Phi-Sobolev inequalities for discrete channels},
  author={Raginsky, Maxim},
  journal={IEEE Transactions on Information Theory},
  volume={62},
  number={6},
  pages={3355--3389},
  year={2016},
  publisher={IEEE}
}

@book{nair2014equivalent,
  title={Equivalent formulations of hypercontractivity using information measures},
  author={Nair, Chandra},
  year={2014},
  publisher={Citeseer}
}

@article{chafai2006binomial,
  title={Binomial-Poisson entropic inequalities and the M/M/$\infty$ queue},
  author={Chafa{\"i}, Djalil},
  journal={ESAIM: Probability and Statistics},
  volume={10},
  pages={317--339},
  year={2006},
  publisher={EDP Sciences}
}

@inproceedings{zhang2002new,
  title={On a new non-Shannon-type information inequality},
  author={Zhang, Zhen and Yang, Jun},
  booktitle={Proceedings IEEE International Symposium on Information Theory,},
  pages={235},
  year={2002},
  organization={IEEE}
}

@article{zhang1998characterization,
  title={On characterization of entropy function via information inequalities},
  author={Zhang, Zhen and Yeung, Raymond W},
  journal={IEEE transactions on information theory},
  volume={44},
  number={4},
  pages={1440--1452},
  year={1998},
  publisher={IEEE}
}

@article{csiszar1967information,
  title={Information-type measures of difference of probability distributions and indirect observation},
  author={Csisz{\'a}r, Imre},
  journal={studia scientiarum Mathematicarum Hungarica},
  volume={2},
  pages={229--318},
  year={1967}
}

@article{ali1966general,
  title={A general class of coefficients of divergence of one distribution from another},
  author={Ali, Syed Mumtaz and Silvey, Samuel D},
  journal={Journal of the Royal Statistical Society: Series B (Methodological)},
  volume={28},
  number={1},
  pages={131--142},
  year={1966},
  publisher={Wiley Online Library}
}

@book{csiszar2011information,
  title={Information theory: coding theorems for discrete memoryless systems},
  author={Csisz{\'a}r, Imre and K{\"o}rner, J{\'a}nos},
  year={2011},
  publisher={Cambridge University Press}
}

@book{Boucheron2013concentration,
author = {Boucheron, Stéphane and Lugosi, Gábor and Massart, Pascal},
year = {2013},
pages = {},
title = {Concentration Inequalities: A Nonasymptotic Theory of Independence},
publisher = {Oxford University Press},
isbn = {9780199535255},
doi = {10.1093/acprof:oso/9780199535255.001.0001}
}

@article{chafai2004entropies,
  title={Entropies, convexity, and functional inequalities, On $\Phi$-entropies and $\Phi$-Sobolev inequalities},
  author={Chafa{\"i}, Djalil},
  journal={Journal of Mathematics of Kyoto University},
  volume={44},
  number={2},
  pages={325--363},
  year={2004},
  publisher={Duke University Press}
}

@book{yeung2012first,
  title={A first course in information theory},
  author={Yeung, Raymond W},
  year={2012},
  publisher={Springer Science \& Business Media}
}

@article{makarychev2002new,
  title={A new class of non-Shannon-type inequalities for entropies},
  author={Makarychev, Konstantin and Makarychev, Yury and Romashchenko, Andrei and Vereshchagin, Nikolai},
  journal={Communications in Information and Systems},
  volume={2},
  number={2},
  pages={147--166},
  year={2002},
  publisher={International Press of Boston}
}

@article{ahlswede1976spreading,
  title={Spreading of sets in product spaces and hypercontraction of the Markov operator},
  author={Ahlswede, Rudolf and G{\'a}cs, Peter},
  journal={The annals of probability},
  pages={925--939},
  year={1976},
  publisher={JSTOR}
}

@inproceedings{lapidoth2018testing,
  title={Testing against independence and a R{\'e}nyi information measure},
  author={Lapidoth, Amos and Pfister, Christoph},
  booktitle={2018 IEEE Information Theory Workshop (ITW)},
  pages={1--5},
  year={2018},
  organization={IEEE}
}

@article{wright2019gleason,
  title={Gleason-type theorems from Cauchy’s functional equation},
  author={Wright, Victoria J and Weigert, Stefan},
  journal={Foundations of Physics},
  volume={49},
  number={6},
  pages={594--606},
  year={2019},
  publisher={Springer}
}

@article{masiha2023f,
  title={f-divergences and their applications in lossy compression and bounding generalization error},
  author={Masiha, Saeed and Gohari, Amin and Yassaee, Mohammad Hossein},
  journal={IEEE Transactions on Information Theory},
  volume={69},
  number={12},
  pages={7538--7564},
  year={2023},
  publisher={IEEE}
}

@article{rudin2021principles,
  title={Principles of mathematical analysis},
  author={Rudin, Walter},
  year={2021}
}

@article{zvarova1974measures,
  title={On measures of statistical dependence},
  author={Zv{\'a}rov{\'a}, Jana},
  journal={{\v{C}}asopis pro p{\v{e}}stov{\'a}n{\'\i} matematiky},
  volume={99},
  number={1},
  pages={15--29},
  year={1974},
  publisher={Mathematical Institute of the Czechoslovak Academy of Sciences}
}

@article{mahdi2018correlation,
  title={A Correlation Measure Based on Vector-Valued $ L\_p $-Norms},
  author={Mahdi Mojahedian, Mohammad and Beigi, Salman and Gohari, Amin and Yassaee, Mohammad Hossein and Aref, Mohammad Reza},
  journal={arXiv e-prints},
  pages={arXiv--1805},
  year={2018}
}

@article{kovavcevic2015entropy,
  title={On the entropy of couplings},
  author={Kova{\v{c}}evi{\'c}, Mladen and Stanojevi{\'c}, Ivan and {\v{S}}enk, Vojin},
  journal={Information and Computation},
  volume={242},
  pages={369--382},
  year={2015},
  publisher={Elsevier}
}

@article{vidyasagar2012metric,
  title={A metric between probability distributions on finite sets of different cardinalities and applications to order reduction},
  author={Vidyasagar, Mathukumalli},
  journal={IEEE Transactions on Automatic Control},
  volume={57},
  number={10},
  pages={2464--2477},
  year={2012},
  publisher={IEEE}
}

@article{liese2006divergences,
  title={On divergences and informations in statistics and information theory},
  author={Liese, Friedrich and Vajda, Igor},
  journal={IEEE Transactions on Information Theory},
  volume={52},
  number={10},
  pages={4394--4412},
  year={2006},
  publisher={IEEE}
}

@book{bauschke2017convex,
  title     = {Convex Analysis and Monotone Operator Theory in Hilbert Spaces},
  author    = {Bauschke, Heinz H. and Combettes, Patrick L.},
  edition   = {2},
  year      = {2017},
  publisher = {Springer},
  address   = {Cham},
  doi       = {10.1007/978-3-319-48311-5}
}

@inproceedings{belghazi2018mutual,
  title={Mutual information neural estimation},
  author={Belghazi, Mohamed Ishmael and Baratin, Aristide and Rajeshwar, Sai and Ozair, Sherjil and Bengio, Yoshua and Courville, Aaron and Hjelm, Devon},
  booktitle={International conference on machine learning},
  pages={531--540},
  year={2018},
  organization={PMLR}
}

@article{reed1960polynomial,
  title={Polynomial codes over certain finite fields},
  author={Reed, Irving S and Solomon, Gustave},
  journal={Journal of the society for industrial and applied mathematics},
  volume={8},
  number={2},
  pages={300--304},
  year={1960},
  publisher={SIAM}
}

@book{huffman2003fundamentals,
  title={Fundamentals of error-correcting codes},
  author={Huffman, W Cary and Pless, Vera S},
  volume={633},
  year={2003},
  publisher={Cambridge university press Cambridge}
}

@article{wangmatrix,
  author = {Wang, Chenyu and Gohari, Amin},
  journal = {Arxiv},
  number = {},
  title = {Matrix $\Phi$-Entropic Measures of Dependence},
  volume = {},
  year = {}
}

@INPROCEEDINGS{wangphient,
  author={Wang, Chenyu and Gohari, Amin},
  booktitle={2025 IEEE International Symposium on Information Theory (ISIT)}, 
  title={On $\Phi$-Entropic Dependence Measures and Non-Local Correlations}, 
  year={2025},
  volume={},
  number={},
  pages={1-5},
  doi={10.1109/ISIT63088.2025.11195250}}

@INPROCEEDINGS{wangribbons,
  author={Wang, Chenyu and Gohari, Amin},
  booktitle={2026 IEEE International Symposium on Information Theory (ISIT)}, 
  title={Ribbons from Independence Structure: Hypercontractivity, $\Phi$-Mutual Information, and Matrix $\Phi$-Entropy}, 
  year={2026},
  volume={},
  number={},
  pages={1-6},
  doi={10.1109/ISIT62367.2026.11653695}}
